\documentclass[aps,onecolumn,nofootinbib,notitlepage,superscriptaddress,preprintnumbers]{revtex4-1}

\usepackage{graphicx}
\usepackage{dcolumn}
\usepackage{bm}
\usepackage{amsmath,amssymb}
\usepackage{float}
\usepackage{multirow}
\usepackage{slashed}
\usepackage{xcolor}
\usepackage{physics}
\usepackage[colorlinks=true, pdfstartview=FitV, bookmarks=true, bookmarksnumbered=true, breaklinks]{hyperref}
\usepackage{mathtools,braket}
\usepackage{soul}
\usepackage{lipsum}  
\usepackage{color}
\definecolor{blue}{rgb}{0.0, 0.0, 1.0}
\definecolor{red}{rgb}{1.0, 0.0, 0.0}
\definecolor{royalblue}{rgb}{0.0, 0.14, 0.4}
\hypersetup{linkcolor=royalblue, citecolor=blue, urlcolor=royalblue}

\usepackage{hyperref}
\hypersetup{colorlinks=true,citecolor=blue,linkcolor=blue,urlcolor=blue}

\usepackage{cleveref}
\usepackage[normalem]{ulem}  

\usepackage{mciteplus}

\newcommand{\nc}{\newcommand}
\nc{\sgn}{\text{sgn}}
\nc{\eff}{\mathrm{eff}}
\nc{\sqM}{\sqrt{M}}
\nc{\NL}{\mathrm{NL}}
\nc{\moc}{\mathcal{M}} 

\allowdisplaybreaks
\usepackage{tikz,xcolor,hyperref}

\newcommand{\Slash}[1]{\ooalign{\hfil/\hfil\crcr$#1$}}

\definecolor{lime}{HTML}{A6CE39}
\DeclareRobustCommand{\orcidicon}{%
	\begin{tikzpicture}
	\draw[lime, fill=lime] (0,0) 
	circle [radius=0.16] 
	node[white] {{\fontfamily{qag}\selectfont \tiny ID}};
	\draw[white, fill=white] (-0.0625,0.095) 
	circle [radius=0.007];
	\end{tikzpicture}
	\hspace{-2mm}
}

\foreach \x in {A, ..., Z}{%
	\expandafter\xdef\csname orcid\x\endcsname{\noexpand\href{https://orcid.org/\csname orcidauthor\x\endcsname}{\noexpand\orcidicon}}
}

\begin{document}
\title{Generalized parton distributions of the kaon and pion within the nonlocal chiral quark model} 

\author{Hyeon-Dong Son\orcidA}
\email[ E-mail: ]{hdson@inha.ac.kr}
\affiliation{Department of Physics,
 Inha University, Incheon 22212, Republic of Korea}
 \author{Parada~T.~P.~Hutauruk\orcidB}
\email[ E-mail: ]{phutauruk@gmail.com}
\affiliation{Department of Physics, Pukyong National University (PKNU), Busan 48513, Korea}
\date{\today}
\begin{abstract}
    In the present study, we explore the properties of generalized parton distributions (GPDs) for the kaon and pion within the framework of the nonlocal chiral quark model (NL$\chi$QM). Valence quark GPDs of the kaon and pion are analyzed with respect to their momentum fraction $x$ and skewness $\xi$ dependencies in the DGLAP and ERBL regions. We observe that the asymmetry of the current quark masses in kaon results in a significant distortion of the quark GPDs in kaon near $\xi=1$, compared to the case of the pion. The quark GPDs of the kaon and pion are evolved to $\mu^2 = 4$ GeV$^2$ and 100 GeV$^2$ by the QCD evolution equation at one-loop order using the \texttt{APFEL++} package. We find that the produced sea quarks and gluons are largely suppressed as $\xi$ becomes nonzero, predominantly confined within the ERBL region. We subsequently examine the polynomiality of the GPDs and numerically obtain the electromagnetic and gravitational form factors of the kaon and pion. For the kaon, gravitational form factor ratios $A_{\bar s/K^+}(0)/A_{s/K^+}(0) = 1.26$ and $D_{\bar s/K^+}(0)/D_{s/K^+}(0) = 1.10$ are reported and compared with results from the recent lattice QCD calculations and other effective models.
\end{abstract}
\maketitle

\section{Introduction}
The pion and kaon are known as the pseudo-Nambu-Goldstone bosons of the spontaneously broken chiral symmetry (SB$\chi$S) of the strong interaction as realized as $SU_f(3)_L\times SU_f(3)_R \to SU_f(3)_V$. Such broken symmetry is understood to be realized by an underlying mechanism of quantum chromodynamics (QCD) at a nonperturbative regime that produces a nonzero fermion spectral density and thus the nonzero quark condensate~\cite{Banks:1979yr}. This scenario effectively provides the quarks with a constituent mass, or, in other words, a dynamical quark mass. On the other hand, the Higgs mechanism yields the current quark masses, which break the chiral symmetry in an explicit manner. These two distinct origins of symmetry breaking contribute to the generation of the hadron masses, depending on their quark contents i.e., heavy or light quark constituents.
Being Nambu-Goldstone bosons, the pion and kaon remain massless in the chiral limit even though the quarks composing them acquire rather large constituent masses via the nonzero chiral quark condensate. Certainly, explicit chiral symmetry breaking plays a crucial role in their mass acquisition owing to the Gell-Mann-Oakes-Renner (GMOR) relation~\cite{Gell-Mann:1968hlm}.

In this context, understanding the internal quark structures of the kaon and pion will provide a hint to scrutinize the role of the explicit chiral symmetry breaking in the nonperturbative aspects of the hadron structure. For instance, the explicit chiral symmetry breaking splits the contribution of the light and strange quark components to the electromagnetic and mechanical properties, such as the charge, mass, pressure, and shear distribution of the kaon. Indeed, the kaon is of great interest since it consists of one valence light-quark with $m_q\simeq 5$ MeV and one valence antistrange quark with $m_s \simeq 100$ MeV, where $m_q$ and $m_s$ are the current quark masses of the light and strange quarks, respectively. By inspecting each quark component of the kaon, we expect that the role of explicit chiral symmetry can be spotlighted in the hadronic properties mentioned above. 

While the electromagnetic properties of a hadron can be studied from the elastic form factors of the photon probe, the mechanical properties of a hadron can be examined from gravitational form factors, accessed indirectly from the generalized parton distributions (GPDs). GPDs are defined by the hadronic matrix elements of QCD bilocal operators along the light cone with nonzero hadron momentum transfer \cite{Muller:1994ses} and unify the description of parton distribution functions (PDFs) and form factors (FFs). Thus, they encode the three-dimensional tomography for the internal structure of hadrons. Remarkably, Lorentz invariance constrains the Mellin moments of a GPD to be finite polynomials in the skewness variable $\xi$ (fraction of longitudinal momentum transfer). This property, known as the polynomiality of GPDs, makes it possible to access the local hadronic matrix elements with QCD operators of arbitrary spin. In this circumstance, the $n=2$ Mellin moments of twist-2 GPDs are related to the matrix elements of the QCD energy-momentum tensor operators. 

Experimentally, the partonic structure of a pseudoscalar meson can 
be probed through the meson-induced Drell-Yan process~\cite{Drell:1970wh,Berger:1979du} and the Sullivan process~\cite{Sullivan:1971kd}. 
The meson GPDs can be accessed via the Sullivan process with 
a real photon emitted, which can be interpreted as a deeply virtual
Compton scattering (DVCS)~\cite{Ji:1996ek} off a virtual meson target 
$\gamma^* \mathcal{M} \to \gamma \mathcal{M}'$~\cite{Amrath:2008vx}.
In the near future, hard exclusive reactions with wide kinematic coverage are expected to be measured at Jefferson Laboratory~\cite{Accardi:2023chb}, Electron-Ion Collider (EIC) at BNL~\cite{Arrington:2021biu}, Electron-Ion Collider in China (EICC)~\cite{Anderle:2021wcy}, J-PARC~\cite{Sawada:2016mao}, and AMBER/COMPASS at CERN~\cite{Adams:2018pwt}. These new experimental results may allow more reliable and robust pion and kaon structures, including GPDs, which will lead us to a deeper understanding of the internal structure of hadron, as discussed in Ref.~\cite{Chavez:2021koz}. 

From the theoretical perspective, since the GPDs are nonperturbative objects, it is rather difficult to compute them directly from QCD. Thus, studies on GPDs using QCD-inspired models that mimic the key properties of QCD (such as S$\chi$SB and confinement) are required.
In the literature, several studies on the pion GPDs have been reported using the Dyson-Schwinger equation (DSE) model~\cite{Mezrag:2022pqk}, the Bethe-Salpeter approach~\cite{Theussl:2002xp}, the Nambu--Jona-Lasinio (NJL) model~\cite{Broniowski:2003rp,Hutauruk:2016sug}, the spectator quark model~\cite{Broniowski:2003rp,Shastry:2023fnc}, the chiral quark models~\cite{Praszalowicz:2001wy,Praszalowicz:2003pr,Hutauruk:2023ccw}, and the BLFQ-NJL model~\cite{Adhikari:2021jrh}. For the kaon structures, the quark distributions were computed within $SU(3)$ NJL model~\cite{Davidson:1994uv,Davidson:2001cc}. Also, a chiral quark model was employed to compute the light-cone distribution amplitudes (DAs)~\cite{Nam:2006au}, and the quark distribution functions~\cite{Nam:2012vm,Hutauruk:2023ccw}, as well as the electromagnetic form factors~\cite{Nam:2007gf}.
In the DSE approach~\cite{Cui:2022bxn,Xu:2023izo,Xing:2023wuk}, the results for the PDFs, electromagnetic and gravitational form factors (EMFFs and GFFs), and DAs were presented. Studies of kaon GPDs have emerged only very recently in the literature and are gaining increasing attention, either mainly focusing on the DGLAP (Dokshitzer-Gribov-Lipatov-Altarelli-Parisi) region of the $x$ variable~\cite{Zhang:2021mtn,Raya:2021zrz,Adhikari:2021jrh} 
or presenting discontinuities along the crossing points~\cite{Xing:2023eed}.
Efforts to understand the partonic structure of the kaon are ongoing from experiments. For instance, Ref~\cite{Bourrely:2023yzi} demonstrated the extraction of the kaon PDFs from meson-induced Drell-Yan and quarkonium production data, by using a statistical model. 
{Along with these studies, lattice QCD results on the pion gravitational form factors were reported in Refs.~\cite{Brommel:2005ee,Brommel:2007zz,Hackett:2023nkr}, as well as the kaon generalized form factors in Ref.~\cite{Delmar:2024vxn}.}
It is worth noting that newer results for the kaon GPDs from the lattice QCD will also be expected in the near future~\cite{Constantinou:2020hdm}. 
 
 Therefore, this study investigates the pion and kaon GPDs within an effective model framework, aiming to gain deeper insight into how explicit chiral symmetry breaking governs their internal structure. {Previously, Praszalowicz and Rosworowski reported studies on pion (generalized) DAs and GPDs using the nonlocal chiral quark model (NL$\chi$QM)~\cite{Praszalowicz:2001wy, Praszalowicz:2003pr}, inspired by the instanton vacuum of QCD at a low energy scale but defined in the Minkowski space-time. Due to the nonlocal quark interaction of the model, the resultant GPDs exhibited continuity at the cross-over line $x=\xi$; however, the nonconservation of the quark vector current led to only approximate preservation of the momentum sum rule.
 Inheriting these features, we extend the NL$\chi$QM to $SU_f$(3) with explicit chiral symmetry breaking by the light and strange current quark masses. We compute the GPDs for the kaon and pion within the model and explore their properties in both DGLAP and ERBL (Efremov-Radyushkin-Brodsky-Lepage) domains, focusing on the role of explicit chiral symmetry breaking. We then evolve the GPDs
 from a nonperturbative model scale $\mu=\mu_0$ to higher values of $\mu$ using one-loop QCD evolution~\cite{Muller:1994ses,Ji:1996nm,Radyushkin:1997ki,Balitsky:1997mj,Radyushkin:1998es,Blumlein:1997pi,Blumlein:1999sc} with the \texttt{APFEL++} numerical package~\cite{Bertone:2022frx} and discuss the effect of the scale evolution.
Additionally, by computing the $n=1$ and $n=2$ Mellin moments of the quark GPDs, we obtain the EMFFs and GFFs of the pion and kaon and discuss their properties. Finally, we compare our results with those of other model predictions~\cite{Hutauruk:2016sug,Adhikari:2021jrh,Raya:2021zrz,Xu:2023izo} 
and available lattice QCD results~\cite{Hackett:2023nkr,Delmar:2024vxn}.}

The present paper is organized as follows: In Sec.~\ref{sec:gpd_def}, we review the general expressions for constructing the twist-2 kaon and pion GPDs. The following section briefly describes the theoretical framework used in this work, NL$\chi$QM. In Sec.~\ref{sec:gpd_nlchqm}, the computation of the kaon and pion GPDs within NL$\chi$QM is described. In Sec.~\ref{sec:results}, we present our numerical results on kaon and pion GPDs. The variables $x$, $\xi$, and $t$ dependencies of the valence quark GPDs are discussed in detail. Also, we discuss the behavior of the GPDs under one-loop QCD evolutions and the characteristics of GFFs of the pion and kaon from 2nd Mellin moments. Finally, Sec.~\ref{sec:sum} is devoted to the summary and conclusion.

\section{Definition of kaon and pion GPDs and their properties} \label{sec:gpd_def}

 The chirality-even twist-2 generalized quark distributions in a charged kaon ($K^+[u\bar{s}]$) can be defined by a Fourier transform of the hadronic matrix element of the quark bilinear operator separated in the light cone distance $\lambda$ as follows,
\begin{align}
\label{eq:gpd_definition_singlet}
    H (x,\xi,t) = \int \frac{d\lambda}{4\pi} \exp \Big[i\lambda xn\cdot {P}\Big] \langle K^{+} (p') \mid \bar{\psi} (-\lambda/2) \Slash{n} \psi (\lambda n/2) \mid K^{+} (p) \rangle.
\end{align}
The variables $x=k^+/ P^+$, $\xi=(p-p')^+/(p+p') =-\Delta^+/ 2P^+$, and $t=-\Delta^2$ represent respectively the longitudinal momentum fraction, skewness parameter, and four-momentum transfer. $P = (p+p')/2$ is the average momentum of the incoming and outgoing hadrons.
The light-cone vector $n$ is introduced, such that $n \cdot P = P^+$ and so forth. Note that the GPD is also dependent on the 
renormalization scale $\mu$ which is suppressed for brevity. 

Based on the expression in Eq.~(\ref{eq:gpd_definition_singlet}), the quark singlet GPDs of the kaon can be separated into its valence $u$ and $\bar s$ parts,
\begin{align}
\label{eq:gpd_definition_quarks}
   \int \frac{d\lambda}{4\pi} \exp \Big[i \lambda x n \cdot P \Big]
    \langle K^+(p') \mid 
    \left\{\begin{array}{c} 
    \bar u(-\lambda n/2) \Slash{n} u(\lambda n/2) \\[8pt]
    \bar s(-\lambda n/2) \Slash{n} s(\lambda n/2)
    \end{array}\right\}
    \mid K^+(p) \rangle 
 =  \left\{\begin{array}{@{\hspace{4pt}}r@{\hspace{4pt}}} 
    2 H^{u/K^+}(x,\xi,t)  \\[8pt]
    2 H^{s/K^+}(x,\xi,t) 
    \end{array}\right\}.
\end{align}
Conventionally, we define 
$H^{\bar s / K^+}(x,\xi,t) \equiv - H^{ s / K^+}(-x,\xi,t)$ 
to reconcile the notation with other works 
\cite{Adhikari:2021jrh,Raya:2021zrz,Zhang:2021mtn}.
Also, note that we multiplied a factor of $2$ to the quark singlet GPDs on the right-hand side of Eq.~(\ref{eq:gpd_definition_quarks}). 

 Due to the Lorentz invariance, $n$-th Mellin moments of the kaon GPDs 
 become an even polynomial in $\xi$ of order $n$ as follows: 
 \begin{align}
     2 \int ^1 _ 0 \;dx x^{n-1} \; H^f (x,\xi,t)
     = \sum^{n}_{m=0,\mathrm{even}} \xi^{m} A^f_{n m} (t).
 \end{align}
 The $t$ dependence of the GPDs Mellin moments takes place in the
 coefficients of $\xi^m$, as the generalized form factors $A^f_{n m} (t)$.
 Note that the polynomiality condition should be satisfied for
 individual quark flavors $f$. 
The simplest examples are given by the $n=1$ Mellin moments, 
\begin{align}\label{eq:n1_Mellin_moments_kaon}
   2 \int^{+1}_{-1} \;dx  \; H^{u/K^+}(x,\xi,t)
    &=A^{u/K^+}_{10}(t), \\ \label{eq:n1_Mellin_moments_kaon_squark}
   2 \int^{+1}_{-1} \;dx \; H^{\bar s/K^+}(x,\xi,t)
    &=A^{\bar s/K^+}_{10}(t),
\end{align}
which are independent of the skewness $\xi$.
The $n=1$ generalized form factors are related to the kaon EMFFs, $F_K(t)$,
by 
\begin{align}\label{eq:EMFF_kaon_n1}
   e_u A^{u/K^+}_{10}(t) + e_{\bar s} A^{\bar s/K^+}_{10}(t) = F_{K^+}(t),
\end{align}
 where $e_u$ and $e_{\bar{s}}$ are the $u$ and antistrange quark charges, respectively. 
The $n=2$ Mellin moments of kaon GPDs are written as
\begin{align}\label{eq:n2_Mellin_moments_kaon}
    2\int^{+1}_{-1} \;dx x \; H^{u/K^+}(x,\xi,t)
    &=A^{u/K^+}_{20}(t) 
    + \xi^2 A^{u/K^+}_{22}(t), \\
    2\int^{+1}_{-1} \;dx x \; H^{\bar s/K^+}(x,\xi,t)
    &=A^{\bar s/K^+}_{20}(t) 
    + \xi^2 A^{\bar s/K^+}_{22}(t).
\end{align}
The $n=2$ generalized form factors $A_{20}(t)$ and 
$A_{22}(t)$ are related to the GFFs, $A(t)$ and $D(t)$ as follows, 
\begin{align}\label{eq:GFFs_kaon_n2}
    A^f_{20}(t) = A^f(t), \qquad    A^f_{22}(t) = D^f(t).
\end{align}
The GFFs $A(t)$ and $D(t)$ are defined by the kaon and pion 
matrix elements of the QCD energy-momentum tensor (EMT) operator. 
Also, $A^f(t=0)$ corresponds to the fraction of longitudinal momentum
carried by a parton $f$ relative to the hadron momentum. Further details about these quantities will be explained in the next section. 

In the forward limit, with $t=0$ and $\xi=0$, the kaon GPDs are governed only by the DGLAP region in $x$ and turn out to be 
the usual singlet quark PDFs:
\begin{align}
    2H^{u/K^+}(x,0,0) &= [u+\bar u]_{K^+}(x), \\
    2H^{\bar s /K^+}(x,0,0) &= [s+\bar s]_{K^+}(x).
\end{align}
Since we consider the isospin symmetric case, we obtain charge conjugation relations for the quark GPDs as follows,
\begin{align}
    H^{u / K^+}(x, \xi, t) &= H^{\bar u / K^-} (x, \xi, t), \\
    H^{\bar s / K^+}(x, \xi, t) &= H^{s / K^-} (x, \xi, t). 
\end{align}

For the pion GPDs in the forward limit, we have relations analogous to the kaon case,
\begin{align}
   2 H^{u/\pi^+}(x,0,0) &= [u+\bar u]_{\pi^+}(x),\\
   2 H^{\bar d / \pi^+}(x,0,0) &= [d+\bar d]_{\pi^+}(x). \\
\end{align}
Also, the charge conjugation relations for the pion can be written as,
\begin{align}
\label{eq:pionPDF}
    H^{\bar d / \pi^+}(x, \xi, t) &= H^{d / \pi^-} (x, \xi, t), \\
    H^{u / \pi^+}(x, \xi, t) &= H^{\bar u / \pi^-} (x, \xi, t), \\ 
    \label{eq:pionPDF2}
    H^{u / \pi^+}(x, \xi, t) &= H^{d / \pi^-} (x, \xi, t), \\
    \label{eq:pionPDF3}
    H^{\bar d / \pi^+}(x, \xi, t)  &= H^{\bar u / \pi^-} (x, \xi, t).
\end{align}
where the last two equations, Eqs.~(\ref{eq:pionPDF2}) and~(\ref{eq:pionPDF3}) are due to the isospin symmetry. 
In addition, the GPDs are even functions of $\xi$ because of 
the time-reversal invariance:
\begin{align}
    H(x,\xi,t) = + H(x,-\xi,t).
\end{align}

\section{Nonlocal chiral quark model}
\label{sec:nlchqm}
The present study is based on the one-loop quark effective action in the large $N_c$ limit, defined in the Minkowski space-time by the following expression:
\begin{align}
\label{eq:action}
    \mathcal{S}_{\mathrm{eff}} = \int \frac{d^4k}{(2\pi)^4}
    \bar \psi(k) (\Slash{k} - \hat m) \psi(k)
    - \int \frac{d^4k}{(2\pi)^4}\frac{d^4p}{(2\pi)^4}
    \bar \psi_f(p) \sqrt{M_f(p)} U_{fg}^{\gamma_5}(p-k)
    \sqrt{M_g(k)}\psi_g(k),
\end{align}
where $\hat{m} = \textrm{diag} (m_u,m_d,m_s)$ is the current quark mass and $f$ and $g$ are the flavor indices. The nonlinear $SU(3)$-flavor chiral field is defined as
\begin{align}
    U^{\gamma_5}_{fg} (x)
    =\exp\left[
        \frac{i}{F_\mathcal{M}}\gamma^5 \lambda^a \mathcal{M}^a
    \right],
\end{align}
which describes the interaction between quarks and the pseudo-Nambu-Goldstone 
bosons. $F_\mathcal{M}$ is the weak decay constant 
of the pseudo-Nambu-Goldstone bosons $\mathcal{M}$. The dynamical quark mass of the light-quark $M_{u,d}=M=350~$ MeV is originally computed by solving the gap equations within the instanton QCD vacuum, with small instanton packing fraction $\bar \rho / \bar R \approx 1/3$, where $\bar \rho \approx 1/3$ fm is the average instanton size and $\bar R \approx 1$ fm is the average distance between the (anti) instantons. 
Also, the momentum dependence of the dynamical quark mass is encoded in the quark form factor $F(k)$. The expression of the dynamical quark mass and the quark form factors in $SU(2)_f$ are respectively given by
\begin{align}
    M(k) &= M F^2(k), \\
    F(k \bar\rho)_{\mathrm{instantons}} &=
    2t \left[
    I_0(t) K_1(t) - I_1(t) K_0(t) - \frac{1}{t}I_1(t)K_1(t) 
    \right]\bigg|_{t=k\bar \rho/2},
    \label{eq:qff_instantons}
\end{align}
which comes from the Fourier transformation of the instanton zero-modes as  discussed in Refs.~\cite{Diakonov:1979nj,Diakonov:1984vw,Diakonov:1985eg,Diakonov:1985fjw}. Note that the QCD instanton vacuum is defined strictly in the Euclidean space. 
In principle, it is possible to make an analytic continuation of the 
effective action from the instanton vacuum from Euclidean to Minkowski space.
However, to avoid complications of the analytic continuation of Eq.~(\ref{eq:qff_instantons}), we adopt the following model for the 
momentum dependence of the dynamical quark mass:
\begin{align}
    M_f(k) &= M_f F^2(k), \\
    F(k) &= \left( \frac{ -\Lambda^2}{ -\Lambda^2 + k^2 + i \epsilon}  \right)^n.
    \label{eq:qff}
\end{align}
As discussed in Ref.~\cite{Praszalowicz:2001wy}, the asymptotic behavior ($k \to \infty$) of the original quark form factor $F(k)$ in Eq.~(\ref{eq:qff_instantons}) reads 
 $F(k) \sim k^{-3}$. Analytic continuation of the half-integer form factor $n=3/2$ in Eq.~(\ref{eq:qff}) from Euclidean to Minkowski space-time is unnecessarily tedious. 
 In Refs.~\cite{Praszalowicz:2001wy,Praszalowicz:2003pr}, the authors 
 investigated the influence of $n$ on the pion distribution amplitudes (DAs), generalized distribution amplitude (GDA), 
 and GPDs by varying $n=1$ to $n=5$. In the GPDs case, it 
 turned out that the results do nott have many characteristic differences.
 
In the first term of the effective action in Eq.~(\ref{eq:action}), we introduced the current quark masses for light and strange quarks as follows 
\begin{align}
    \hat m = \left[\begin{array}{ccc}
    m_u & 0 & 0 \\
    0 & m_d & 0 \\ 
    0 & 0 & m_s
    \end{array}\right].
\end{align}
Throughout this work, we consider the isospin symmetry, $m_u = m_d$, and treat the current mass of the strange quark $m_s$ separately. Similar extensions to $SU(3)$ flavor of the chiral quark models were utilized to study the light-cone distribution amplitudes~\cite{Nam:2006au}, electromagnetic form factors~\cite{Nam:2007gf}, and quark distributions~\cite{Hutauruk:2023ccw, Nam:2012vm} of the pion and kaon.


\section{Kaon GPDs within the Nonlocal Chiral Quark Soliton Model}\label{sec:gpd_nlchqm}
\begin{figure*}
    \centering
    \includegraphics[width=8cm]{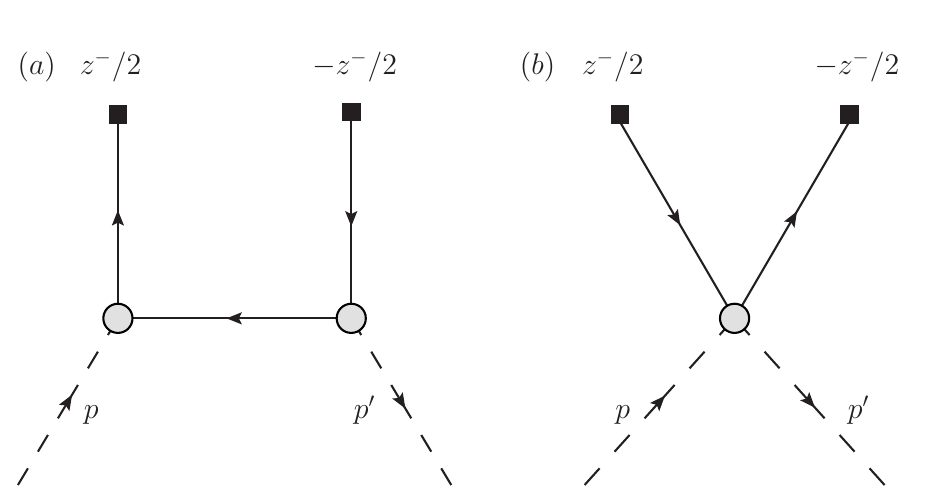}
    \caption{Feynman diagrams for the quark GPDs within the model, Eq.~(\ref{eq:action}). The rectangular points are the positions of the quark operators and gray blobs represent the quark-meson couplings (vertices) which are dependent on the momenta of the connected quark lines. Note that, for the antiquark case, the direction of the internal line should be flipped.}
    \label{fig:gpd-diagrams}
\end{figure*}

In this section, we compute the matrix elements presented in Eq.~(\ref{eq:gpd_definition_quarks}), using the effective quark model in Eq.~(\ref{eq:action}) to obtain the quark GPDs:
\begin{align}
\label{eq:GPD_MTE}
    &\int   \frac{d\lambda}{4\pi}e^{i \lambda x n \cdot P}
 \langle K^+(p') | 
 \bar \psi (-\lambda n/2) \gamma^+ \psi (\lambda /2) 
 | K^+(p) \rangle =  2H^{u/K^+}(x,\xi,t)-2H^{\bar s/K^+}(-x,\xi,t).
\end{align}
In the current study, we work in a Lorentz frame where the transverse components of the average kaon momentum become zero, $\vec {P}_\perp=0$. In this frame, two independent 4-momenta $P$ and $\Delta$ are represented in the light-cone coordinates as follows
\begin{align}
    P&=\left(P^+,\frac{-t+4m_K^2}{4P^+},\vec 0_\perp\right),\\
    \Delta&=\left(-2\xi P^+,\xi \frac{-t+4m_K^2}{2P^+}, \vec{\Delta}_\perp \right),
\end{align}
where the transverse momentum transfer reads,
\begin{align}
    \vec {\Delta}_\perp^2 = (1-\xi^2)(-t+4 m_K^2)-4m_K^2.
\end{align}
The condition $\vec {\Delta}_\perp^2>0$ imposes the following constraint on the skewness variable $\xi$:
\begin{align}
\label{eq:xi_range}
 -\sqrt{\frac{-t}{-t+4m_K^2}} \le \xi \le \sqrt{\frac{-t}{-t+4m_K^2}}.
\end{align}
In this work, we consider only the positive values of $\xi$, since the GPDs are an even function of $\xi$. The matrix elements can be computed straightforwardly by using the functional method.  
We obtain various quark one-loop integrals $\sim \mathcal{O}(N_c)$, leading order contributions in the large $N_c$ limit. In Fig.~\ref{fig:gpd-diagrams}, the diagrammatic representations of the quark matrix elements are shown.
Note that we disregard the subleading meson-loop contributions. 

The valence-quark GPDs for each flavor $u$ and $\bar s$ 
can be written in terms of two contributions 
corresponding to the left ($\mathcal{T}_{(a)}$) and
right ($\mathcal{T}_{(b)}$) Feynman diagrams as shown in Fig.~\ref{fig:gpd-diagrams}. The expressions for the $u$ and $\bar s$ valence-quark GPDs for the kaon are given as
\begin{align}
\label{eq:gpds_loop_integral1}
    H^{u/K^+}(x,\xi,t) &= \frac{i N_c}{8 (2\pi)^4 F_K^2}
    \int d^2 k_\perp \int^\infty_{-\infty} dk^- \left [ 
    \mathcal{T}_{(a)}^{u\bar s} (k-\Delta/2,k-P,k+\Delta/2)
    +\mathcal{T}_{(b)}^{u} (k-\Delta/2,k+\Delta/2)
    \right]\bigg|_{k^+=xP^+}, \\
    -H^{\bar s/K^+}(-x,\xi,t) &=\frac{i N_c}{8 (2\pi)^4 F_K^2}
    \int d^2 k_\perp \int^\infty_{-\infty} dk^- \left [ 
    \mathcal{T}_{(a)}^{\bar s u } (k-\Delta/2,k+P,k+\Delta/2)
    +\mathcal{T}_{(b)}^{\bar s} (k-\Delta/2,k+\Delta/2)
    \right]\bigg|_{k^+=xP^+},
    \label{eq:gpds_loop_integral2}
\end{align}
where, for brevity, the integrands of the $\mathcal{T}_{(a)}^{fg}$ and $\mathcal{T}_{(b)}^{f}$ in Eqs.~(\ref{eq:gpds_loop_integral1}) and~(\ref{eq:gpds_loop_integral2}) are respectively defined as
\begin{align} 
\label{eq:gpds_loop_integrand1}
    \mathcal{T}_{(a)}^{fg} (k_1,k_2,k_3)&=
    \mathrm{Tr}_D \left[
    \gamma^+ \frac{\sqrt{M_f(k_1)}}{\Slash{k}_1 - m_f- M_f (k_1)+i\epsilon}
    \gamma^5
    \frac{\sqrt{M_g(k_2)}}{\Slash{k}_2 - m_g- M_g (k_2)+i\epsilon}
    \gamma^5
    \frac{\sqrt{M_f(k_3)}}{\Slash{k}_3 - m_f- M_f (k_3)+i\epsilon}\right],\\
    \mathcal{T}_{(b)}^f (k_1,k_2)&= \mathrm{Tr}_D \left[
    \gamma^+ \frac{\sqrt{M_f(k_1)}}{\Slash{k}_1 -m_f- M_f(k_1)+i\epsilon}
    \frac{\sqrt{M_f(k_2)}}{\Slash{k}_2 - m_f- M_f (k_2)+i\epsilon}\right].
    \label{eq:gpds_loop_integrand2}
\end{align}

Finally, by evaluating the integrals over the quark loop momenta $k^-$ and $k_\perp$, we obtain the quark GPDs for the kaon, which can be written in terms of three separate terms
\begin{align}\label{eq:gpd_uquark_Is}
    H^{u/K^+}(x, \xi, t) = I_1^{u/K^+}(x,\xi,t) \Theta(x-\xi) 
    + I_2^{u/K^+}(x,\xi,t) \Theta (\xi-|x|)
    +I_3^{u/K^+}(x,\xi,t) \Theta (\xi-|x|), \\ 
    - H^{\bar s /K^+}(-x, \xi, t) = I_1^{\bar s/K^+}(x,\xi,t) \Theta(-x-\xi) 
    + I_2^{\bar s/K^+}(x,\xi,t) \Theta (\xi-|x|)
    +I_3^{\bar s/K^+}(x,\xi,t) \Theta (\xi-|x|), 
    \label{eq:gpd_squark_Is}
\end{align}
where the integral of $\mathcal{T}_{(a)}$ provides functions $I_1$ 
and $I_2$, and integral of $\mathcal{T}_{(b)}$ correspond to $I_3$.
In the above expressions, the functions $I_1$, $I_2$, and $I_3$ are nonvanishing at specified regions by the Heaviside step functions $\Theta$. This is due to the fact that the imaginary positions of the poles in $k^-$ integrals depend on the combinations of $x-\xi$, $x\pm1$, and $x+\xi$. Of course, the functions vanish outside the region $-1< x < 1$. A detailed calculation and full analytic expressions of $I_1$, $I_2$, and $I_3$ can be found in \Cref{apdx:gpds_analytic_results}. Note that $I_1$ and $I_2$, originating from diagrams of type (a) in Fig.~\ref{fig:gpd-diagrams}, contribute to both the ERBL ($|x|\le \xi$) and the DGLAP($x \ge \xi$) regions, while diagram (b), $I_3$, only participates in the ERBL region. It is worth noting that the GPDs become zero in the region $x<-\xi$. $I^f_3$ is always an odd-function of $x$, regardless of its flavor:  $I^f_3(x,\xi,t)=-I^f_3(-x,\xi,t)$.
For the calculation of the pion GPDs, we simply replace the strange current-quark mass $m_s$ in the expressions of the kaon GPDs with $m_d$. 
Due to the isospin-symmetry $m_d=m_u$, we have exact relations
between the quark and antiquark part of $I_1$, $I_2$ and $I_3$,
\begin{align}
    I^u_1(x,\xi,t)&= -  I^{\bar d}_1(-x,\xi,t), \label{eq:pion_ud_relation_I1}\\ 
    I^u_2(x,\xi,t)&= -  I^{\bar d}_2(-x,\xi,t), \label{eq:pion_ud_relation_I2}\\
    I^u_3(x,\xi,t)&=   I^{\bar d}_3(x,\xi,t). \label{eq:pion_ud_relation_I3}
\end{align}
Thus, the expression for the pion GPDs is written as  
\begin{align}
    H^{I=0}(x,\xi,t) &= H^{u/\pi^+}(x,\xi,t) - H^{\bar d/\pi^+} (-x,\xi,t) \cr
    &= I_1^{u/\pi^+}(x,\xi,t) \Theta(x-\xi) - I_1^{u/\pi^+}(-x,\xi,t) \Theta(-x-\xi) \cr
    &+I_2^{u/\pi^+}(x,\xi,t) \Theta (\xi-|x|) - I_2^{u/\pi^+}(-x,\xi,t) \Theta (\xi-|x|) \cr
    &+2 I_3^{u/\pi^+}(x,\xi,t) \Theta (\xi-|x|), \label{eq:pion_gpd_singlet}\\
    H^{I=1}(x,\xi,t) &= H^{u/\pi^+}(x,\xi,t) + H^{\bar d/\pi^+} (-x,\xi,t) \cr
     &=I_1^{u/\pi^+}(x,\xi,t) \Theta(x-\xi) + I_1^{u/\pi^+}(-x,\xi,t) \Theta(-x-\xi) \cr
    &+I_2^{u/\pi^+}(x,\xi,t) \Theta (\xi-|x|) +  I_2^{u/\pi^+}(-x,\xi,t) \Theta (\xi-|x|),
     \label{eq:pion_gpd_isovector}
\end{align}
where $I=0$ and $I=1$ stand for the isosinglet and isovector of GPDs, respectively. 

\section{Numerical result and discussion}
\label{sec:results}
\subsection{Model parameters}\label{sec:model_parameters}
Before presenting the numerical results for the kaon GPDs in the NL$\chi$QM model, we briefly discuss how we determine the model parameters for our numerical calculations. In this work, we take the dynamical quark mass $M=350~$MeV, adapted from the instanton vacuum~\cite{Diakonov:1979nj,Diakonov:1984vw,Diakonov:1985eg,Diakonov:1985fjw}, as done in~\cite{Praszalowicz:2003pr}. For the current mass of the $u$-quark, we choose $m_u=5~$MeV, a typical value used in studies of meson properties of the model. 
Also, the physical masses of the pseudoscalar mesons are taken $m_\pi = 140~$ MeV and $m_K=494~$ MeV, based on the experimental data. For simplicity, we take $n=1$ for the momentum dependence of the dynamical quark mass in Eq.~(\ref{eq:qff}). The numerical results are expected to be mildly dependent on the choice of $n$, as comprehensively discussed in Refs.~\cite{Praszalowicz:2001wy,Praszalowicz:2003pr}.
To fix two parameters $\Lambda$ and $m_s$, we consider two conditions as follows
\begin{align}\label{eq:parameter_c1}
    \int^1_0 \;du \; \phi_{(\pi^+,K^+)}(u) &=1, \\
    F_{(\pi^+,K^+)}(t=0) &=1,
    \label{eq:parameter_c2}
\end{align}
where $\phi_{(\pi^+,K^+)}(u)$ are the light-cone distribution amplitudes (DA) 
of the pion and kaon, and $F_{(\pi^+,K^+)}(t)$ are the EM form factors, 
defined in Eq.~(\ref{eq:EMFF_kaon_n1}).
The analytic results for the meson DAs are given in \Cref{apdx:DAs}. 
By exploring these conditions with the model parameters  $\Lambda$ and $m_s$, we encounter an interesting correlation 
of these two values to simultaneously obtain the correct normalization of the DAs and the electromagnetic form factors. It turns out that one cannot choose an arbitrary value of $m_s$, given a value of $\Lambda$.
 
Note that both the electric charge (through $n=1$ Mellin moment of quark-GPDs at zero momentum 
transfer) and the meson decay constants (squared) are represented as an identical quark-loop integral in the chiral limit. Thus, the logarithmic divergence of the quark loop-momentum integral for the EM charge can be perfectly absorbed into the decay constant in the normalization factor, $1/F^2$, for example, see Refs. \cite{Diakonov:1997vc,Nam:2007gf}. 
However, if the nonzero current quark masses are introduced, 
two integrals receive distinct forms of mass corrections. Due to this, 
one has to be careful when choosing the mass parameters of the model. 
For the pion case, it is pointed out that the parameters should be selected in a way that the produced physical quantities follow the 
GMOR relation (and to absorb additional
higher order mass corrections), to properly compute the pion gravitational form factors \cite{Son:2014sna}. 
A systematic study for such relations of internal model parameters and 
physical quantities, especially for the kaon,
is needed and should be treated in detail elsewhere. 

Here, we simply present the selected values of parameters. 
We varied the values of $\Lambda$ from $1100$ MeV to $1200$ MeV and determined that $m_s\sim130$~MeV to fit both conditions in Eqs.~(\ref{eq:parameter_c1}) and~(\ref{eq:parameter_c2}) simultaneously.
For numerical calculations, we use the parameter set 
$(\Lambda, m_q, m_s)= (1150,5,133)~$MeV and obtained meson decay constants are 
$(F_\pi,F_K) =(94.3,108.6)~$MeV, which are comparable with 
the experimental values, $(F_\pi,F_K) =(93,113)~$MeV.

\subsection{Kaon and pion GPDs }
Here we present our numerical results for the kaon GPDs and the pion GPDs calculated in the NL$\chi$QM model with the choice of the parameter explained above. It is worth noting that the characteristic properties of the quark GPDs for the kaon and pion discussed in this section do not depend much on the choice of parameter sets. 

\subsubsection{Portraits of the quark GPDs in kaon and pion}
\begin{figure}   
    \centering
    \includegraphics[width=8.6cm]{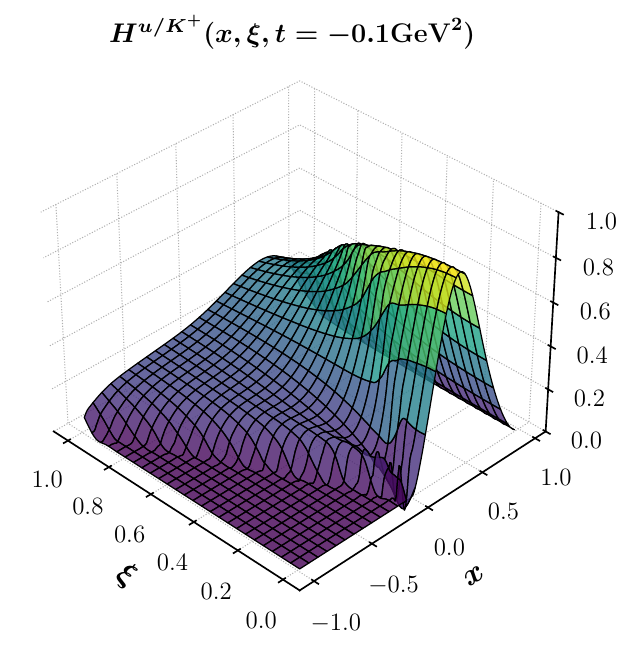}
        \hspace*{0.2 cm}
    \includegraphics[width=8.6cm]{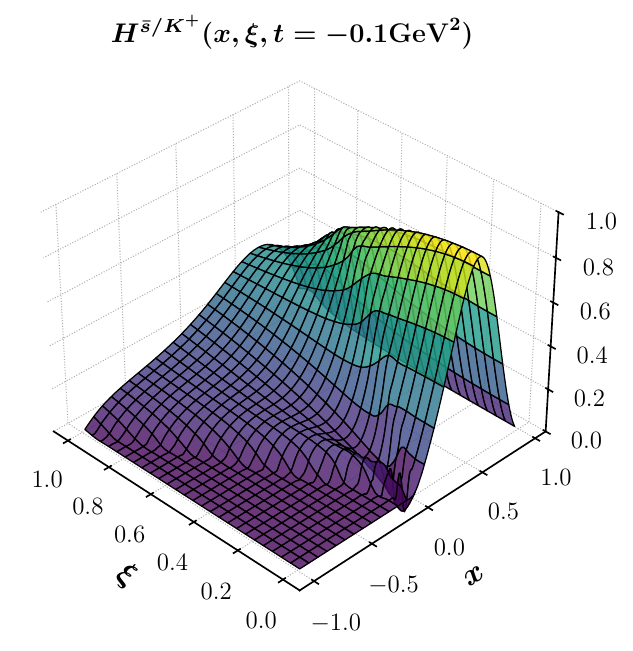}
    \caption{Three-dimensional portrait of $u$-quark (left panel) and $\bar s$-quark (right panel) GPDs for the $K^+$ at a fixed momentum transfer $t=-0.1$GeV$^2$. }
     \label{fig:3d_quark_gpds}
\end{figure}

As an illustration, we plot the $u$-quark and $\bar s$-quark GPDs for the $K^+$ at $t=-0.1$ GeV$^2$ in Fig.~\ref{fig:3d_quark_gpds}. As stated in Eq.~(\ref{eq:xi_range}), the positiveness of the momentum transfer $\Delta_\perp^2 \ge 0$ restricts the physically possible value of $\xi$. However, in the present work, we release the constraint and consider the unphysical range. Besides that, we consider only the positive values of $\xi$, since the GPDs are even functions of the skewness. 
In Fig.~\ref{fig:3d_quark_gpds}, one observes that the $u$- and $\bar {s}$-quark distributions look similar in shape, but they have rather different behavior in $t-$dependencies, where the values of $t$ for $H^{\bar{s}/K}$ are larger than that for $H^{u/K}$, as expected.

\begin{figure}    
    \centering
    \includegraphics[width=7.6cm]{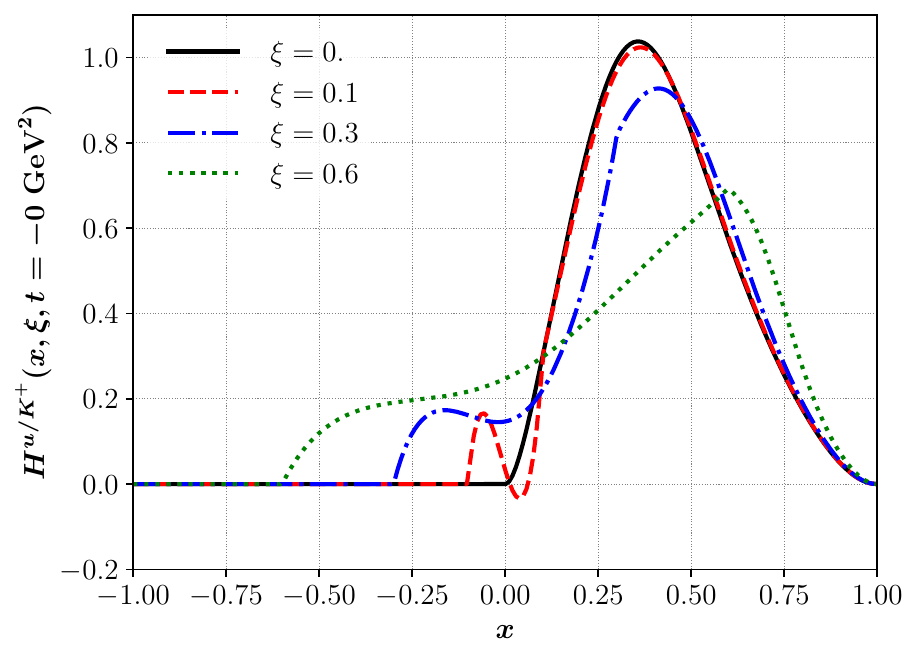}
    \includegraphics[width=7.6cm]{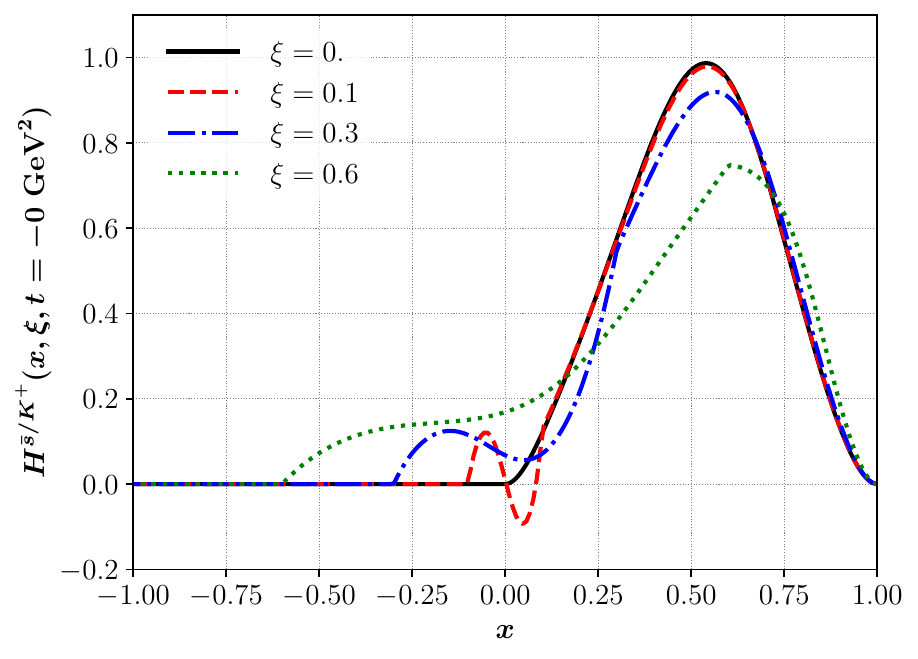}
    \includegraphics[width=7.6cm]{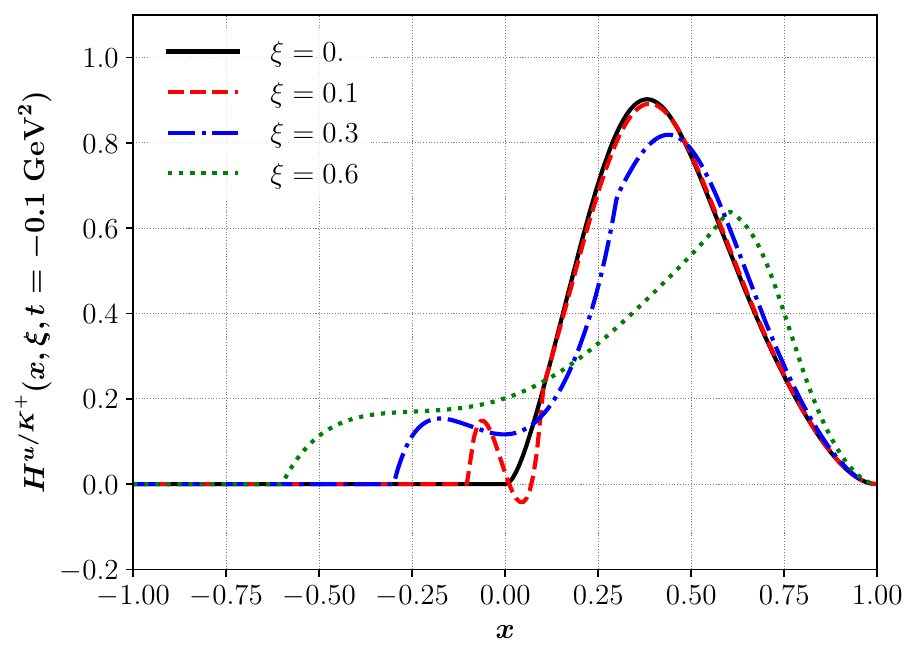}
    \includegraphics[width=7.6cm]{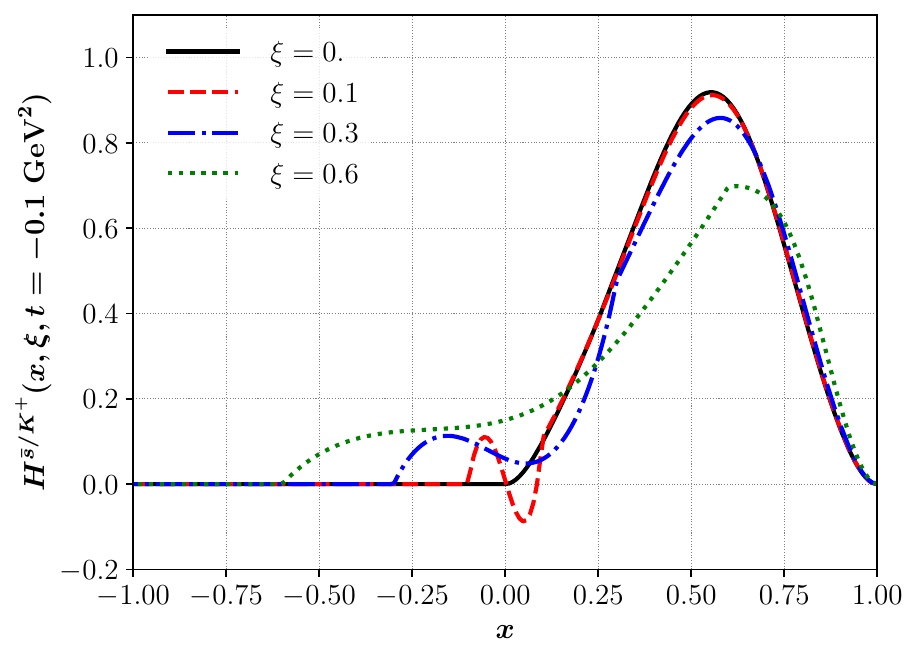}
    \includegraphics[width=7.6cm]{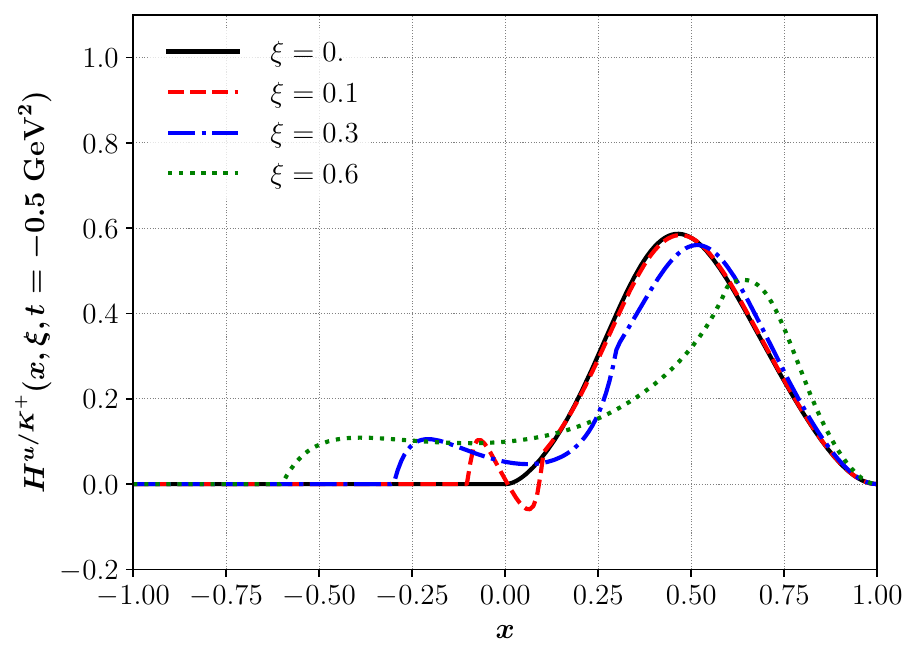}
    \includegraphics[width=7.6cm]{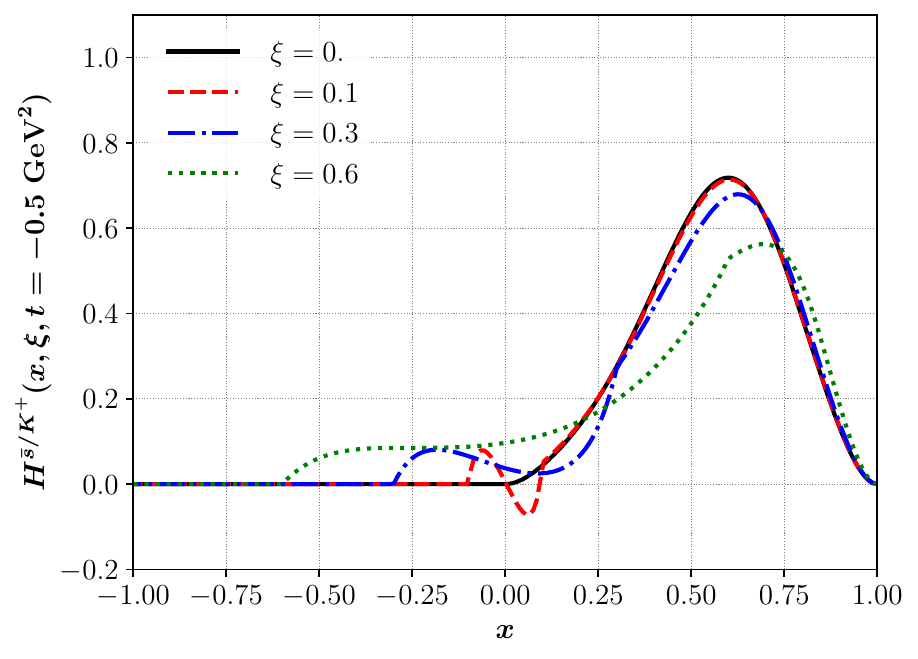}
    \includegraphics[width=7.6cm]{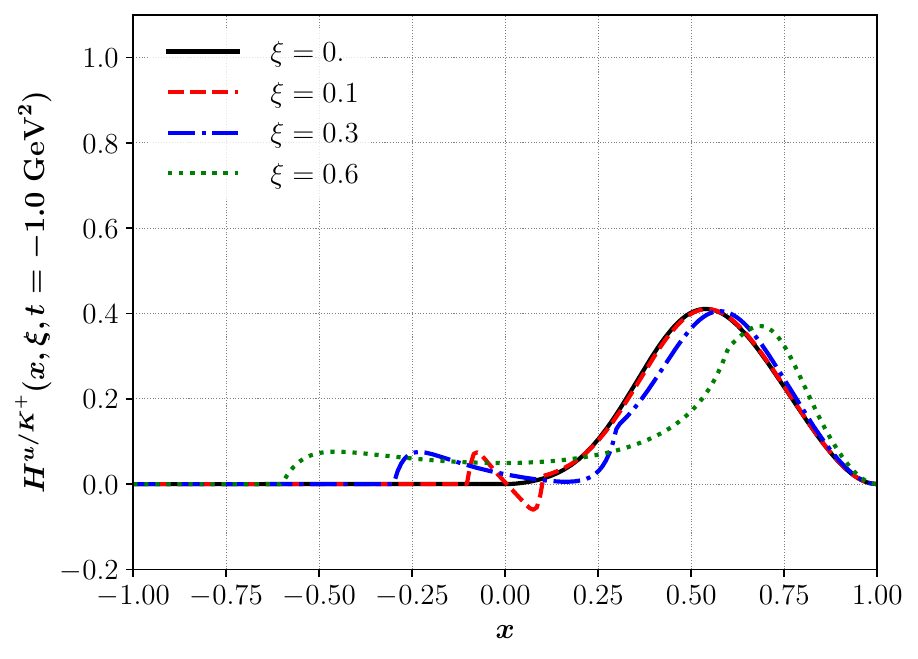}
    \includegraphics[width=7.6cm]{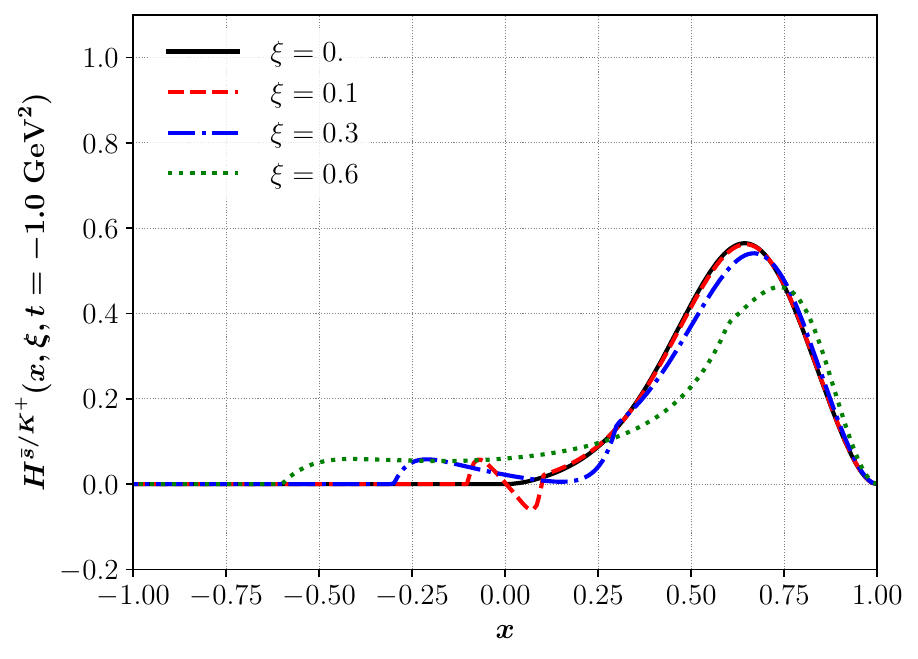}
    \caption{The $u$-quark (left panel) and $\bar s$-quark (right panel)
    GPDs in $K^+$, with momentum transfer values $t=(0,-0.1,-0.5,-1)$ GeV$^2$ (from top to bottom). The GPDs for different $\xi$ values are distinguished by $\xi=0$ (black, solid),
    $\xi=0.1$ (red, long-dashed), $\xi=0.3$ (blue, dot-dashed), and $\xi=0.6$ (green, dashed).}
    \label{fig:2d_quark_gpds}
\end{figure}

In Fig.~\ref{fig:2d_quark_gpds}, we show the quark GPDs of the kaon for $t=0, -0.1, -0.5,$ and $-1~$GeV$^2$ from top to bottom panels. The $u$-quark and $\bar s$-quark GPDs in kaon are shown in the left and right panels of Fig.~\ref{fig:2d_quark_gpds}, respectively. One sees clearly that the $u$-quark distributions decrease much faster than the $\bar s$-quark distributions as $-t$ becomes larger. This difference in tendency on the momentum transfer implies that the characteristic size of $\bar s-$quark is smaller than that of $u$ quark in a charged kaon, $K^+$.
Since the size and the position of the peaks change as $-t$ gets larger, we conclude that the $t$-dependence of the GPDs are nonfactorized from 
$x$ and $\xi$. 
\begin{figure}   
    \centering
    \includegraphics[width=8.6cm]{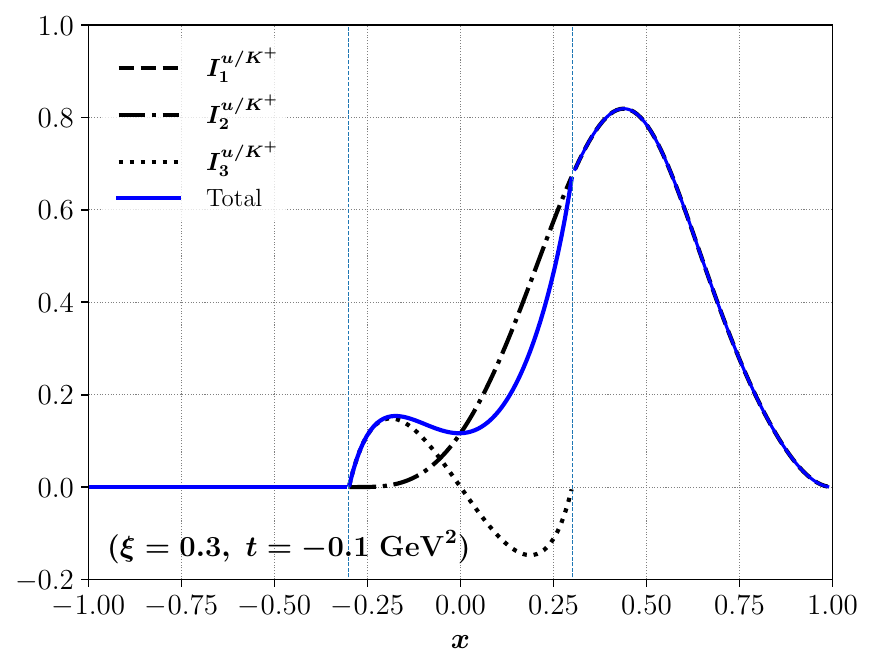}
    \includegraphics[width=8.6cm]{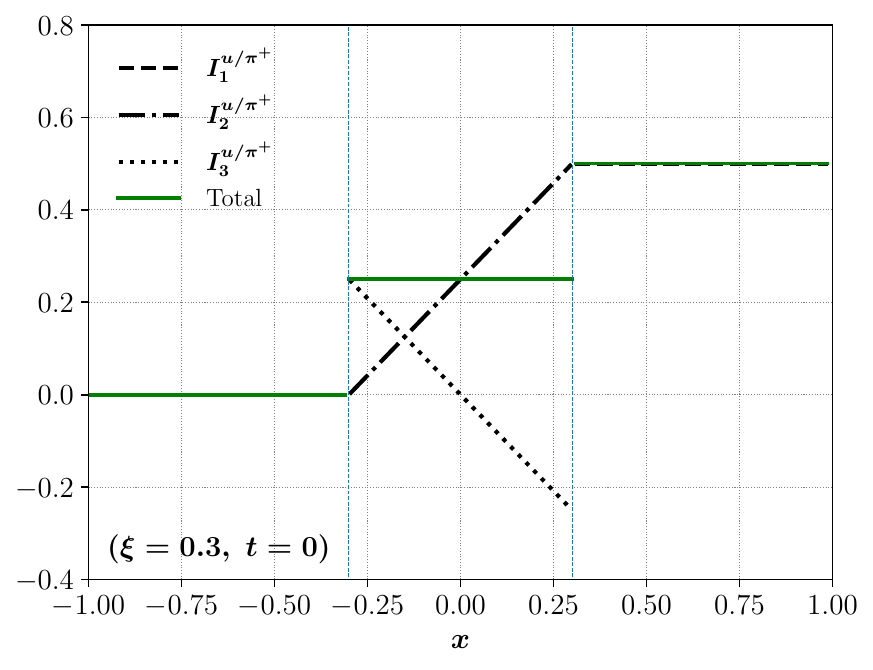}
    \caption{Contributions of functions of $I_1^u$, $I_2^u$, and $I_3^u$ to the quark GPD $H^{u/K^+}(x,\xi=0.3,t=-0.1~$GeV$^2)$ (left panel). 
    Results from Eqs.~(\ref{eq:diagrams_toy_I1})--(\ref{eq:diagrams_toy_I3}) 
    are depicted for comparison (right panel). }
     \label{fig:diagram_contributions}
\end{figure}
It is interesting to notice that the kaon GPDs at negative ERBL region $ -\xi \le x <0 $ develop a shoulderlike structure, as $\xi$ becomes larger. This behavior is mainly due to the contribution of the $I_3$ term, as indicated in Fig.~\ref{fig:diagram_contributions}, where each contribution of $I_1^u$, $I_2^u$, and $I_3^u$ are depicted separately. Here we emphasize again that, among those three terms, only $I_2^u$ and $I_3^u$ terms contribute to the ERBL region. One notices the increment of $I_3^u$ near $x \sim -\xi/2$. 

It is instructive to consider the case where 
the momentum dependence of the dynamical quark mass is turned off, $F(k)=1$
\textit{i.e.} the local chiral quark model (L$\chi$QM).
At the zero-momentum transfer $t=0$, the simple analytical results for the quarks in the massless pion can be obtained ~\cite{Polyakov:1999gs} as\footnote{Note that the definition for $\xi$ in Ref.~\cite{Polyakov:1999gs} has a factor of 2, which is rather different from that in this work. If we take $\xi \to \xi/2 $ and consider the $\bar d-$quark distributions, $I_i^u (x, \xi, t) = - I_i^{\bar d} (-x, \xi, t)$, we obtain the same result for 
the isoscalar GPDs in the pion as shown in Eqs.~(50)-(52) of Ref.~\cite{Polyakov:1999gs}.}
\begin{align}\label{eq:diagrams_toy_I1}
    I_1^u &= \frac{1}{2} \Theta(x-\xi), \\
    \label{eq:diagrams_toy_I2}
    I_2^u &= \left(\frac{x}{4\xi}+\frac{1}{4}\right)\Theta(\xi-|x|), \\
    \label{eq:diagrams_toy_I3}
    I_3^u &=  -\frac{x}{4\xi} \Theta(\xi-|x|),
\end{align}
and are depicted on the right panel of Fig.~\ref{fig:diagram_contributions}.
In Ref.~\cite{Polyakov:1999gs}, it was discussed that the $n=2$ Mellin moment of the isosinglet GPDs for the massless pion vanishes at $\xi=1$ due to the chiral symmetry. In Eqs.~(\ref{eq:diagrams_toy_I1})--(\ref{eq:diagrams_toy_I3}), the cancellation of $I_2^u$ and $I_3^u$ ensures such property $H(x,\xi=1,0)=0$~\cite{Polyakov:1999gs}.
\begin{figure}   
    \centering
    \includegraphics[width=8.5cm]{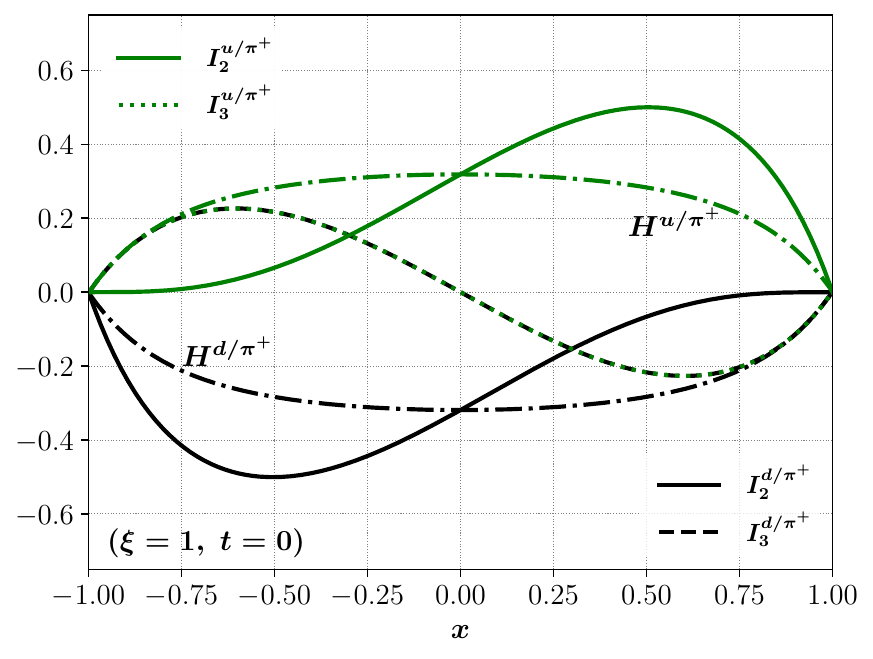}
    \includegraphics[width=8.7cm]{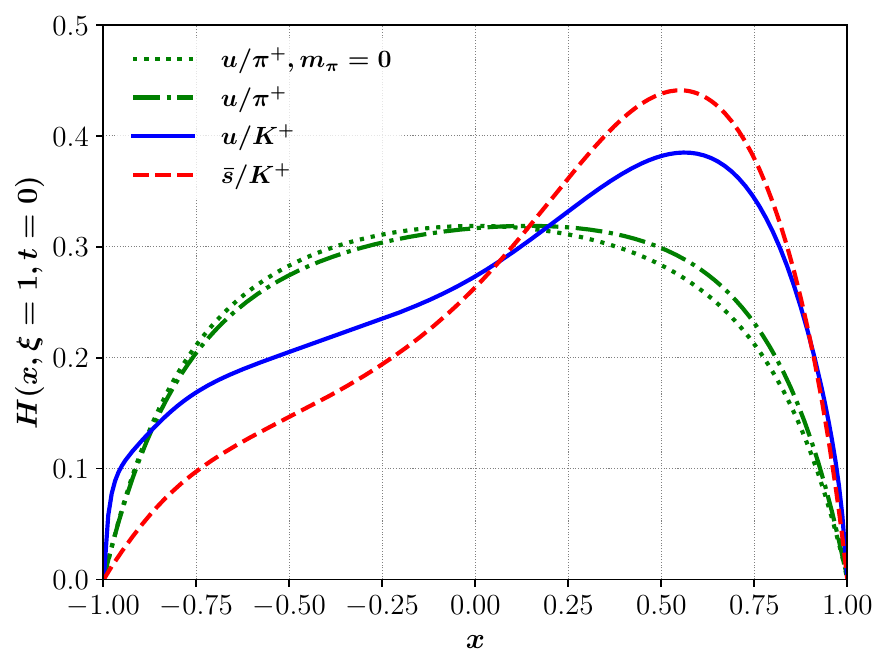}
    \caption{ Pion and kaon GPDs at $\xi=1$ and $t=0~\mathrm{GeV}^2$ are depicted. The left panel shows the contributions of $I_2$ and $I_3$ to the $u$- and $d$- quark GPDs of the pion in the chiral limit. 
    On the right panel, the pion and kaon quark GPDs are compared. With the physical pion mass, the GPD becomes slightly asymmetric (green dot-dashed line). The $u$- and $\bar s$-quark GPDs in kaon are much asymmetric 
    compared to the pion case.
    }
     \label{fig:xi1}
\end{figure}

Figure~\ref{fig:xi1} shows the numerical results of the pion and kaon GPDs at $\xi=1$ and $t=0$. One observes that, in the chiral limit, the 
pion $u-$quark GPD becomes an even function of $x$.
From Eqs.~(\ref{eq:pion_ud_relation_I2}) and (\ref{eq:pion_ud_relation_I3}), we see that $H^{d/\pi^+}(x,1,0) = - I^{u/\pi^+}_2(-x,1,0) + I^{u/\pi^+}_3(x,1,0)$, 
since $I_1(x,1,0)$ vanishes.  
Finally, we observe that the isoscalar GPD of the pion vanishes, $H^{I=0}(x,1,0)=H^{u/\pi^+(x,1,0)}+H^{d/\pi^+(x,1,0)}=H^{u/\pi^+(x,1,0)}-H^{\bar d/\pi^+}(-x,1,0)=0$, in the chiral limit, as shown in the left panel of Fig.~\ref{fig:xi1}.
This ensures the requirement of the chiral symmetry that the second Mellin moment at $\xi=0$ should vanish in the chiral limit, $ \int^1_{-1} dx\;x\; H^{I=0}(x,1,0)=0$ \cite{Polyakov:1999gs}. 
When the physical pion mass is considered, the $u$-quark GPD becomes modestly asymmetric, as displayed by the green dot-dashed curve in the right panel of Fig.~\ref{fig:xi1}. Consequently, both the isoscalar GPD $xH^{I=0}(x,1,0)$ and its Mellin moment have nonzero values.
Note that the asymmetry of the $u$-quark and $\bar s$-quark GPDs in the kaon are more pronounced, which is related to the $\mathcal{O}(m_K^2)$ correction on the second Mellin moment. This feature will be discussed further in the following section, in the context of the GFFs.

Here we remind the readers that the simple model, where the 
momentum dependence of the dynamical quark mass is ignored $F(k) = 1$, results in a discontinuous quark GPD at the cross-over points $x=\pm \xi$, and 
a nonzero value at the end-point $x=\pm 1$ as indicated in the 
right panel of Fig.~\ref{fig:diagram_contributions}. The detailed analytic expressions for the kaon and pion GPDs with constant dynamical quark mass are provided in \Cref{apdx:constant_mass}. Interestingly, the numerical results look similar to those from Ref.~\cite{Xing:2023eed}, although we do not explicitly show them in the present paper.
In the NL$\chi$QM, respecting the momentum dependence of 
the quarks, the quark GPDs are continuous over the entire region of $x$. Still, their first derivatives are discontinuous at the cross-over points $x=\pm \xi$.
This can be seen more clearly by inspecting different contributions of $I_1$, $I_2$, and $I_3$ to the quark distributions. We found that $I_3$ vanishes at the point $x=+ \xi$, and the contributions of $I_1+I_2$ are nonzero and smooth at $x=+\xi$, while at $x=-\xi$, $I_2$ and $I_3$ approach zero in different degrees, as illustrated in the right panel of Fig.~\ref{fig:diagram_contributions}.

\subsubsection{One-loop evolution of the GPDs}
Identifying the valence quark GPDs computed within the NL$\chi$QM 
as the initial state of the QCD evolution at a low renormalization point, we are in a position to evaluate the GPDs at a higher scale. {To achieve that, it is important to determine the reference scale of the model $\mu_0$. For instance, in the instanton model~\cite{Diakonov:1985eg}, the scale is naturally given as the inverse of the average instanton size $\mu_0= 1/\bar \rho \approx 600~$MeV. However, it is difficult to specify the initial scale in a rigorous way within the current framework. One may consider the practical strategy proposed in Ref.~\cite{Davidson:1994uv}, determining the initial scale in a way that the evolved quark distributions reproduce the quark momentum fraction extracted from experimental data. Note that the value found in Ref.~\cite{Davidson:1994uv} was as small as $\mu_0=312$~MeV. This procedure was adopted in various studies. For instance, in Ref.~\cite{Praszalowicz:2003pr}, the authors discuss the one-loop evolution of the pion PDFs by taking $\mu_0=350$ and $\mu_0=450~$MeV. Similarly, in Ref.~\cite{Nam:2012vm,Hutauruk:2023ccw}, NLO QCD evolutions of the pion and kaon PDFs were carried out between $\mu_0=400$ and $\mu_0=500~$MeV. It is worth noting that at such a small initial scale, the effect of the loop corrections on the QCD evolution is rather significant. However, in this work, we do not focus on describing or predicting the measurable quantities: We aim to sketch the qualitative behavior of the valence quark GPDs as well as the produced gluon and sea quarks under the QCD evolution.
In this manner, we present the numerical results with initial scale $\mu_0 =$ 0.33 GeV, where the effect of the one-loop QCD evolution looks dramatic. 
At the end, we discuss the effect of the model initial scale, by varying $\mu_0$ up to 1~GeV.
}

\begin{figure}   
    \centering
    \includegraphics[width=17cm]{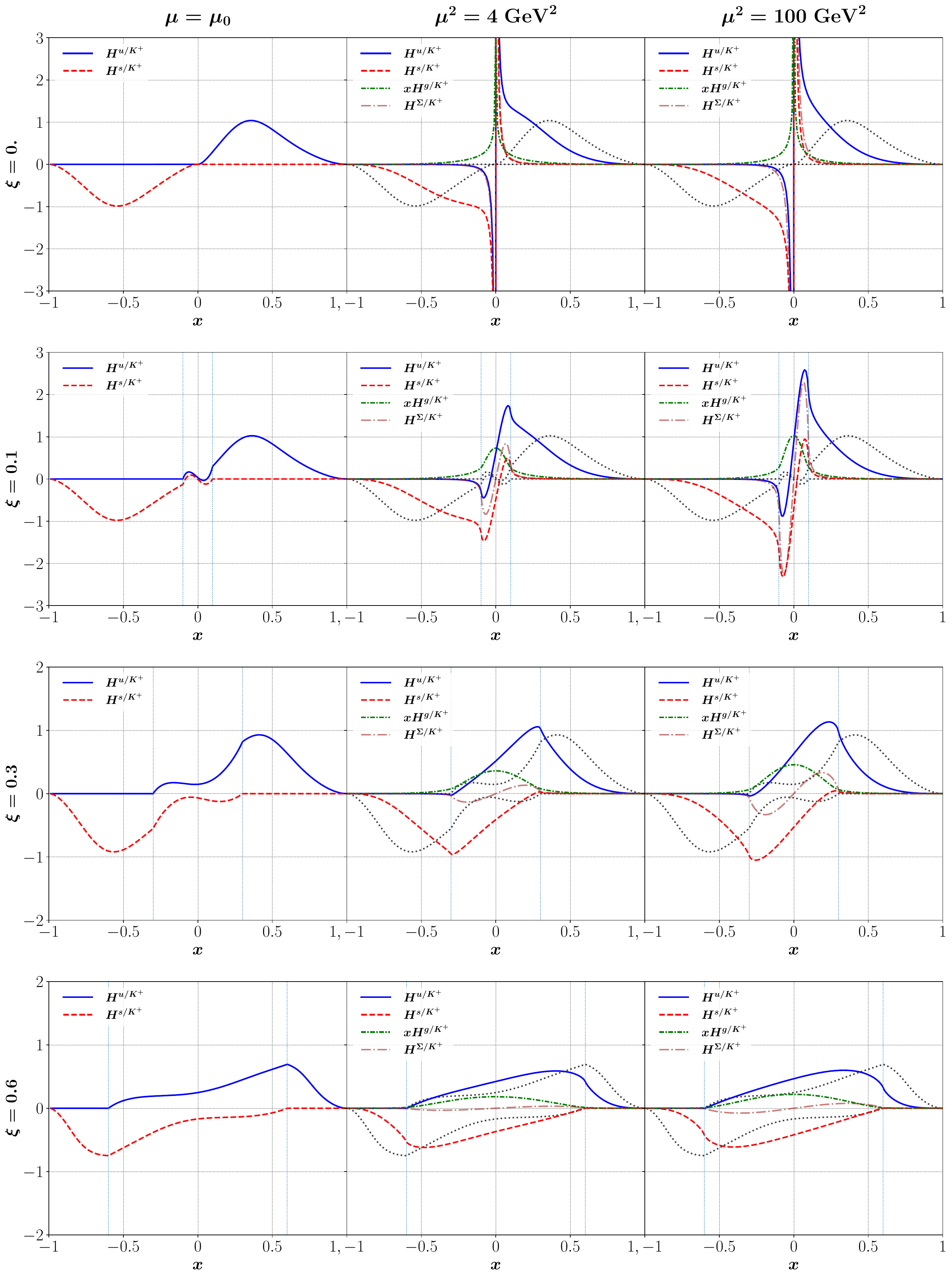}
    \caption{The kaon GPDs after evolving from the initial scale
    $\mu_0 =$ 0.33~GeV (left panel) to $\mu^2 =$ 4 GeV$^2$ (middle panel), and $\mu^2 =$ 100 GeV$^2$ (right panel).}
     \label{fig:kaon_gpds_evolution}
\end{figure}

The perturbative evolution of GPDs at one loop (LO) was studied in Refs.~\cite{Muller:1994ses,Ji:1996nm,Radyushkin:1997ki,Balitsky:1997mj,Radyushkin:1998es,Blumlein:1997pi,Blumlein:1999sc}, and followed by the next leading order (NLO) analyses~\cite{Belitsky:1998gc,Belitsky:1999gu,Belitsky:1999fu,Belitsky:1999hf,Braun:2019qtp}. The numerical techniques for the LO GPD evolution have been developed by authors in Refs.~\cite{Vinnikov:2006xw,Bertone:2022frx} and implemented in \texttt{APFEL++} package in Ref.~\cite{Bertone:2017gds}.
We use these tools to evolve the kaon and pion GPDs computed in the present formalism from the initial model scale $\mu_0 =$ 0.33 GeV to $\mu^2 =$ 4 and 100 GeV$^2$, with the reference strong coupling $\alpha_s(\mu=m_c)=0.35$. 
With this configuration, the valence quarks carry approximately $50\%$ of the pion momentum at $\mu^2=27$ GeV$^2$. The results for the kaon GPDs at $\mu =\mu_0 =$ 0.33 GeV, $\mu^2 = $ 4 GeV$^2$, and $\mu^2 =$ 100 GeV$^2$ for various values of $\xi$ are shown in Fig.~\ref{fig:kaon_gpds_evolution}.

In the left panel of Fig.~\ref{fig:kaon_gpds_evolution}, we show the 
results for the quark GPDs of the kaon $\xi =$ 0, 0.1, 0.3, and 0.6. In the second and third columns, the kaon GPDs at renormalization scales $\mu^2=4~$GeV$^2$ and 
$\mu^2=10~$GeV$^2$ are plotted, respectively. Here, we introduced the notation $\Sigma$ for the summation of all sea quarks: For instance, $\Sigma/ \pi^+$ represents $(\bar u+ d+ (s+ \bar s)+ (c+ \bar c)+ (b+ \bar b))/\pi^+$.
Let us describe the effect of the evolution in a qualitative manner. 
First, in general, the produced sea quarks and gluons are suppressed
and become finite at $x=0$ as $\xi$ grows. Noticeably, the sea-quark GPDs are mostly contained within the ERBL region. 
Second, at small $\xi$, the $u$- and $\bar s$-quark GPDs present 
sharp peaks near $x=\xi$ in the ERBL region. The peaks are tempered as $\xi$ increases.
At larger $\xi$, the shape of the $u-$ and $\bar s$-quark GPDs 
does not change much under evolution and the size of the produced 
sea quarks is suppressed. 
The GPDs are continuous at the crossing points, $x=\pm \xi$, but present kinks as discussed in Ref.~\cite{Bertone:2022frx}. 

\begin{figure}   
    \centering
    \includegraphics[width=17cm]{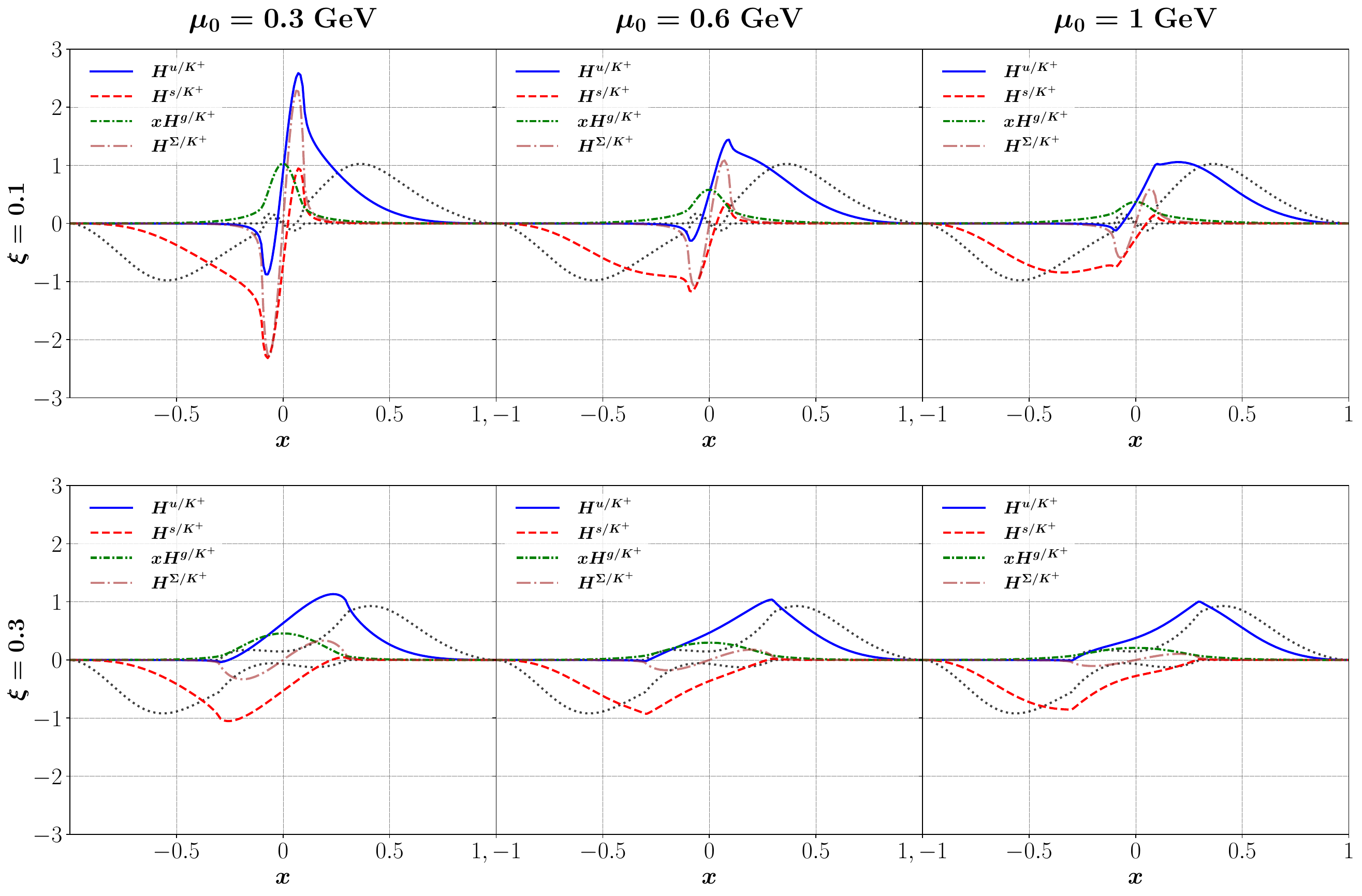}
    \caption{The kaon GPDs after being evolved to $\mu^2=$100 GeV$^2$ from the initial scales
    $\mu_0 =$ 0.33~GeV (left panel) to $\mu_0 =$ 0.6 GeV (middle panel), and $\mu_0 =$ 1 GeV (right panel).}
     \label{fig:kaon_gpds_evolution_mu0}
\end{figure}

Finally, in Fig~\ref{fig:kaon_gpds_evolution_mu0}, we present the evolved kaon GPDs to $\mu^2=$100 GeV$^2$, by varing the initial scale $\mu_0=(0.33,0.6,1.0)$ GeV (from left to right) for $\xi=0.1$ (first row) and $\xi=0.3$ (second row).
For $\xi=0.1$, considerable deviation between the results with $\mu=0.3$~GeV
and $\mu=0.6$~GeV is observed.  
Indeed, the impact of the initial scale turns out to be significant,
 especially at small $\xi$ values which necessitate the NLO analysis. 

 The scale evolutions of the pion GPDs also present similar behavior 
as the kaon one and we do not present them here for brevity.

\subsubsection{Envelope functions and forward limit}
In the DVCS process, the Compton form factors (CFFs) are directly extracted from the crosssection. At the leading order approximation in the strong coupling $\alpha_s$, the CFFs are written as the convolution of the GPDs and the perturbative quark and antiquark lines in $x$, 
\begin{align}
    \mathcal{H}^{a}(\xi,t) = e_a^2 \int^1_{-1} dx H^a(x, \xi, t) 
    \left(
    \frac{1}{-\xi+x - i\epsilon}-\frac{1}{+\xi+x - i\epsilon}
    \right),
\end{align}
where $a$ stands for quarks, $a=u, d, s, ... $.
Therefore, directly from the experimental measurement, only the 
GPDs along the cross-over line $x=\xi$ are accessible 
through the imaginary part of the CFFs. Note that the 
accessible flavor structures depend on the type of the hard exclusive reactions 
and target hadrons. 

\begin{figure}   
    \centering
    \includegraphics[width=6.1cm]{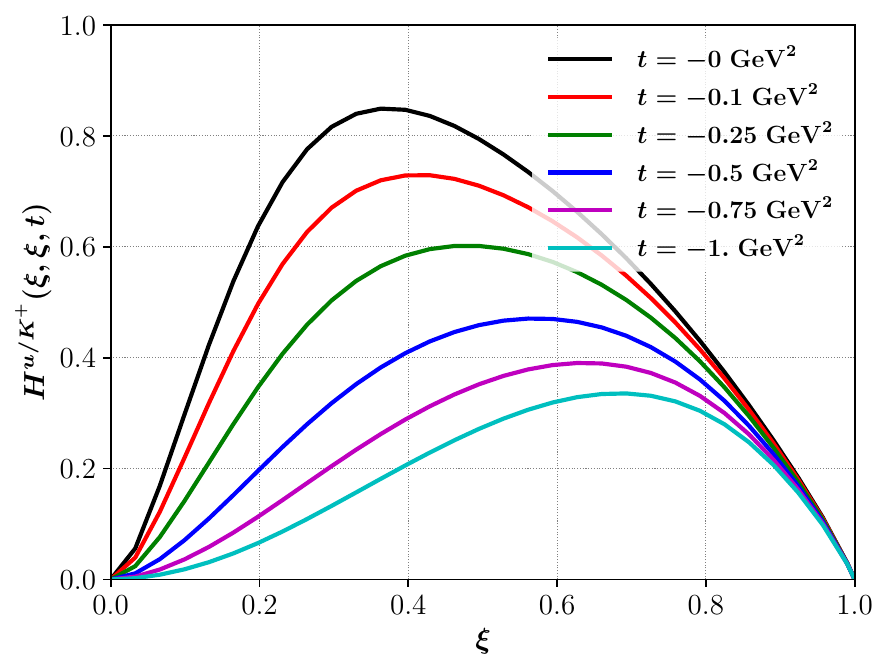}
    \includegraphics[width=5.9cm]{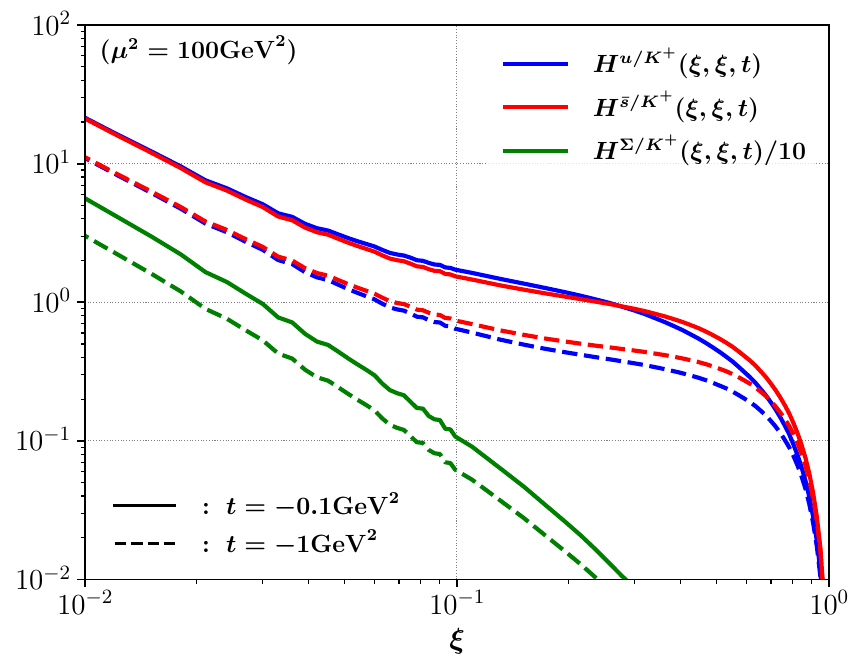}
    \includegraphics[width=5.6cm]{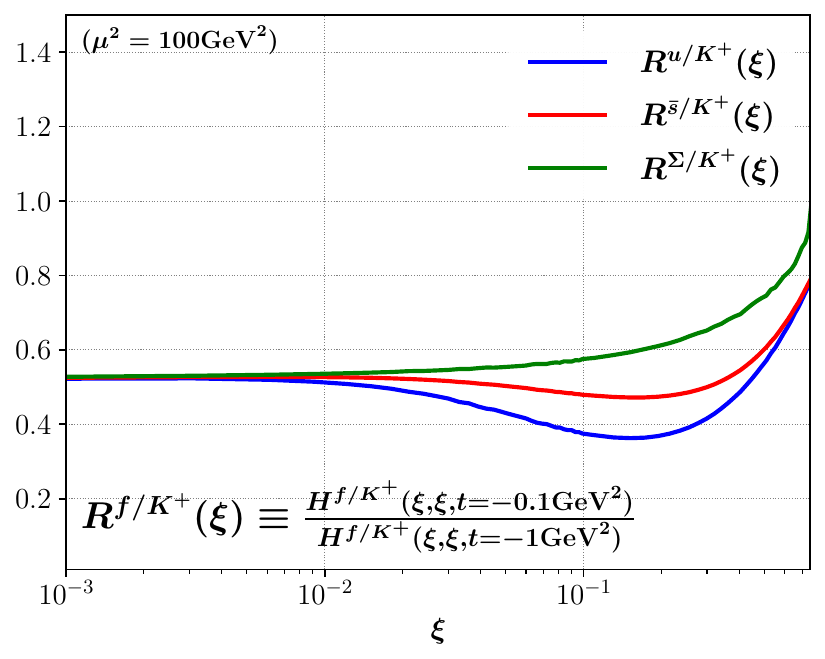}
    \caption{Quark GPDs of the kaon along the cross-over line $x=\xi$. The left panel shows the $t$ dependencies of the $u$-quark GPD of the kaon, varying its value from $0$ to $-1~$GeV$^2$ (curves from top to bottom), evaluated at the model scale $\mu=\mu_0$. The middle panel presents the results of the LO evolution, for $\mu = 100~$GeV$^2$ with $t=-0.1~$GeV$^2$ (solid lines) and $t=-1~$GeV$^2$ (dashed lines). In the right panel, the ratio of the GPDs at different $t$ values is depicted, $R^{f/K^+}(\xi) \equiv H^{f/K^+}(\xi,\xi,t=-0.1~\mathrm{GeV}^2)/H^{f/K^+}(\xi,\xi,t=-1~\mathrm{GeV}^2)$.  }
     \label{fig:cross}
\end{figure}

In Fig.~\ref{fig:cross}, we graphically demonstrate how the $u$-quark GPDs in kaon behave along the cross-over line $x=\xi$ with varying the momentum transfer from $t=0$ to $t=-1$GeV$^2$. Since the model does not exhibit the sea quarks, all GPDs become zero at $x=\xi=0$, as shown in the left panel of Fig.~\ref{fig:cross}. Other distributions such as $\bar s$-quark GPDs in kaon or the pion GPDs have qualitatively the same behavior.  
Authors of Ref.~\cite{Shastry:2023fnc} reported that the LO QCD evolution of the valence quark GPDs from the initial model scale to $\mu=2~$GeV strongly affects the $\xi$ dependence of the CFFs. In this study, we performed the LO evolution of the envelope functions $H(\xi,\xi,t)$ from the model scale $\mu = 0.33~$GeV to $\mu^2 = 100~$GeV$^2$, which are shown graphically in the middle panel of Fig.~\ref{fig:cross}. The envelope functions at $t=-0.1~$GeV$^2$ and $t=-1~$GeV$^2$ are represented in solid and dashed lines, respectively. 
In the right panel of Fig.~\ref{fig:cross}, we depicted the ratio $R^{f/K^+}(\xi) \equiv H^{f/K^+}(\xi,\xi,t=-0.1~\mathrm{GeV}^2)/H^{f/K^+}(\xi,\xi,t=-1~\mathrm{GeV}^2)$.
The results indicate that the $t$-dependencies of the GPDs are nonfactorized 
for $\xi \approxeq 0.01$.

\begin{figure}    
    \centering
    \includegraphics[width=7.3cm]{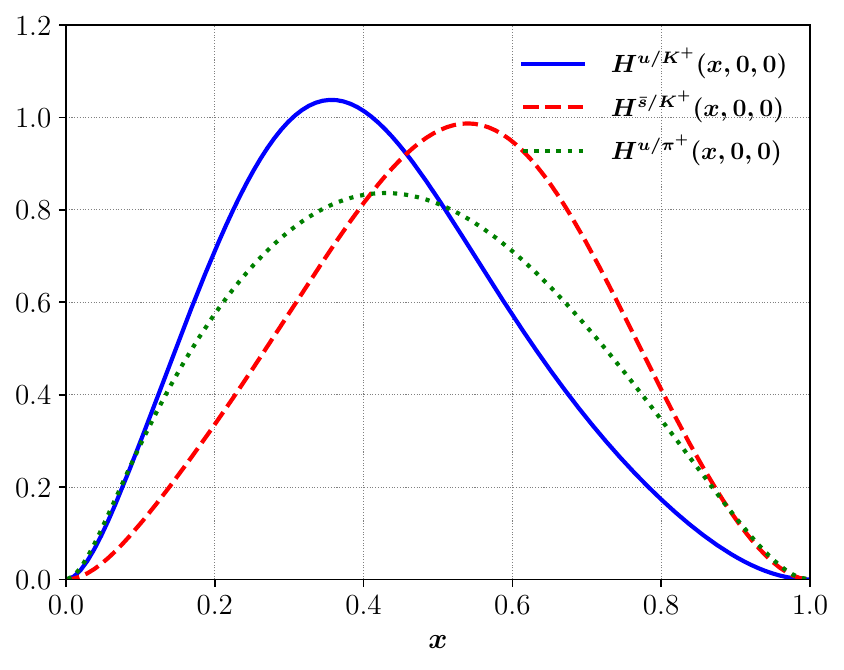}
    \includegraphics[width=7.3cm]{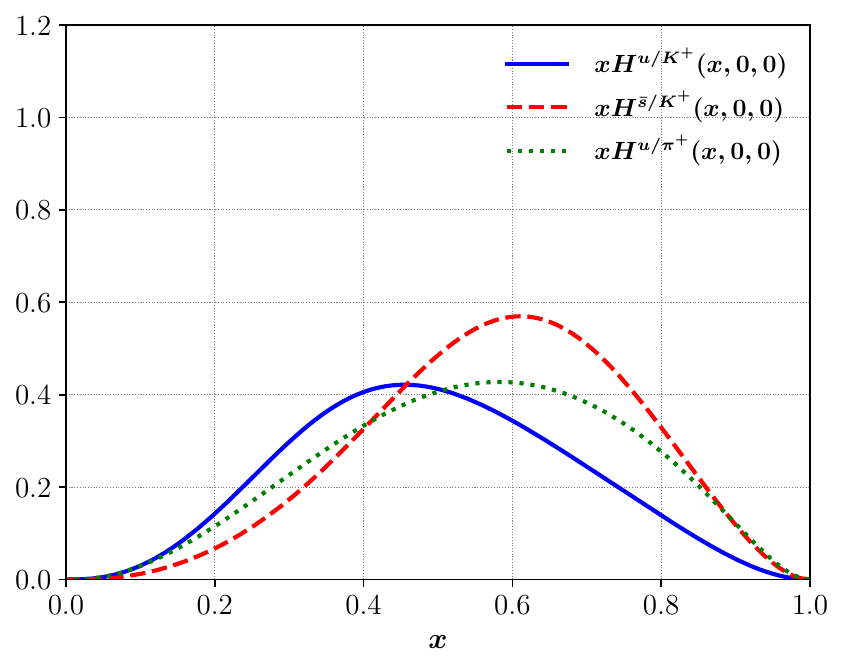}
    \caption{Quark GPDs in the forward limit ($t=0$ and $\xi=0$) evaluated at the model scale $\mu=\mu_0$, which correspond to the valence quark distribution functions. 
    The $u$-quark and $\bar s$-quark 
    PDFs in $K^+$ are shown in the blue solid and red long-dashed lines, respectively. For comparison, the valence $u$-quark distributions in $\pi^+$ are represented by the green dashed line. On the right panel, the PDFs are weighted by $x$.}
\label{fig:valence_quark_pdfs}
\end{figure}
In Fig.~\ref{fig:valence_quark_pdfs}, the quark GPDs in $K^+$ (as well as the $\pi^+$ GPDs for comparison) are plotted in the forward limit $-t=0~$GeV$^2$ and $\xi=0$. In this limit, the GPDs become the valence quark distributions. In Fig.~\ref{fig:valence_quark_pdfs}, the valence $u$-quark, $\bar s$-quark distributions in $K^+$, and 
$u$-quark distribution in $\pi^+$ are presented in blue solid, red dashed, and green dotted lines, respectively. Also, the corresponding $x$-weighted quark distributions $x f(x)$ are shown in the right panel of Fig.~\ref{fig:valence_quark_pdfs}. 
The integral of the right curves corresponds to the momentum fraction, namely $M_2$, of each valence quark in the pion and kaon, which is expressed by
\begin{align}
    M_2^{u/K^+} = 2\int^1_0 \; dx \; x H^{u/K^+}(x,0,0),\\
    M_2^{\bar s/K^+} = 2\int^1_0 \; dx \; x H^{\bar s/K^+}(x,0,0),\\
    M_2^{u/\pi^+} =2 \int^1_0 \; dx \; x H^{u/ \pi^+}(x,0,0).
\end{align}
Note that, in general, the QCD momentum sum rule reads
$\sum_{a=q,\bar q, g} M^a_2 = 1$, where $a$ denotes quarks and gluon ($a=u, d, s, ..., g$). 
On the other hand, in the current model, the quarks are recognized as constituent quarks,  
which are the only effective degrees of freedom. Thus, the momentum sum rule must be 
satisfied by these quarks. Since the quark-meson interactions of the model 
are nonlocal, Eq.~(\ref{eq:action}), the gauge invariance requires 
additional terms for the vector and axial currents as explored in previous studies~\cite{Broniowski:1999bz,Dorokhov:2000gu,Kim:2004hd}, followed by investigations of hadronic matrix elements~\cite{Nam:2006au,Nam:2007gf,Nam:2012vm}. Instead, we have chosen to simplify our approach by fixing the model-cutoff parameter $\Lambda$ to achieve the correct normalization of meson distribution amplitudes and the vector form factors, as explained in Sec. \ref{sec:model_parameters}. {Consequently, the momentum sum rule is slightly broken 
and we obtain $M_2^{u/K^+}\sim 0.41$, $M_2^{\bar s /K^+} \sim 0.52$, and $M_2^{K^+}\sim 0.93$. For the pion, we obtain $M_2^{\pi^+}\sim 0.93$.}  We would like to note that the result of the current work can be compared with experimental data for the ratio of the $u$-quark distributions
 in kaon and pion \cite{Saclay-CERN-CollegedeFrance-EcolePoly-Orsay:1980fhh}. 
 We observed that the current model underestimates the ratio for $x \gtrsim 0.4$, as reported already in Ref.~\cite{Hutauruk:2023ccw}.

\subsection{EMFFs and GFFs of the kaon and pion}\label{sec:FFs}
As presented in Sec. \ref{sec:gpd_def}, $n=1$ and $n=2$ Mellin moments of the kaon GPDs are related to the electromagnetic (EM) form factors and GFFs of the kaon, respectively. 
For convenience, let us recall Eqs.~(\ref{eq:n1_Mellin_moments_kaon})-(\ref{eq:GFFs_kaon_n2}).
First, for $n=1$ case,
\begin{align}
\label{eq:ffgpd}
   2 \int^{+1}_{-1} \;dx  \; H^{u/K^+}(x,\xi,t)
    &=A^{u/K^+}_{10}(t), \\
  2  \int^{+1}_{-1} \;dx \; H^{\bar s/K^+}(x,\xi,t)
    &=A^{\bar s/K^+}_{10}(t).
\end{align}
Kaon EM form factor is obtained by summing up the generalized form factors given above, multiplied by the corresponding quark charges,
\begin{align}\label{eq:n=1GFF}
   e_u A^{u/K^+}_{10}(t) + e_{\bar s} A^{\bar s/K^+}_{10}(t) = F_{K^+}(t).
\end{align}

The $n=2$ Mellin moments read,
\begin{align}\label{eq:n2_moments}
    2\int^{+1}_{-1} \;dx x \; H^{u/K^+}(x,\xi,t)
    &=A^{u/K^+}_{20}(t) 
    + \xi^2 A^{u/K^+}_{22}(t), \\
    2\int^{+1}_{-1} \;dx x \; H^{\bar s/K^+}(x,\xi,t)
    &=A^{\bar s/K^+}_{20}(t) 
    + \xi^2 A^{\bar s/K^+}_{22}(t).
\end{align}
The generalized form factors $A_{20}$ and $A_{22}$ are related to the GFFs $A(t)$ and $D(t)$, which characterize the mechanical structure of a hadron, such as mass, pressure, and shear stress~\cite{Polyakov:2018zvc}. 
In general, hadron matrix elements of the QCD energy-momentum tensor (EMT) operator define the GFFS. For the kaon case,
it is written as follows
\begin{align}\label{eq:kaon_GFFs}
    \langle K^+(p') | \hat T^{\mu\nu}(0) | K^+(p) \rangle 
 = 2 P^\mu P^\nu A_{K^+}(t) 
 + \frac{1}{2}(q^\mu q^\nu - g^{\mu\nu}q^2)D_{K^+}(t).
\end{align}
Note that the mass form factor $A(t)$ is constrained by the 
conservation of the EMT operator, such that $A(t=0)=1$. 
On the other hand, $D(t)$ is not controlled by any conservation law. It represents purely internal properties such as pressure and shear force, respectively through spatial isotropic ($\delta_{ij}$) and shear ($\hat r_i \hat r_j - \delta_{ij}/3$) components of the hadron matrix elements in Eq.~(\ref{eq:kaon_GFFs})~\cite{Polyakov:2002yz}.
Authors in Ref.~\cite{Perevalova:2016dln} postulated that the $D$-term ($D(t=0)$) should be negative due to 
the local stability condition. This statements was confirmed using different approaches for the nucleon~\cite{Ji:1997gm,Goeke:2007fp,Cebulla:2007ei,Jung:2013bya,Pasquini:2014vua}, as well as experimental studies~\cite{Burkert:2018bqq}. Also, the lattice QCD results of the nucleon gravitational form factors were presented in Refs.~\cite{LHPC:2007blg,Hackett:2023rif} and references therein.
For the massless pion, it was predicted that $D(0)=-1$ in the soft pion limit~\cite{Voloshin:1982eb,Leutwyler:1989tn}. 
Chiral perturbation theory (ChPT) predicted that the $D$-term of 
a pseudo-Goldstone boson (here denoted as $\mathcal{M}$) deviates 
from $-1$ by order of $m_\mathcal{M}^2$ at the order of $p^2$ and 
$m_K^2$ and one has~\cite{Donoghue:1991qv}
\begin{align}
  A(0)+D(0) = 16 \frac{m_\mathcal{M}^2}{F^2}(L_{11}-L_{13}) +
  (\text{Meson\;Loops}),
\end{align}
 where the gravitational low-energy-constants of $L_{11}, L_{12}$, and $L_{13}$ are introduced. It is predicted that, including the meson-loop contributions~\cite{Donoghue:1991qv,Hudson:2017xug}, 
\begin{align} \label{eq:ChPT_GFFs_pion}
    A_\pi(0)+D_\pi(0) \approx 0.03, \\ 
    A_K(0)+D_K(0) \approx 0.23.
    \label{eq:ChPT_GFFs_kaon}
\end{align}

The GFFs can be further decomposed into the valence quark ones as follows
\begin{align}
    A_{K^+}(t) &= A_{u/K^+}(t)+A_{\bar s/K^+}(t), \\
    D_{K^+}(t) &= D_{u/K^+}(t)+D_{\bar s/K^+}(t).
\end{align}
We note that the physical interpretation of these quark-flavor decompositions requires careful consideration. The $00$- and $ii$-components of the EMT matrix elements are related to the mass and isotropic pressure spatial densities, respectively. 
When making the flavor decomposition of Eq.~(\ref{eq:kaon_GFFs}), since the EMT operator for individual parton ($a$) is not conserved, additional Lorentz structure $g^{\mu\nu}$ appears with a form factor $\bar c_{a}(t)$. Thus, mass and pressure distributions of specific parton components must be 
accompanied by contributions from the form factor $\bar c_{a}(t)$.
A recent study showed that the nonconservation of the flavor-specific EMT significantly influences the mechanical properties of hadrons~\cite{Won:2023ial}. The analysis within the chiral quark soliton model revealed that the $\bar c$ form factors of the individual quark flavors redistribute the proton's mechanical properties, including its pressure and mass distributions, while the total energy-momentum tensor remains conserved by constituent quarks.

Now, we relate the GFFs $A(t)$ and $D(t)$
to the generalized form factors $A_{20}(t)$ and 
$A_{22}(t)$, obtained from $n=2$ Mellin moments of the valence-quark GPDs, 
\begin{align}
    A^{f/\mathcal{M}}_{20}(t) &= A_{f/\mathcal{M}}(t), \\
    A^{f/\mathcal{M}}_{22}(t) &= D_{f/\mathcal{M}}(t).
\end{align}
We emphasize that the momentum fraction $\langle x \rangle_{f/\mathcal{M}}$ and the `mass fraction, $m_{f/\mathcal{M}}$' of the individual parton are different and thus these two quantities should be understood with care 
concerning the relation~\cite{Won:2023ial}
\begin{align}
   A^{f/\mathcal{M}}_{20}(0) =  \langle x \rangle_{f/\mathcal{M}}
   \ne m_{f/\mathcal{M}} = (A_{f/\mathcal{M}}(0) + \bar c_{f/\mathcal{M}} (0))m_\mathcal{M}.
\end{align}
Two quantities become identical when we sum $\bar{c}$ overall quark flavors $f$, since $\sum_f \bar c_f (t)=0$. For the pion case, the momentum and mass fraction become identical since the isovector EMT is vanishing due to the isospin symmetry. A detailed analysis of the kaon GFFs will be presented in a separate publication.

To obtain the EMFFs and GFFs from the GPDs, we compute the $n=1$ Mellin moments by varying $\xi$ in the range $\xi \in [0.0, 1]$. We confirm that the numerical result for the EM form factors do not depend on the $\xi$ values. Also, we compute the $n=2$ Mellin moments and fitted the form factors by using Eq.~(\ref{eq:n2_moments}). It shows that our results are stable, regardless of the choice of $\xi$ values. 
We carry out this process for multiple values of $-t$ in the range $=(0,1)$~GeV$^2$ to obtain the $t$-dependence of the form factors. 

\begin{figure}
    \centering
    \includegraphics[width=8.cm]{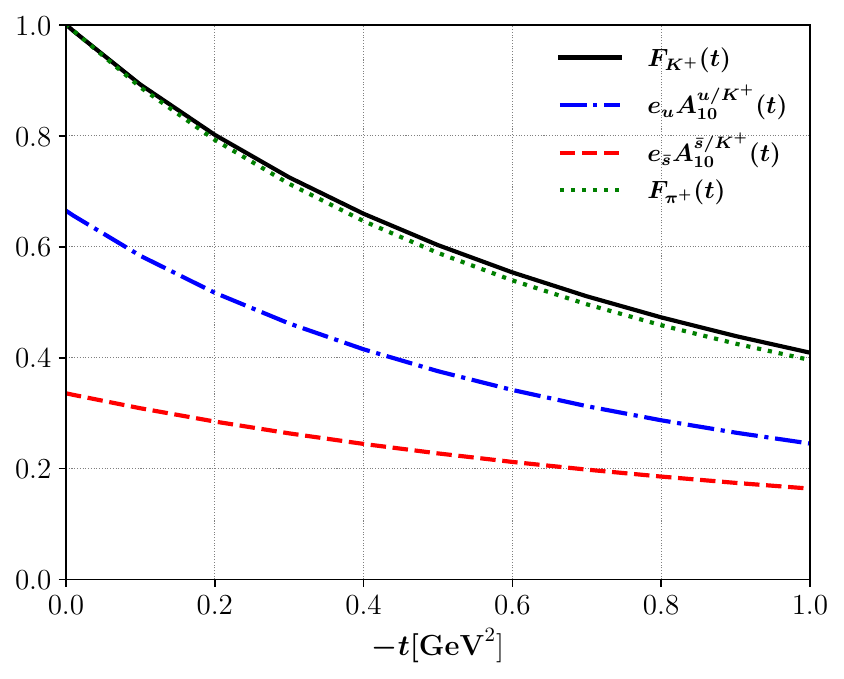}
    \caption{
    Generalized form factors $A_{10}$ of $u$ (blue dashed-dotted line) and $\bar s$ (red dashed line) quark (multiplied by the quark charges) and total electromagnetic form factor (black solid line) of the kaon ($K^+$) are depicted. Also, the pion electromagnetic form factor (green dotted line) is shown for comparison.
    One observes that the $u$-quark form factor is much stiffer than that of the $\bar s$-quark in a $K^+$. Numerical results are at the model scale $\mu=\mu_0$.
    }
    \label{fig:vFFs}
\end{figure}

Figure~\ref{fig:vFFs} shows the EM form factors of the kaon and pion.
The solid black curve represents the total kaon EM form factor $F_{K^+}(t)$.
The dash-dotted blue and dashed red lines show the $u$- and $\bar s$-quark contributions $e_u A^{u/K^+}{10}(t)$ and $e_{\bar s} A^{\bar{s}/K^+}{10}(t)$, respectively.
The dotted green line represents the pion EM form factor $F{\pi^+}(t)$.
To discuss the kaon and pion form factors, we determine the radius of the charge distributions from the slope of each form factor 
at $t=0$, defined by $\sqrt{r^2}_F \equiv (-6 \; dF(t)/dt)^{-1/2}|_{t=0}$. We obtain the EM radii of the kaon and pion,
\begin{align}
    \sqrt{r^2}_{F_{K^+}}=0.526~\mathrm{fm} ~\mathrm{and}~
     \sqrt{r^2}_{F_{\pi^+}}=0.544~\mathrm{fm},
\end{align}
respectively, which underestimate the averaged values given in the Review of Particle Physics \cite{ParticleDataGroup:2024cfk}, 
$\sqrt{r^2}_{F_{K^+}}=0.560 \pm 0.031$~fm and $\sqrt{r^2}_{F_{\pi^+}}=0.659 \pm 0.004$~fm.  In Ref.~\cite{Nam:2007gf}, it was argued that the nonlocal contributions to the EM form factors enhance the charge radius of the mesons by $\sim10\%$. Such contributions come from the additional terms in the conserved vector current of the model. For the kaon, however, the $1/N_c$ suppressed meson fluctuations contribute more significantly to the charge radius. 

\begin{figure}
    \centering
    \includegraphics[width=7.6cm]{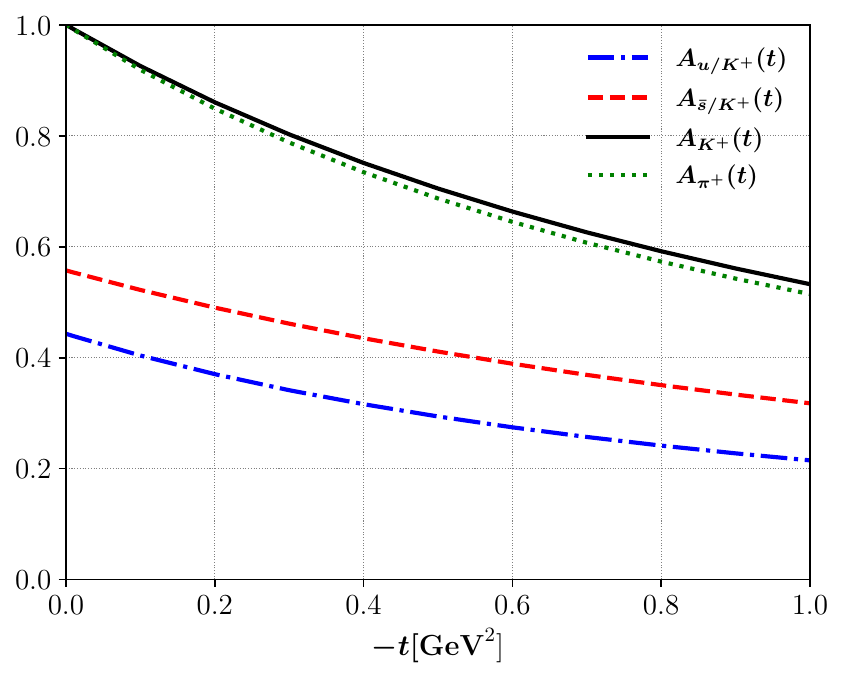}
    \includegraphics[width=7.6cm]{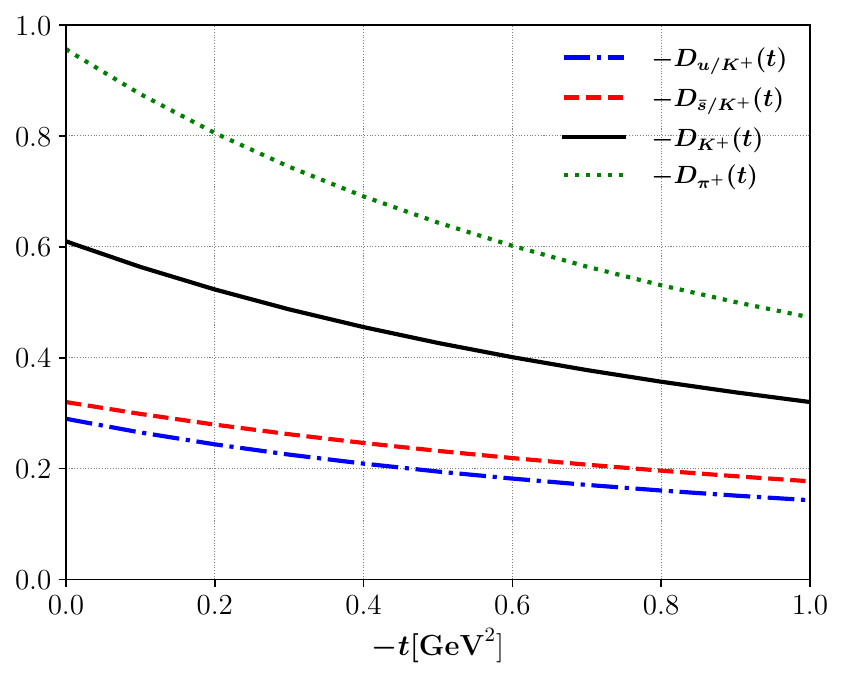}
    \caption{Gravitational form factors of the kaon ($K^+$) $A_{K^+}(t)$ (left panel) and $D(t)$ (right panel), normalized by $A_{K^+}(t=0)$. Note that the form factors of the pion are represented by the green dotted line. Numerical results are at the model scale $\mu=\mu_0$.}
    \label{fig:GFFs}
\end{figure}

Now we turn to analyzing the GFFs. Figure~\ref{fig:GFFs} presents our numerical results for the kaon and pion GFFs, with the mass form factors $A(t)$ and $D(t)$ displayed in the left and right panels, respectively. The solid black line represents the total kaon mass form factor $A_{K^+}(t)$, while the dotted green line shows the pion mass form factor $A_{\pi^+}(t)$. Our numerical results show that the pion and kaon form factors are very close to each other. This similarity can be partially understood from ChPT, where the mass form factor at leading order is given by $A(t) = 1-2L_{12}t/F^2 + \mathcal{O}(E^4)$ ~\cite{Donoghue:1991qv}, independent of the meson mass.  At $t=0$, we obtain the values: 
\begin{align}
    A_{u/K^+}(0) = 0.412, ~\mathrm{and}~   A_{\bar s /K^+}(0) = 0.518,
\end{align}
and thus $A_{K^+} = 0.930$, which correspond to the $M_2$ values presented in the previous section. 
Reference.~\cite{Nam:2007gf} found that the nonlocal contributions affect 
the ratio of the strange to light quark vector form factors by only $5\%$, 
even though the nonlocal contribution to the vector form factor is quite sizeable $\sim 30\%$.
Thus, concerning the deficiency of the momentum sum rule by $7\%$, 
it is reasonable to present the ratio
\begin{align}
    A_{\bar s /K^+}(0) / A_{u/K^+}(0) = 1.257,
\end{align}
as a prediction of our study.
A rough estimation of the nonlocal correction to this ratio reduces the value by 
about $0.02$, since the nonlocal contribution to the vector form factor is flavor-independent 
at zero momentum transfer~\cite{Nam:2007gf}.

We obtain the $D$-term form factors,
\begin{align}
    D_{u/K^+}(0) = -0.270, ~\mathrm{and}~   D_{\bar s /K^+}(0) = -0.297,
\end{align}
and the ratio of the form factors,
\begin{align}
    D_{\bar s /K^+}(0)/D_{u/K^+}(0) = 1.1.
\end{align}
Also, we present the sum of $A_{K^+}(0)$ and $D_{K^+}(0)$,
\begin{align}
    A_{K^+}(0) + D_{K^+}(0) = 0.363,
\end{align}
which is comparable to the ChPT result of Eq.~(\ref{eq:ChPT_GFFs_kaon}).
For the pion case, we obtain that
\begin{align}
    A_{\pi^+}(0) + D_{\pi^+}(0) = 0.04.
\end{align}
Finally, we exhibit the slopes of the 
kaon and pion GFFs:
\begin{align}
    \sqrt{r^2}_{A_K}=0.416~\mathrm{fm},~\mathrm{and}~
    \sqrt{r^2}_{A_\pi}=0.436~\mathrm{fm},
\end{align}
for the mass form factors and
\begin{align}
    \sqrt{r^2}_{D_K}=0.328~\mathrm{fm},~\mathrm{and}~
    \sqrt{r^2}_{D_\pi}=0.436~\mathrm{fm},
\end{align}
for the $D$-term form factors.
Note that the radii of the $D$-term form factors are 
significantly underestimated due to the lack of the meson-loop 
contributions. For comparison, we also present the results of the L$\chi$QM, in which the momentum dependence of the dynamical quark mass is switched off, $F(k)=1$. In this model, the momentum sum rule is satisfied exactly and the light and strange quarks share:
\begin{align}
    A_{u/K^+}(0) = 0.477, ~\mathrm{and}~   A_{\bar s /K^+}(0) = 0.532.
\end{align}
Also, the $D$-term form factors are 
\begin{align}
    D_{u/K^+}(0) = -0.351, ~\mathrm{and}~   D_{\bar s /K^+}(0) = -0.421.
\end{align}
Summing over the quark flavors, we obtain $ A_{K^+}(0) + D_{K^+}(0)=0.237$, 
which agrees well with the predictions of ChPT in Ref.~\cite{Donoghue:1991qv}. However, the local model results exhibit strong dependence on the particular parameter set used. Furthermore, the slopes of all form factors in the local model are systematically suppressed, with values about 20\% smaller than in the NL$\chi$QM.

In Table~\ref{tab:gff_form_factors_comparisons}, we provide a comparison of our results to
other theoretical works by Donoghue and Leutwyler (ChPT)~\cite{Donoghue:1991qv}, Hutauruk \textit{et al.} (BSE-NJL model)~\cite{Hutauruk:2016sug}, 
Adhikari \textit{et al.} (BLFQ-NJL model)~\cite{Adhikari:2021jrh}, and 
Xu~\textit{et al.} (Dyson-Schwinger equation (DSE) model)~\cite{Xu:2023izo},
as well as the lattice QCD results~\cite{Hackett:2023nkr,Delmar:2024vxn}.
While the pion GFFs were first explored within lattice QCD in Refs.~\cite{Brommel:2005ee,Brommel:2007zz}, those studies did not clearly determine the D-term. Recent lattice calculations have 
reported the pion and kaon GFFs~\cite{Hackett:2023nkr,Delmar:2024vxn}. Our results for $A_{\pi^+}(0)+D_{\pi^+}(0)$ and $A_{\bar{s}/K^+}(0)/A_{u/K^+}(0)$ agree with these lattice QCD results in Refs.~\cite{Hackett:2023nkr,Delmar:2024vxn}, respectively. Note that the lattice results are evaluated at $\mu=2$~GeV, but specific details such as the renormalization scheme and lattice parameters may differ.
While the predictions for the kaon and pion $D$-term are consistent across different approaches, the strange- to light-quark ratios of the form factors show notable deviations.

\begin{table}[t]
    \caption{Comparison of the current results on the GFFs 
    with other theoretical and lattice QCD results. 
    }
        \centering
        \begin{tabular}{c@{\hskip 0.05in}c@{\hskip 0.05in}c@{\hskip 0.05in}c@{\hskip 0.05in}c@{\hskip 0.05in}c@{\hskip 0.05in}}
        \hline\hline \\ [-2ex]
       & Approach & $A_{K^+}(0)+D_{K^+}(0)$ & $A_{\bar s /K^+}(0)/A_{u /K^+}(0)$ 
       & $D_{\bar s /K^+}(0)/D_{u /K^+}(0)$ & $A_{\pi^+}(0)+D_{\pi^+}$(0) 
       \\[0.5ex]
             \hline \\[-2ex]
       Present work 1 & NL$\chi$QM& {0.363} & {1.257} & {1.100} & {0.040} \\[1.2ex]
       Present work 2 & L$\chi$QM & {$1-0.763=0.237$} & {$0.532/0.477=1.115$} & {$-0.421/-0.351=1.199$} & $1-0.974=0.026$ \\[1.2ex]
        
        Donoghue \& Leutwyler~\cite{Donoghue:1991qv} & ChPT($E^2$) &
        $1-0.77=0.23$ & $\cdot \cdot \cdot$ & $\cdot \cdot \cdot$ & $1-0.97=0.03$ \\[1.2ex]
        Hutauruk \textit{et al}.~\cite{Hutauruk:2016sug} & BSE-NJL & 
        $\cdot \cdot \cdot$ & $0.58/0.42=1.38$ & $\cdot \cdot \cdot$ & $\cdot \cdot \cdot$ \\[1.2ex]
        Adhikari \textit{et al}.~\cite{Adhikari:2021jrh} & BLFQ-NJL &
        $\cdot \cdot \cdot$ & $0.58/0.44=1.32$ & $\cdot \cdot \cdot$ & $\cdot \cdot \cdot$ \\[1.2ex]
        Xu~\textit{et al}. \cite{Xu:2023izo} & DSE &
         $1-0.767=0.233$ & $0.613/0.387=1.584$ & $-0.436/-0.331=1.32$ & $1-0.97=0.03$ \\[1.2ex]
        Hackett~\textit{et al}. \cite{Hackett:2023nkr} & MIT-Lattice &
        $\cdot \cdot \cdot$ &  $\cdot \cdot \cdot$ &  $\cdot \cdot \cdot$ & $\approx 0.10$ \\[1.2ex]
        Delmar~\textit{et al}. \cite{Delmar:2024vxn} & ETMC-Lattice &
        $\cdot \cdot \cdot$ & $\approx 1.3$ &  $\cdot \cdot \cdot$ & $\cdot \cdot \cdot$ \\[1.2ex]
           \hline\hline
        \end{tabular}
        \label{tab:gff_form_factors_comparisons}
    \end{table}

\section{Summary and conclusion}
\label{sec:sum}
As a summary, in this work, we investigated the valence quark GPDs of the kaon and pion within the framework of the NL$\chi$QM. We then derived the analytic expressions for the $u$- and $\bar s$-quark GPDs in the charged kaon ($K^+$) and discussed their properties in terms of the diagrammatic contributions of the model. Fixing the model parameters for numerical calculation, we observed that there was a narrow window for $m_s$, given cutoff $\Lambda$ values, to correctly reproduce the normalization conditions for the meson DAs and the electromagnetic form factors. 

In the numerical results, we discussed the prominent role of the contribution from the two-meson vertex of the model through $I_3$ in shaping the ERBL region $-\xi < x < \xi$. Such an aspect is compared with the results of a simpler local model, where the momentum dependence of the quark is absent. 
Also, the quark GPDs for the pion and kaon at $\xi=1$ and $t=0$ are evaluated 
in which sizable flavor asymmetries of $\mathcal{O}(m_{\pi,K}^2)$ were found. 

The one-loop QCD scale evolution is applied from the current model results at $\mu_0 = 0.33~$GeV, to higher scales $\mu^2=4~$GeV$^2$ and $\mu^2=100~$GeV$^2$. It is observed that the effect of the evolution to the shape of the valence quark GPDs is diminishing as the skewness values are increased. Also, the produced gluon and sea-quark GPDs are suppressed in size. 

Additionally, through the first two Mellin moments of the GPDs,
we obtain the kaon and pion electromagnetic and gravitational form factors. Noticeably, we obtain the ratio of the $\bar s$ and $u$ quark $D$-term form factors in the kaon, $D_{\bar s/K^+}(0)/D_{u / K^+}(0)=1.10$, which are compared to the result of Ref.~\cite{Xu:2023izo}, $D_{\bar s/K^+}(0)/D_{u / K^+}(0)=1.32$. The form factors at zero momentum transfer are stable within the model parameters. However, the slopes (or the radii) are mildly dependent on the cut-off value $\Lambda$, as expected. 

In the present work, we found that the momentum sum rules for kaon and pion are not satisfied ($M_2^{K^+} \approx 0.93$, and $M_2^{\pi^+} \approx 0.93$), which is attributed to the model's nonlocal interaction. Previous works have addressed similar issues of broken gauge invariance in nonlocal chiral quark models~\cite{Broniowski:1999bz,Dorokhov:2000gu,Kim:2004hd,Nam:2007gf,Hutauruk:2023ccw}. The application of these approaches to calculations of hadronic matrix elements on 
the light cone remains for future investigation.

\section*{Acknowledgements}
 HDS expresses gratitude to Wen-Chen Chang, Yongwoo Choi, Hyun-Chul Kim, and Kirill Semenov-Tyan-Shanskiy for fruitful discussions and encouragement. 
 The authors also acknowledge Seung-il Nam for providing technical details of his works. 
 H.-D. Son was supported by the National Research Foundation of Korea(NRF) grant funded by the Korea government(MSIT) (RS-2023-00210298). The work of P.T.P.H. was supported by the National Research Foundation of Korea (NRF) grants funded by the Korean government (MSIT) Nos. 2018R1A5A1025563, 2022R1A2C1003964, and 2022K2A9A1A0609176, and the PUTI Q1 Research Grant from the University of Indonesia (UI) under contract No. NKB-442/UN2.RST/HKP.05.00/2024.

\begin{appendix}
\section{Analytical results}
\label{appendix:analytic_results}
Here we provide the complete analytical results of the present work. We utilize the analytical method as demonstrated in Refs.~\cite{Praszalowicz:2001wy,Praszalowicz:2003pr}. The notations and definitions used in the present work are adapted from the references. 
While the authors of Refs.~\cite{Praszalowicz:2001wy,Praszalowicz:2003pr} considered the strict chiral limit $m_f=0$, in the present work,
 we extend this approach to the case of 
 nonzero current quark masses in $SU(3)$-flavor, $m_q$ and $m_s$.

\subsection{Kaon DAs}
\label{apdx:DAs}
As in chiral limit, $m_\pi \approx 0$ the quark kaon and pion GPDs are connected to the pion and kaon DAs, respectively. This is so-called the \textit{soft pion theorem}.  Here we begin with the light-cone DA of the kaon, which is defined by the following 
\begin{align}
    \phi_{K^+}(u) = 
    \frac{1}{i \sqrt{2} F_K}
    \int^\infty_{-\infty} \frac{d \tau}{\pi}
    e^{- i \tau (2u-1) n \cdot P}
    \langle 0 | \bar \psi(n \tau) \gamma^+ \gamma^5 \psi(-n \tau)
    | K^+(P) \rangle.
\end{align}
Using the model of Eq.~(\ref{eq:action}), we have an expression for the kaon DA in the following quark one-loop integral:
\begin{align}
    \phi_{K^+}(u) = i \frac{N_c}{F_K^2}
    \int \frac{d^4 k }{(2\pi)^4} \delta(k^+ - u P^+) 
    \mathrm{Tr}\left[ \Slash{n}\gamma^5
     \frac{\sqrt{M_u(k)}}{\Slash{k}-m_u-M_u(k)+i\epsilon} \gamma^5
     \frac{\sqrt{M_s(k-P)}}{\Slash{k}-\Slash{P}-m_s-M_s(k-P)i\epsilon}
    \right].
\end{align}
Here, we emphasize that we follow the analytic procedure presented in Ref.~\cite{Praszalowicz:2001wy}. The quark denominator with momentum $k_i$ and with flavor, $f$ can be written as follows
\begin{align}
    k_i^2 - \bar M_f(k_i)^2 + i \epsilon 
    &= k_i ^2 - m_f^2 - 2m_f M_f F(k_i)^2 - M_f^2 F(k_i)^4 + i \epsilon \cr 
    &=\Lambda^2 \zeta_i^{-4n}
    \left(
     \zeta_i^{4n+1} + \left(1-\frac{m_f^2}{\Lambda^2}\right) \zeta_i^{4n}
     -\frac{2m_f M_f}{\Lambda^2} \zeta_i^{2n} - \frac{M_f^2}{\Lambda^2}
    \right) \equiv \Lambda^2 \zeta_i^{-4n} G_f(\zeta_i),
\end{align}
where the dynamical quark mass is defined as $\bar M_f(k) \equiv m_f + M_f(k)$. Also, we introduce the new variable $\Lambda^2 \zeta_i \equiv k_i^2 - \Lambda^2 + i \epsilon$. Then, we need to solve the $4n+1$-order equation $G_f(\zeta_i)=0$ to find the solutions $z_i^f$, 
\begin{align}
    G_f(\zeta_i) = \prod^{4n+1}_{i=1} (\zeta_i - z_i^f). 
\end{align}
After performing the $k^-$ integral, we obtain the final expression 
for the kaon DA and one has
\begin{align}
    \phi_{K^+}(u)&=\frac{N_c\sqrt{M_u M_s}}{(2\pi)^2 F_K^2}
    \int^\infty_0 d \kappa_\perp^2\; (u-1)^n \sum^{4n+1}_{i=1} (z_i^u)^n \phi^u_i
    \left[\prod^{4n+1}_{k=1}
    (\kappa_\perp^2 +u(u-1)\tilde m_K^2 +1 + u z_i^u+(1-u)z_k^s) 
    \right]^{-1} \cr 
    &\times \bigg\{
    u(\kappa_\perp^2 +u(u-1)\tilde m_K^2+1 + u z_i^u)^{3n} M_u \left(1+ \frac{m_u}{M_u} (z_i^u)^{2n}\right)  \cr
    &+(1-u)^{2n+1}(z_i^u)^{2n}(\kappa_\perp^2+1+u(u-1)\tilde m_K^2+u z_i^u)^n M_s \cr
   &\times \left(1+ \frac{m_s}{M_s}(u-1)^{-2n}(\kappa_\perp^2+u(u-1)\tilde m_K^2+1+u z_i^u)^{2n}\right)    
    \bigg\},
\end{align}
where the scaled variables $\kappa_\perp = k_\perp / \Lambda$ 
and $\tilde m_K \equiv m_K/\Lambda$ are introduced. We define the function $\phi_i^f$, coming from the quark denominator
$ 1/(k_i^2 - \bar M_f(k_i)^2 + i \epsilon )$ in which the poles in 
$k^-$ integral is evaluated as follows
\begin{align}
    \phi_i^f &= \prod^{4n+1}_{k=1,k \ne i} \frac{1}{z_i^f - z_k^f}.
\end{align}
The authors of Ref.~\cite{Praszalowicz:2003pr} discussed the analytic properties of $\phi_i^f$ for the pion. Note that to ensure the convergence of the quark loop integrals, the solutions should satisfy a condition,
\begin{align}
    \sum^{4n+1}_{i=1} (z_i^f)^m \phi_i^f 
    = \left\{
    \begin{array}{c}
    0 \quad \mathrm{for} \quad m<4n,\\
    1 \quad \mathrm{for} \quad m=4n,
    \end{array}
    \right.
\end{align}
We check numerically that the above relation holds for both the light and strange quarks in the current model.
For the pion DA, we simply replace the kaon masses with the pion masses and the strange quark masses with the light quark masses, $m_K \to m_\pi$, $m_s \to m_u$, and $M_s \to M_u$, respectively. 

\subsection{Results for meson DAs}
For the numerical computation of the kaon and pion DAs,
we use $M_s=M_u=M=350~$MeV from the instanton model,
and $m_u=5~$MeV as our input parameters. 
To determine the cutoff $\Lambda$ and
the strange quark mass $m_s$, we vary these two parameters
to reproduce the meson decay constants $F_{K^+}$ and $F_{\pi^+}$
from the normalization of the meson DAs. 
\begin{figure}
    \centering
    \includegraphics[width=8.5cm]{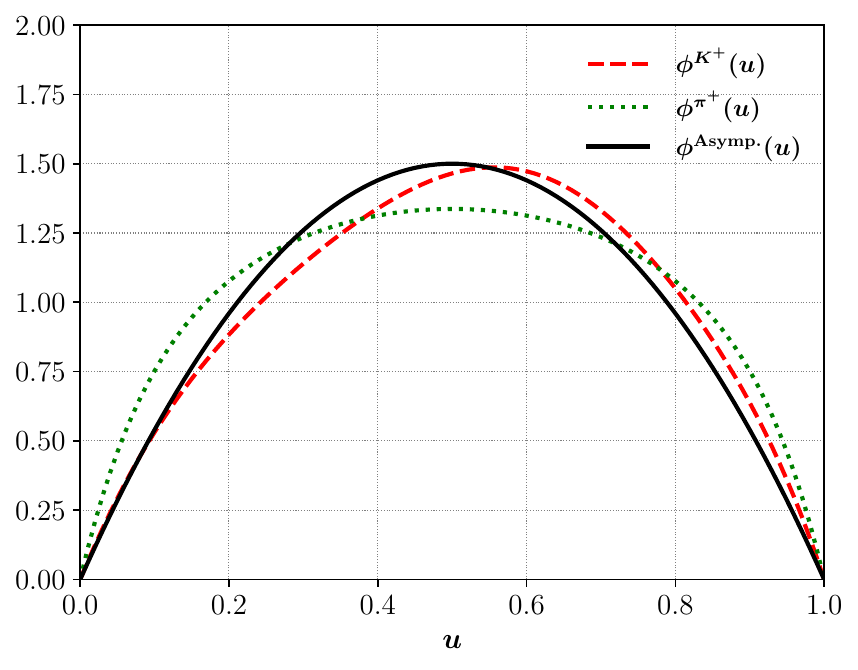}
    \caption{Meson light-cone distribution amplitudes in the nonlocal 
    chiral quark model. Along with the numerical results of current work,  the asymptotic DA $\phi^{\mathrm{Asymp.}}=6u(1-u)$ is shown in the black solid curve.}
    \label{fig:DAs}
\end{figure}

In Fig.~\ref{fig:DAs}, we show the numerical results for the kaon 
and pion DAs as well as the asymptotic meson DA  $\phi^{\mathrm{Asymp.}}=6u(1-u)$. We observe that the kaon DA is skewed toward $u>1/2$, due to asymmetric quark masses. 
The results shown here are identical to 
the `MIA1' results in Ref.~\cite{Nam:2006au}, apart from 
small differences due to different choices of input parameters and 
cutoff value $\Lambda$. In Ref.~\cite{Nam:2006au}, a detailed 
analysis of the pion and kaon DAs within the nonlocal chiral quark model was reported. Further discussions on fixing the model parameters are given in Sec.~\ref{sec:model_parameters}.

\subsection{Kaon generalized quark distributions}\label{apdx:gpds_analytic_results}

In the present work, we consider only $\xi>0$, owing to the 
even nature of the quark GPDs in $\xi$. In this section, we show explicit expressions 
of the integrands in Eqs.~(\ref{eq:gpds_loop_integral1})-(\ref{eq:gpds_loop_integrand2}). 
For convenience, we introduce the scaled variables:
\begin{align}
    \eta = \frac{P^+ k^-}{ \Lambda^2}, \quad
    \vec{\kappa}_\perp = \frac{\vec{k}_\perp}{\Lambda}, \quad
    \tau = \frac{\tilde t}{4\Lambda^2}, \quad 
    \vec{\delta}_\perp = \frac{\vec {\Delta}}{2 \Lambda}. \quad 
\end{align}
The $k^-$ integral can be evaluated by closing the contours of either the upper or lower half-plane. We make such a choice to minimize the number of poles enclosed in the contour to reduce the computational load in the numerical calculation. In computing 
integrals of $\mathcal{T}_{(a)}$, we observe that there are three type of poles,
which come from different quark propagators. 
Here we denote them as $\eta_i^{(u1)}$, $\eta_i^{(\bar s)}$, and $\eta_i^{(u2)}$ where $i=[1,4n+1]$. For the $u$-quark part, $\mathcal{T}^{u\bar s}_{(a)}$, we find 
\begin{align}
    \eta_i^{(u1)} &= - \xi \tau + \frac{1}{x+\xi}
    ((\vec{\kappa}_\perp - \vec{\delta}_\perp)^2 +1 + z_i^{u}
    -i \epsilon \; \sgn(x+\xi), \\
    \eta_i^{(\bar s)} &= - \tau + \frac{\vec {\kappa}_\perp^2 +1 + z_i^{\bar s}}{x+\xi}
     -i \epsilon \; \sgn(x-1),\\
    \eta_i^{(u2)} &=  \xi \tau + \frac{1}{x-\xi}
    ((\vec{\kappa}_\perp + \vec{\delta}_\perp)^2 +1 + z_i^{u}
    -i \epsilon \; \sgn(x-\xi). 
\end{align}
Hence, we observe that the integral vanishes for $x<-\xi$. 
For $x > \xi$, we close the contour in the upper-half plane, taking the poles $\eta_i^{(\bar s)}$. For $-\xi < x < \xi$, we enclose the lower-half-plane to take into account the poles from $\eta_i^{(u1)}$.
These two separate contributions correspond to $I_1^{u/K^+}(x,\xi,t)$ 
and $I_2^{u/K^+}(x,\xi,t)$, respectively. For the $\bar s$-quark GPDs, one can obtain the expressions $I_1^{\bar s/K^+}(x,\xi,t)$ and $I_2^{\bar s/K^+}(x,\xi,t)$ in a similar way. Evaluating the integral with $\mathcal{T}_{(b)}$ is simpler, since we have 
two poles
\begin{align}
    \eta_i^{(f1)} &= - \xi \tau + \frac{1}{x+\xi}
    ((\vec{\kappa}_\perp - \vec{\delta}_\perp)^2 +1 + z_i^{f}
    -i \epsilon \; \sgn(x+\xi), \\
    \eta_i^{(f2)} &=  \xi \tau + \frac{1}{x-\xi}
    ((\vec{\kappa}_\perp + \vec{\delta}_\perp)^2 +1 + z_i^{f}
    -i \epsilon \; \sgn(x-\xi), 
\end{align}
and obtain nonzero contribution for $\sgn(x-\xi)\sgn(x+\xi)<0$. 
For a detailed discussion of the pole description, see Ref.~\cite{Praszalowicz:2001wy}. Then, the transverse momenta $\kappa_\perp$ should be integrated. The angular integral between $\vec{\kappa}_\perp$ and $\vec{\delta}_\perp$ are to be evaluated by making a change to the variable $\lambda = e^{i \theta} $. We choose the closed contour $C(0,1)$ in the complex plane of $\lambda$ and take into account the poles within the circle, which is done numerically. Finally, we arrive at the expression, as in Eqs.~(\ref{eq:gpd_uquark_Is}) and ~(\ref{eq:gpd_squark_Is}) 
\begin{align*}
    H^{u/K^+}(x, \xi, t) = I_1^{u/K^+}(x,\xi,t) \Theta(x-\xi) 
    + I_2^{u/K^+}(x,\xi,t) \Theta (\xi-|x|)
    +I_3^{u/K^+}(x,\xi,t) \Theta (\xi-|x|), \\ 
    - H^{\bar s /K^+}(-x, \xi, t) = I_1^{\bar s/K^+}(x,\xi,t) \Theta(-x-\xi) 
    + I_2^{\bar s/K^+}(x,\xi,t) \Theta (\xi-|x|)
    +I_3^{\bar s/K^+}(x,\xi,t) \Theta (\xi-|x|),
\end{align*}
where
\begin{align}
    I_1^{u / K^+} (x, \xi, t) &=-\frac{i N_c M_s M_u}
    {2(2\pi)^3 F_K^2}
    \int^\infty_0 d \kappa_\perp^2\; \int_{C(0,1)} d\lambda
    \sum^{4n+1}_{i=1}  \phi^{s}_i(-1)^n  [(x-\xi)\lambda]^{6n+1} 
    (C \lambda^2 + f^s_i \lambda + C)^n
    (C \lambda^2 + g^s_i \lambda + C)^n
    \cr
    & \left[\prod^{4n+1}_{k=1}
      (C \lambda^2 + F^{\bar s u}_{ik} \lambda + C )
      (C \lambda^2 + G^{\bar s u}_{ik} \lambda + C )
    \right]^{-1} h_{u}^{(1)},\\
    I_2^{u / K^+} (x, \xi, t) &=\frac{i N_c M_s M_u}
    {2(2\pi)^3 F_K^2}
    \int^\infty_0 d \kappa_\perp^2\; \int_{C(0,1)} d\lambda
    \sum^{4n+1}_{i=1}  \phi^{u}_i(-1)^n  [(x+\xi)\lambda]^{7n+1} 
    (A \lambda^2 + b^u_i \lambda + A)^n
    (z_i^u)^n
    \cr
    & \left[\prod^{4n+1}_{k=1}
      (A \lambda^2 + B^{ u}_{ik} \lambda + A )
      (C \lambda^2 + D^{u \bar s}_{ik} \lambda + C )
    \right]^{-1} h_{u}^{(2)},\\
    I_1^{\bar s / K^+} (x, \xi, t) &=\frac{i N_c M_s M_u}
    {2(2\pi)^3 F_K^2}
    \int^\infty_0 d \kappa_\perp^2\; \int_{C(0,1)} d\lambda
    \sum^{4n+1}_{i=1}  \phi^{u}_i(-1)^n  [(-x-\xi)\lambda]^{6n+1} 
    (\tilde{C} \lambda^2 + \tilde{f}^u_i \lambda + \tilde{C})^n
    (\tilde{C} \lambda^2 + \tilde{g}^u_i \lambda + \tilde{C})^n
    \cr
    & \left[\prod^{4n+1}_{k=1}
      (\tilde{C} \lambda^2 + \tilde{F}^{u \bar s}_{ik} \lambda + \tilde{C} )
      (\tilde{C} \lambda^2 + \tilde{G}^{u \bar s}_{ik} \lambda + \tilde{C} )
    \right]^{-1} h_{\bar {s}}^{(1)},\\
    I_2^{\bar s / K^+} (x, \xi, t) &=-\frac{i N_c M_s M_u}
    {2(2\pi)^3 F_K^2}
    \int^\infty_0 d \kappa_\perp^2\; \int_{C(0,1)} d\lambda
    \sum^{4n+1}_{i=1}  \phi^{s}_i(-1)^n  [(-x+\xi)\lambda]^{7n+1} 
    (\tilde{A} \lambda^2 + \tilde{b}^s_i \lambda + \tilde{A})^n
    (z_i^s)^n
    \cr
    & \left[\prod^{4n+1}_{k=1}
      (\tilde{A} \lambda^2 + \tilde{B}^{s}_{ik} \lambda + \tilde{A} )
      (\tilde{C} \lambda^2 + \tilde{D}^{\bar s u}_{ik} \lambda + \tilde{C} )
    \right]^{-1} h_{\bar{s}}^{(2)}.
\end{align}
We defined the following functions for convenience, 
\begin{align}
   h_{\bar s}^{(1)}&= \mu_{us}^2(-x-\xi)
   \left(\frac{\tilde C \lambda^2 + \tilde g^{u}_i \lambda + \tilde C }{(-x-1)\lambda}\right)^{2n}
   \left(1 + \frac{m_s}{M_s}\left(\frac{\tilde C \lambda^2 + \tilde f^u_i \lambda + \tilde C}{(-x-1)\lambda}\right)^{2n}\right)
   \left(1+ \frac{m_u}{M_u} (z_i^u)^{2n}\right) \cr
  & +\mu_{ss}^2 (1+x) (z_i^u)^{2n}
   \left(1 + \frac{m_s}{M_s}\left(\frac{\tilde C \lambda^2 + \tilde f^u_i \lambda + \tilde C}{(-x-1)\lambda}\right)^{2n}\right)
   \left(1 + \frac{m_s}{M_s}\left(\frac{\tilde C \lambda^2 + \tilde g^u_i \lambda + \tilde C}{(-x-1)\lambda}\right)^{2n}\right) \cr
  & +\mu_{us}^2(-x+\xi) \left(\frac{\tilde C \lambda^2 + \tilde f^{u}_i \lambda + \tilde C }{(-x-1)\lambda}\right)^{2n}
    \left(1 + \frac{m_s}{M_s}\left(\frac{\tilde C \lambda^2 + \tilde g^u_i \lambda + \tilde C}{(-x-1)\lambda}\right)^{2n}\right)
    \left(1+ \frac{m_u}{M_u} (z_i^u)^{2n}\right) \cr
 &   + (z_i^u)^{2n}
    \left(\frac{\tilde C \lambda^2 + \tilde g^{u}_i \lambda + \tilde C }{(-x-1)\lambda}\right)^{2n}
    \left(\frac{\tilde C \lambda^2 + \tilde f^{u}_i \lambda + \tilde C }{(-x-1)\lambda}\right)^{2n}\cr
   &\times \left[
    (-x-1)\delta_\perp^2 + \frac{\xi^2-1}{-x-1}\kappa_\perp^2 + 
    \frac{\xi^2-x^2}{-x-1}(1+z_i^u)+ \xi \sqrt{\kappa_\perp^2 \delta_\perp^2} \frac{\lambda^2+1}{\lambda}
    \right],
\end{align}

\begin{align}
   h_{\bar s}^{(2)}&= \mu_{su}^2(-x-\xi)\left(\frac{\tilde A \lambda^2 + \tilde b^{\bar s}_i \lambda + \tilde A }{(-x+\xi)\lambda}\right)^{2n}
   \left(1 + \frac{m_u}{M_u}\left(\frac{\tilde C \lambda^2 + \tilde d^{\bar s}_i \lambda + \tilde C}{(-x+\xi)\lambda}\right)^{2n}\right) 
   \left(1+ \frac{m_s}{M_s} (z_i^{\bar s})^{2n}\right) \cr
  &  +\mu_{ss}^2(1+x) \left(\frac{\tilde C \lambda^2 + \tilde d^{\bar s}_i \lambda + \tilde C }{(-x+\xi)\lambda}\right)^{2n}
    \left(1 + \frac{m_s}{M_s} (z_i^{\bar s})^{2n} \right)
    \left(1 + \frac{m_s}{M_s}\left(\frac{\tilde A \lambda^2 + \tilde b^{\bar s}_i \lambda + \tilde A }{(-x+\xi)\lambda}\right)^{2n} \right)
     \cr
  &+\mu_{su}^2 (-x+\xi) (z_i^{\bar s})^{2n}
   \left(1 + \frac{m_s}{M_s}\left(\frac{\tilde A \lambda^2 + \tilde b^{\bar s}_i \lambda + \tilde A}{(-x+\xi)\lambda}\right)^{2n}\right)
   \left(1 + \frac{m_u}{M_u}\left(\frac{\tilde C \lambda^2 + \tilde d^{\bar s}_i \lambda + \tilde C}{(-x+\xi)\lambda}\right)^{2n}\right) \cr
 & + (z_i^{\bar s})^{2n}
    \left(\frac{\tilde A \lambda^2 + \tilde b^{\bar s}_i \lambda + \tilde A }{(-x+\xi)\lambda}\right)^{2n}
    \left(\frac{\tilde C \lambda^2 + \tilde d^{\bar s}_i \lambda + \tilde C }{(-x+\xi)\lambda}\right)^{2n}\cr
   &\times \left[
    (1-\xi)(\xi^2-x^2)\tau+ (\xi-1)\delta_\perp^2  + (1+\xi) \kappa_\perp^2
    +(x+\xi)(1+z_i^{\bar s}) -x \sqrt{\kappa_\perp^2 \delta_\perp^2} \frac{\lambda^2+1}{\lambda}
    \right],
\end{align}

\begin{align}
   h_{u}^{(1)}&= \mu_{us}^2(x-\xi)
   \left(\frac{ C \lambda^2 +  g^{\bar s}_i \lambda +  C }{(x-1)\lambda}\right)^{2n}
   \left(1 + \frac{m_u}{M_u}\left(\frac{ C \lambda^2 +  f^{\bar s}_i \lambda +  C}{(x-1)\lambda}\right)^{2n}\right)
   \left(1+ \frac{m_s}{M_s} (z_i^{\bar s})^{2n}\right) \cr
  & +\mu_{uu}^2 (1-x) (z_i^{\bar s})^{2n}
   \left(1 + \frac{m_u}{M_u}\left(\frac{ C \lambda^2 +  f^{\bar s}_i \lambda +  C}{(x-1)\lambda}\right)^{2n}\right)
   \left(1 + \frac{m_u}{M_u}\left(\frac{ C \lambda^2 +  g^{\bar s}_i \lambda +  C}{(x-1)\lambda}\right)^{2n}\right) \cr
  & +\mu_{us}^2(x+\xi) \left(\frac{ C \lambda^2 +  f^{\bar s}_i \lambda +  C }{(x-1)\lambda}\right)^{2n}
    \left(1 + \frac{m_u}{M_u}\left(\frac{ C \lambda^2 +  g^{\bar s}_i \lambda +  C}{(x-1)\lambda}\right)^{2n}\right)
    \left(1+ \frac{m_s}{M_s} (z_i^{\bar s})^{2n}\right) \cr
 &   + (z_i^{\bar s})^{2n}
    \left(\frac{ C \lambda^2 +  g^{\bar s}_i \lambda +  C }{(x-1)\lambda}\right)^{2n}
    \left(\frac{ C \lambda^2 +  f^{\bar s}_i \lambda +  C }{(x-1)\lambda}\right)^{2n}\cr
   &\times \left[
    (x-1)\delta_\perp^2 + \frac{\xi^2-1}{x-1}\kappa_\perp^2 + 
    \frac{\xi^2-x^2}{x-1}(1+z_i^{\bar s})+ \xi \sqrt{\kappa_\perp^2 \delta_\perp^2} \frac{\lambda^2+1}{\lambda}
    \right],
\end{align}

\begin{align}
   h_{u}^{(2)}&= \mu_{su}^2(x-\xi)\left(\frac{ A \lambda^2 +  b^{u}_i \lambda +  A }{(x+\xi)\lambda}\right)^{2n}
   \left(1 + \frac{m_s}{M_s}\left(\frac{ C \lambda^2 +  d^u_i \lambda +  C}{(x+\xi)\lambda}\right)^{2n}\right) 
   \left(1+ \frac{m_u}{M_u} (z_i^u)^{2n}\right) \cr
  &  +\mu_{uu}^2(1-x) \left(\frac{ C \lambda^2 +  d^{u}_i \lambda +  C }{(x+\xi)\lambda}\right)^{2n}
    \left(1 + \frac{m_u}{M_u} (z_i^u)^{2n} \right)
    \left(1 + \frac{m_u}{M_u}\left(\frac{ A \lambda^2 +  b^{u}_i \lambda +  A }{(x+\xi)\lambda}\right)^{2n} \right)
     \cr
  &+\mu_{su}^2 (x+\xi) (z_i^{u})^{2n}
   \left(1 + \frac{m_u}{M_u}\left(\frac{ A \lambda^2 +  b^u_i \lambda +  A}{(x+\xi)\lambda}\right)^{2n}\right)
   \left(1 + \frac{m_s}{M_s}\left(\frac{ C \lambda^2 +  d^u_i \lambda +  C}{(x+\xi)\lambda}\right)^{2n}\right) \cr
 & + (z_i^u)^{2n}
    \left(\frac{ A \lambda^2 +  b^{u}_i \lambda +  A }{(x+\xi)\lambda}\right)^{2n}
    \left(\frac{ C \lambda^2 +  d^{u}_i \lambda +  C }{(x+\xi)\lambda}\right)^{2n}\cr
   &\times \left[
    (1-\xi)(\xi^2-x^2)\tau+ (\xi-1)\delta_\perp^2  + (1+\xi) \kappa_\perp^2
    +(-x+\xi)(1+z_i^u) +x \sqrt{\kappa_\perp^2 \delta_\perp^2} \frac{\lambda^2+1}{\lambda}
    \right].
\end{align}

Also, we obtain the expression for $I^{f/K^+}_3(x,\xi,t)$, 
for both quark flavors $f = (u, \bar s) $,
\begin{align}
    I_3^{f/K^+} (x, \xi, t) &= (-1)^n \frac{i N_c M_f^2}{4(2\pi)^3 F_K^2}
    \int^\infty_0 d \kappa_\perp^2\; \int_{C(0,1)} d\lambda
    \sum^{4n+1}_{i=1}  \phi^f_i \cr
    & \left[\prod^{4n+1}_{k=1}
   A \lambda^2 + B^f_{ik} \lambda + A 
    \right]^{-1} 
     [(x+\xi)z_i^f \lambda (A \lambda^2 + b_i^f+A)]^n \cr
   &  \times \left[(x+\xi)[(x+\xi)z_i^f \lambda]^{2n}+(x-\xi)(A \lambda^2 + b_i^f \lambda + A)^{2n}
    +2x \frac{m_f}{M_f}(A \lambda^2 + b_i^f \lambda + A)^{2n}
    \right]. \cr
\end{align}

In the above expressions, we introduced the following notation:
\begin{align}
    &\mu_{fg}^2 = \frac{M_f M_g}{\Lambda^2},  \cr 
    & A = 2x\sqrt{\kappa_\perp^2 \delta_\perp^2} ,\\ 
   & C = (1-x) \sqrt{\kappa_\perp^2 \delta_\perp^2}, \\
   &b_i^u = 2 \xi [(x^2-\xi^2)\tau + \kappa_\perp^2 + \delta_\perp^2 + 1] + (\xi - x ) z_i^u ,\\
   &B^u_{ik} =  2 \xi [(x^2-\xi^2) \tau + \kappa_\perp^2 + \delta_\perp^2 +1]+(\xi-x) z_i^u + (x + \xi ) z_k^u,\\
   & d_i^u = (x-1)[(1-\xi)(x+\xi)\tau + \delta_\perp^2]
    - (\kappa_\perp^2 +1)(\xi+1 ) + z_i^u(x-1) ,\\
    & D_{ik}^{u \bar s} = (x-1)[(1-\xi)(x+\xi)\tau + \delta_\perp^2]
    - (\kappa_\perp^2 +1)(\xi+1 ) + z_i^u(x-1) - z_k^{\bar s}(x+\xi),\\
     & f_i^{\bar s} = (x-1)[(x+\xi)(1-\xi)\tau + \delta_\perp^2]-(1+\xi)(\kappa_\perp^2+1)-(x+\xi)z_i^{\bar s},\\
   & F_{ik}^{\bar s u} = (x-1)[(x+\xi)(1-\xi)\tau + \delta_\perp^2]-(1+\xi)(\kappa_\perp^2+1)-(x+\xi)z_i^{\bar s} + (x-1)z_k^u,\\
    & g_i^{\bar s} = (x-1)[(\xi-x)(1+\xi)\tau-\delta_\perp^2]
    +(1-\xi)(\kappa_\perp^2+1)+z_i^{\bar s}(x-\xi), \\
    & G_{ik}^{\bar s u}=(x-1)[(\xi-x)(1+\xi)\tau-\delta_\perp^2]
    +(1-\xi)(\kappa_\perp^2+1)+z_i^{\bar s}(x-\xi) + (1-x)z_k^u, \\
    & \tilde{A} = -2x\sqrt{\kappa_\perp^2 \delta_\perp^2}, \\ 
   & \tilde C = (1+x) \sqrt{\kappa_\perp^2 \delta_\perp^2}, \\
   &\tilde b_i^{\bar{s}} = 2 \xi [(x^2-\xi^2)\tau + \kappa_\perp^2 + \delta_\perp^2 + 1] + (\xi +x ) z_i^{\bar s}, \\
   &\tilde B^{\bar{s}}_{ik} =  2 \xi [(x^2-\xi^2) \tau + \kappa_\perp^2 + \delta_\perp^2 +1]+(\xi+x) z_i^{\bar s} + ( \xi-x  ) z_k^{\bar s},\\
   &\tilde  d_i^{\bar{s}} = (-x-1)[(1-\xi)(-x+\xi)\tau + \delta_\perp^2]
    - (\kappa_\perp^2 +1)(\xi+1 ) + z_i^{\bar{s}}(-x-1) ,\\
    &\tilde  D_{ik}^{ \bar s u} = (-x-1)[(1-\xi)(-x+\xi)\tau + \delta_\perp^2]
    - (\kappa_\perp^2 +1)(\xi+1 ) + z_i^{\bar{s}}(-x-1) - z_k^u(-x+\xi),\\
 &\tilde  f_i^{u} = (-x-1)[(-x+\xi)(1-\xi)\tau + \delta_\perp^2]-(1+\xi)(\kappa_\perp^2+1)-(-x+\xi)z_i^u,\\
   &\tilde  F_{ik}^{ u \bar s} = (-x-1)[(-x+\xi)(1-\xi)\tau + \delta_\perp^2]-(1+\xi)(\kappa_\perp^2+1)-(-x+\xi)z_i^u + (-x-1)z_k^{\bar s},\\
    &\tilde  g_i^{u} = (-x-1)[(\xi+x)(1+\xi)\tau-\delta_\perp^2]
    +(1-\xi)(\kappa_\perp^2+1)+z_i^u(-x-\xi), \\
    &\tilde  G_{ik}^{u \bar s }=(-x-1)[(\xi+x)(1+\xi)\tau-\delta_\perp^2]
    +(1-\xi)(\kappa_\perp^2+1)+z_i^u(-x-\xi) + (1+x)z_k^{\bar s} .
\end{align}
The above expressions might look overwhelming, 
but objects with tilde are obtained by taking $x$ to $-x$ 
and switching the flavors $u \leftrightarrow \bar s$. 
The final integral over $\kappa_\perp$ is to be done numerically. 
In the present study, we only consider the simplest case $n=1$ for the 
momentum dependence of the dynamical quark mass. However, one can easily evaluate higher $n$ cases, by using the above formulas. From studies in Ref.~\cite{Praszalowicz:2003pr},
the dependence of the quark GPDs on the choice of $n$ is expected to be small. 
For the pion in the chiral limit $m_s = m_u = 0$,
one may easily check that the above results are identical
to those of Ref.~\cite{Praszalowicz:2003pr}. 

\section{Results in local chiral quark model}
In this section, we present the results for the 
kaon GPDs in the local chiral quark model (L$\chi$QM), 
\begin{align}\label{eq:local_action}
    S_{\mathrm{eff}} = \int \frac{d^4k}{(2\pi)^4}
    \bar \psi(k) (\Slash{k} - \hat m) \psi(k)
    - \int \frac{d^4k}{(2\pi)^4}\frac{d^4p}{(2\pi)^4}
    \bar \psi_f(p) \sqrt{M_f M_g} U_{fg}^{\gamma_5}(p-k)    \psi_g(k),
\end{align}
in which we took $F(k) \to 1 $ in the 
effective action of NL$\chi$QM, as in Eq.~(\ref{eq:action}). 
Computation of the matrix elements in this model is almost 
identical to that in the NL$\chi$QM, yet much simpler.

\subsection{Meson DAs}
Kaon DA in the local model can be computed straightforwardly and written as
follows:
\begin{align}
    \phi_{K^+}(u) = \frac{N_c\sqrt{M_u M_s} \Lambda}{(2\pi)^2 F_K^2}
    \int^\infty_0 d\kappa_\perp^2
    \frac{u \bar \mu_u + (1-u) \bar \mu_s}
    {\kappa_\perp^2 +u(u-1)\tilde m_K^2 
    + u \bar \mu_u^2 + (1-u) \bar \mu_s^2 
    },
\end{align}
where $\bar \mu_ u = (m_u+M_s)/\Lambda$, $\bar \mu_ s = (m_s+M_s)/\Lambda$, 
and $\tilde m_K = m_K/\Lambda$. 
For the pion DA, the result is obtained by letting $(m_s,M_s,m_K) \to (m_u,M_u,m_\pi)$ from the above expression. 
\begin{align}
    \phi_{\pi^+}(u) = \frac{N_cM_u (m_u + M_u) }{(2\pi)^2 F_\pi^2}
    \int^\infty_0 d\kappa_\perp^2
    \frac{1}
    {\kappa_\perp^2 +u(u-1)\tilde m_\pi^2 + \mu_u^2
    }.
\end{align}

For the pion in the chiral limit $(m_u, m_\pi)=0$,  
\begin{align}
    \phi_{\pi^+}(u) = \frac{N_c M_u^2 }{(2\pi)^2 F_\pi^2}
    \int^\infty_0 dk_\perp^2
    \frac{1}{k_\perp^2 + M_u^2 }.
\end{align}
The meson DAs are logarithmically divergent in $k_\perp \to \infty$, 
and thus one needs to introduce a regularization method, for instance, 
Pauli-Villars regularization. 
Note that in the chiral limit, the pion DA is independent of $u$. 
Once the current quark mass and the pion mass are introduced, 
$u$ dependence of the DA appears mildly in the form of $u(u-1)$,
symmetric with respect to $u=1/2$. 
Still, the result does not vanish at the endpoints, $u=0$ and $u=1$. 
For the kaon, we observe that the DA becomes more asymmetric due to
a rather large meson mass and difference of the light and strange 
current quark masses. 
In this section, we do not provide the numerical results 
on the meson DAs in the local model. We only use them 
to obtain the model parameters $m_s$, $m_u$, and $M_{\mathrm{PV}}$ 
(Pauli-Villars mass) from the normalization of them. 
Typically, for the form factors presented in Sec.~\ref{sec:FFs}, 
we used $M=350~$MeV and $m_u=5~$MeV as input and determined 
$M_{\mathrm{PV}}=560~$MeV, $m_s=130~$MeV, 
which obtains the meson decay constants $F_K=103.6~$MeV and $F_\pi=94.5~$MeV.
\subsection{Kaon GPDs in local chiral quark model}
\label{apdx:constant_mass}
The diagrammatic structure of the kaon GPDs in the local model
is same as Eqs.~(\ref{eq:gpd_uquark_Is}) and (\ref{eq:gpd_squark_Is}), 
which are again represented in the following loop integrals:
\begin{align}
    I_1^{u/K^+}(x,\xi,t) 
    &= \frac{N_c M_u M_s}{2(2\pi)^3 F^2} \int^\infty_0 d \kappa_\perp^2 
    \int_{C(0,1)}\frac{d\lambda}{i \lambda} (x-1)
    [F_0+ C (\lambda + \lambda^{-1})]^{-1}
    [G_0+ C (\lambda + \lambda^{-1})]^{-1}\cr
   & \left[
        (x+\xi) \bar{\mu}_{su}^2 - (x-1) \bar{\mu}_{uu}^2 + (x-\xi) \bar{\mu}_{su}^2
        + \frac{\xi^2-x^2}{x-1} \bar{\mu}_{ss}^2 + (x-1) \delta_\perp^2
        + \frac{\xi^2-1}{x-1} \kappa_\perp^2
        + \xi \sqrt{\kappa_\perp^2 \delta_\perp^2} (\lambda + \lambda^{-1})
    \right], \cr \\
    I_2^{u/K^+}(x,\xi,t) 
    &= -\frac{N_c M_u M_s}{2(2\pi)^3 F_K^2} \int^\infty_0 d \kappa_\perp^2 
    \int_{C(0,1)}\frac{d\lambda}{i \lambda} (x+\xi)
    [D_0+ C (\lambda + \lambda^{-1})]^{-1}
    [B_0^u+ A (\lambda + \lambda^{-1})]^{-1}\cr
   & \left[
        (x+\xi) \bar{\mu}_{su}^2 - (x-1) \bar{\mu}_{uu}^2 + (x-\xi) \bar{\mu}_{su}^2  + (\xi-x)\bar{\mu}_{uu}^2\right. \cr
    & \left.   + (1-\xi)(\xi^2-x^2)\tau 
        +(\xi-1)\delta_\perp^2 + (1+\xi) \kappa_\perp^2
        + x \sqrt{\kappa_\perp^2 \delta_\perp^2} (\lambda + \lambda^{-1})
    \right],\\
    I_1^{\bar s/K^+}(x,\xi,t) 
    &= -\frac{N_c M_u M_s}{2(2\pi)^3 F_K^2} \int^\infty_0 d \kappa_\perp^2 
    \int_{C(0,1)}\frac{d\lambda}{i \lambda} (-x-1)
    [\tilde F_0+ \tilde C (\lambda + \lambda^{-1})]^{-1}
    [\tilde G_0+ \tilde C (\lambda + \lambda^{-1})]^{-1}\cr
   & \left[
        (-x+\xi) \bar{\mu}_{su}^2 - (-x-1) \bar{\mu}_{ss}^2 + (-x-\xi) \bar{\mu}_{su}^2
        + \frac{\xi^2-x^2}{-x-1} \bar{\mu}_{uu}^2 + (-x-1) \delta_\perp^2
        + \frac{\xi^2-1}{-x-1} \kappa_\perp^2
        + \xi \sqrt{\kappa_\perp^2 \delta_\perp^2} (\lambda + \lambda^{-1})
    \right],\cr \\
    I_2^{\bar s/K^+}(x,\xi,t) 
    &= \frac{N_c M_u M_s}{2(2\pi)^3 F_K^2} \int^\infty_0 d \kappa_\perp^2 
    \int_{C(0,1)}\frac{d\lambda}{i \lambda} (-x+\xi)
    [\tilde D_0+ \tilde C (\lambda + \lambda^{-1})]^{-1}
    [\tilde B_0^s+ \tilde A (\lambda + \lambda^{-1})]^{-1}\cr
   & \left[
        (-x-\xi) \bar{\mu}_{su}^2 - (-x-1) \bar{\mu}_{ss}^2 + (\xi-x) \bar{\mu}_{su}^2  + (\xi+x)\bar{\mu}_{ss}^2\right. \cr
    & \left.   + (1-\xi)(\xi^2-x^2)\tau 
        +(\xi-1)\delta_\perp^2 + (1+\xi) \kappa_\perp^2
        - x \sqrt{\kappa_\perp^2 \delta_\perp^2} (\lambda + \lambda^{-1})
    \right],\\
    I_3^{f/K^+} (x,\xi,t) 
    & = - \frac{N_c M_f (m_f+M_f)}{2 (2\pi)^3 F_K^2} \int^\infty_0 d \kappa_\perp^2     \int_{C(0,1)}\frac{d\lambda}{i \lambda}
    [B_0^f + A (\lambda + \lambda^{-1})]^{-1}.
\end{align}
In the above expressions, we introduced the following conventions:
\begin{align}
     &A = 2x\sqrt{\kappa_\perp^2 \delta_\perp^2} ,\\ 
     &C = (1-x) \sqrt{\kappa_\perp^2 \delta_\perp^2}, \\
    &\bar{\mu}_f^2 = \frac{(m_f+M_f)^2}{\Lambda^2}, \\
    &\bar{\mu}_{fg}^2 = \frac{(m_f+M_f)(m_g+M_g)}{\Lambda^2}, \\
    &B_0^f = 2\xi [(x^2-\xi^2)\tau + \kappa_\perp^2 + \delta_\perp^2 + 
    \bar{\mu}_f^2 ],\\
    &D_0 =(x-1)[(x+\xi)(1-\xi)\tau + \delta_\perp^2]
    -(1+\xi)\kappa_\perp^2 + x(\bar{\mu}_u^2-\bar{\mu}_s^2)
    - \bar{\mu}_u^2(1+\xi\frac{\bar{\mu}_s^2}{\bar{\mu}_u^2}),\\
    &F_0 = D_0, \\
    &G_0 = (x-1) [(\xi-x)(1+\xi)\tau - \delta_\perp^2]
    +(1-\xi) \kappa_\perp^2 + x (\bar{\mu}_s^2- \bar{\mu}_u^2 )
    + \bar{\mu}_u^2(1-\xi \frac{\bar{\mu}_s^2}{\bar{\mu}_u^2}),\\
     &\tilde{A} = -2x\sqrt{\kappa_\perp^2 \delta_\perp^2}, \\ 
     &\tilde C = (1+x) \sqrt{\kappa_\perp^2 \delta_\perp^2}, \\
    &\tilde D_0 = (-x-1) [(-x+\xi)(1-\xi)\tau + \delta_\perp^2]
    -(1+\xi) \kappa_\perp^2 - x (\bar{\mu}_s^2- \bar{\mu}_u^2 )
    - \bar{\mu}_s^2(1+\xi \frac{\bar{\mu}_u^2}{\bar{\mu}_s^2}),\\
   &\tilde F_0 = \tilde D_0,\\
    &\tilde G_0 = (-x-1) [(\xi+x)(1+\xi)\tau - \delta_\perp^2]
    +(1-\xi) \kappa_\perp^2 + x (\bar{\mu}_s^2- \bar{\mu}_u^2 )
    + \bar{\mu}_s^2(1-\xi \frac{\bar{\mu}_s^2}{\bar{\mu}_s^2}).
\end{align}
The functions have UV divergences in the $\kappa_\perp$ integral, thus one has to introduce a suitable regularization scheme with 
finite cutoff. For the numerical computation, we introduce the Pauli-Villars regularization with single subtraction, for simplicity. 
Also, note that $\Lambda$ here is an arbitrary parameter with dimension of mass and the result should not be dependent on the choice of it.  

\end{appendix}

\bibliographystyle{elsarticle-num}
\bibliography{Kaon_GPD_NLChQM}

\begin{thebibliography}{10}
\expandafter\ifx\csname url\endcsname\relax
  \def\url#1{\texttt{#1}}\fi
\expandafter\ifx\csname urlprefix\endcsname\relax\def\urlprefix{URL }\fi
\expandafter\ifx\csname href\endcsname\relax
  \def\href#1#2{#2} \def\path#1{#1}\fi

\bibitem{Banks:1979yr}
T.~Banks, A.~Casher, {Chiral Symmetry Breaking in Confining Theories}, Nucl. Phys. B 169 (1980) 103--125.
\newblock \href {https://doi.org/10.1016/0550-3213(80)90255-2} {\path{doi:10.1016/0550-3213(80)90255-2}}.

\bibitem{Gell-Mann:1968hlm}
M.~Gell-Mann, R.~J. Oakes, B.~Renner, {Behavior of current divergences under SU(3) x SU(3)}, Phys. Rev. 175 (1968) 2195--2199.
\newblock \href {https://doi.org/10.1103/PhysRev.175.2195} {\path{doi:10.1103/PhysRev.175.2195}}.

\bibitem{Muller:1994ses}
D.~M\"uller, D.~Robaschik, B.~Geyer, F.~M. Dittes, J.~Ho\v{r}ej\v{s}i, {Wave functions, evolution equations and evolution kernels from light ray operators of QCD}, Fortsch. Phys. 42 (1994) 101--141.
\newblock \href {http://arxiv.org/abs/hep-ph/9812448} {\path{arXiv:hep-ph/9812448}}, \href {https://doi.org/10.1002/prop.2190420202} {\path{doi:10.1002/prop.2190420202}}.

\bibitem{Drell:1970wh}
S.~D. Drell, T.-M. Yan, {Massive Lepton Pair Production in Hadron-Hadron Collisions at High-Energies}, Phys. Rev. Lett. 25 (1970) 316--320, [Erratum: Phys.Rev.Lett. 25, 902 (1970)].
\newblock \href {https://doi.org/10.1103/PhysRevLett.25.316} {\path{doi:10.1103/PhysRevLett.25.316}}.

\bibitem{Berger:1979du}
E.~L. Berger, S.~J. Brodsky, {Quark Structure Functions of Mesons and the Drell-Yan Process}, Phys. Rev. Lett. 42 (1979) 940--944.
\newblock \href {https://doi.org/10.1103/PhysRevLett.42.940} {\path{doi:10.1103/PhysRevLett.42.940}}.

\bibitem{Sullivan:1971kd}
J.~D. Sullivan, {One pion exchange and deep inelastic electron - nucleon scattering}, Phys. Rev. D 5 (1972) 1732--1737.
\newblock \href {https://doi.org/10.1103/PhysRevD.5.1732} {\path{doi:10.1103/PhysRevD.5.1732}}.

\bibitem{Ji:1996ek}
X.-D. Ji, {Gauge-Invariant Decomposition of Nucleon Spin}, Phys. Rev. Lett. 78 (1997) 610--613.
\newblock \href {http://arxiv.org/abs/hep-ph/9603249} {\path{arXiv:hep-ph/9603249}}, \href {https://doi.org/10.1103/PhysRevLett.78.610} {\path{doi:10.1103/PhysRevLett.78.610}}.

\bibitem{Amrath:2008vx}
D.~Amrath, M.~Diehl, J.-P. Lansberg, {Deeply virtual Compton scattering on a virtual pion target}, Eur. Phys. J. C 58 (2008) 179--192.
\newblock \href {http://arxiv.org/abs/0807.4474} {\path{arXiv:0807.4474}}, \href {https://doi.org/10.1140/epjc/s10052-008-0769-1} {\path{doi:10.1140/epjc/s10052-008-0769-1}}.

\bibitem{Accardi:2023chb}
A.~Accardi, et~al., {Strong interaction physics at the luminosity frontier with 22 GeV electrons at Jefferson Lab}, Eur. Phys. J. A 60~(9) (2024) 173.
\newblock \href {http://arxiv.org/abs/2306.09360} {\path{arXiv:2306.09360}}, \href {https://doi.org/10.1140/epja/s10050-024-01282-x} {\path{doi:10.1140/epja/s10050-024-01282-x}}.

\bibitem{Arrington:2021biu}
J.~Arrington, et~al., {Revealing the structure of light pseudoscalar mesons at the electron\textendash{}ion collider}, J. Phys. G 48~(7) (2021) 075106.
\newblock \href {http://arxiv.org/abs/2102.11788} {\path{arXiv:2102.11788}}, \href {https://doi.org/10.1088/1361-6471/abf5c3} {\path{doi:10.1088/1361-6471/abf5c3}}.

\bibitem{Anderle:2021wcy}
D.~P. Anderle, et~al., {Electron-ion collider in China}, Front. Phys. (Beijing) 16~(6) (2021) 64701.
\newblock \href {http://arxiv.org/abs/2102.09222} {\path{arXiv:2102.09222}}, \href {https://doi.org/10.1007/s11467-021-1062-0} {\path{doi:10.1007/s11467-021-1062-0}}.

\bibitem{Sawada:2016mao}
T.~Sawada, W.-C. Chang, S.~Kumano, J.-C. Peng, S.~Sawada, K.~Tanaka, {Accessing proton generalized parton distributions and pion distribution amplitudes with the exclusive pion-induced Drell-Yan process at J-PARC}, Phys. Rev. D 93~(11) (2016) 114034.
\newblock \href {http://arxiv.org/abs/1605.00364} {\path{arXiv:1605.00364}}, \href {https://doi.org/10.1103/PhysRevD.93.114034} {\path{doi:10.1103/PhysRevD.93.114034}}.

\bibitem{Adams:2018pwt}
B.~Adams, et~al., {Letter of Intent: A New QCD facility at the M2 beam line of the CERN SPS (COMPASS++/AMBER)} (8 2018).
\newblock \href {http://arxiv.org/abs/1808.00848} {\path{arXiv:1808.00848}}.

\bibitem{Chavez:2021koz}
J.~M.~M. Ch\'avez, V.~Bertone, F.~De~Soto~Borrero, M.~Defurne, C.~Mezrag, H.~Moutarde, J.~Rodr\'\i{}guez-Quintero, J.~Segovia, {Accessing the Pion 3D Structure at US and China Electron-Ion Colliders}, Phys. Rev. Lett. 128~(20) (2022) 202501.
\newblock \href {http://arxiv.org/abs/2110.09462} {\path{arXiv:2110.09462}}, \href {https://doi.org/10.1103/PhysRevLett.128.202501} {\path{doi:10.1103/PhysRevLett.128.202501}}.

\bibitem{Mezrag:2022pqk}
C.~Mezrag, {An Introductory Lecture on Generalised Parton Distributions}, Few Body Syst. 63~(3) (2022) 62.
\newblock \href {http://arxiv.org/abs/2207.13584} {\path{arXiv:2207.13584}}, \href {https://doi.org/10.1007/s00601-022-01765-x} {\path{doi:10.1007/s00601-022-01765-x}}.

\bibitem{Theussl:2002xp}
L.~Theussl, S.~Noguera, V.~Vento, {Generalized parton distributions of the pion in a Bethe-Salpeter approach}, Eur. Phys. J. A 20 (2004) 483--498.
\newblock \href {http://arxiv.org/abs/nucl-th/0211036} {\path{arXiv:nucl-th/0211036}}, \href {https://doi.org/10.1140/epja/i2003-10174-3} {\path{doi:10.1140/epja/i2003-10174-3}}.

\bibitem{Broniowski:2003rp}
W.~Broniowski, E.~Ruiz~Arriola, {Impact parameter dependence of the generalized parton distribution of the pion in chiral quark models}, Phys. Lett. B 574 (2003) 57--64.
\newblock \href {http://arxiv.org/abs/hep-ph/0307198} {\path{arXiv:hep-ph/0307198}}, \href {https://doi.org/10.1016/j.physletb.2003.09.009} {\path{doi:10.1016/j.physletb.2003.09.009}}.

\bibitem{Hutauruk:2016sug}
P.~T.~P. Hutauruk, I.~C. Cloet, A.~W. Thomas, {Flavor dependence of the pion and kaon form factors and parton distribution functions}, Phys. Rev. C 94~(3) (2016) 035201.
\newblock \href {http://arxiv.org/abs/1604.02853} {\path{arXiv:1604.02853}}, \href {https://doi.org/10.1103/PhysRevC.94.035201} {\path{doi:10.1103/PhysRevC.94.035201}}.

\bibitem{Shastry:2023fnc}
V.~Shastry, W.~Broniowski, E.~Ruiz~Arriola, {Off-shellness in generalized parton distributions and form factors of the pion}, Phys. Rev. D 108~(11) (2023) 114024.
\newblock \href {http://arxiv.org/abs/2308.09236} {\path{arXiv:2308.09236}}, \href {https://doi.org/10.1103/PhysRevD.108.114024} {\path{doi:10.1103/PhysRevD.108.114024}}.

\bibitem{Praszalowicz:2001wy}
M.~Praszalowicz, A.~Rostworowski, {Pion light cone wave function in the nonlocal NJL model}, Phys. Rev. D 64 (2001) 074003.
\newblock \href {http://arxiv.org/abs/hep-ph/0105188} {\path{arXiv:hep-ph/0105188}}, \href {https://doi.org/10.1103/PhysRevD.64.074003} {\path{doi:10.1103/PhysRevD.64.074003}}.

\bibitem{Praszalowicz:2003pr}
M.~Praszalowicz, A.~Rostworowski, {Pion generalized distribution amplitudes in the nonlocal chiral quark model}, Acta Phys. Polon. B 34 (2003) 2699--2730.
\newblock \href {http://arxiv.org/abs/hep-ph/0302269} {\path{arXiv:hep-ph/0302269}}.

\bibitem{Hutauruk:2023ccw}
P.~T.~P. Hutauruk, S.-i. Nam, {Updated analyses of gluon distribution functions for the pion and kaon from the gauge-invariant nonlocal chiral quark model}, Phys. Rev. D 109~(5) (2024) 054040.
\newblock \href {http://arxiv.org/abs/2302.05566} {\path{arXiv:2302.05566}}, \href {https://doi.org/10.1103/PhysRevD.109.054040} {\path{doi:10.1103/PhysRevD.109.054040}}.

\bibitem{Adhikari:2021jrh}
L.~Adhikari, C.~Mondal, S.~Nair, S.~Xu, S.~Jia, X.~Zhao, J.~P. Vary, {Generalized parton distributions and spin structures of light mesons from a light-front Hamiltonian approach}, Phys. Rev. D 104~(11) (2021) 114019.
\newblock \href {http://arxiv.org/abs/2110.05048} {\path{arXiv:2110.05048}}, \href {https://doi.org/10.1103/PhysRevD.104.114019} {\path{doi:10.1103/PhysRevD.104.114019}}.

\bibitem{Davidson:1994uv}
R.~M. Davidson, E.~Ruiz~Arriola, {Structure functions of pseudoscalar mesons in the SU(3) NJL model}, Phys. Lett. B 348 (1995) 163--169.
\newblock \href {https://doi.org/10.1016/0370-2693(95)00091-X} {\path{doi:10.1016/0370-2693(95)00091-X}}.

\bibitem{Davidson:2001cc}
R.~M. Davidson, E.~Ruiz~Arriola, {Parton distributions functions of pion, kaon and eta pseudoscalar mesons in the NJL model}, Acta Phys. Polon. B 33 (2002) 1791--1808.
\newblock \href {http://arxiv.org/abs/hep-ph/0110291} {\path{arXiv:hep-ph/0110291}}.

\bibitem{Nam:2006au}
S.-i. Nam, H.-C. Kim, A.~Hosaka, M.~M. Musakhanov, {The Leading-twist pion and kaon distribution amplitudes from the QCD instanton vacuum}, Phys. Rev. D 74 (2006) 014019.
\newblock \href {http://arxiv.org/abs/hep-ph/0605259} {\path{arXiv:hep-ph/0605259}}, \href {https://doi.org/10.1103/PhysRevD.74.014019} {\path{doi:10.1103/PhysRevD.74.014019}}.

\bibitem{Nam:2012vm}
S.-i. Nam, {Parton-distribution functions for the pion and kaon in the gauge-invariant nonlocal chiral-quark model}, Phys. Rev. D 86 (2012) 074005.
\newblock \href {http://arxiv.org/abs/1205.4156} {\path{arXiv:1205.4156}}, \href {https://doi.org/10.1103/PhysRevD.86.074005} {\path{doi:10.1103/PhysRevD.86.074005}}.

\bibitem{Nam:2007gf}
S.-i. Nam, H.-C. Kim, {Electromagnetic form factors of the pion and kaon from the instanton vacuum}, Phys. Rev. D 77 (2008) 094014.
\newblock \href {http://arxiv.org/abs/0709.1745} {\path{arXiv:0709.1745}}, \href {https://doi.org/10.1103/PhysRevD.77.094014} {\path{doi:10.1103/PhysRevD.77.094014}}.

\bibitem{Cui:2022bxn}
Z.~F. Cui, M.~Ding, J.~M. Morgado, K.~Raya, D.~Binosi, L.~Chang, F.~De~Soto, C.~D. Roberts, J.~Rodr\'\i{}guez-Quintero, S.~M. Schmidt, {Emergence of pion parton distributions}, Phys. Rev. D 105~(9) (2022) L091502.
\newblock \href {http://arxiv.org/abs/2201.00884} {\path{arXiv:2201.00884}}, \href {https://doi.org/10.1103/PhysRevD.105.L091502} {\path{doi:10.1103/PhysRevD.105.L091502}}.

\bibitem{Xu:2023izo}
Y.-Z. Xu, M.~Ding, K.~Raya, C.~D. Roberts, J.~Rodr\'\i{}guez-Quintero, S.~M. Schmidt, {Pion and kaon electromagnetic and gravitational form factors}, Eur. Phys. J. C 84~(2) (2024) 191.
\newblock \href {http://arxiv.org/abs/2311.14832} {\path{arXiv:2311.14832}}, \href {https://doi.org/10.1140/epjc/s10052-024-12518-x} {\path{doi:10.1140/epjc/s10052-024-12518-x}}.

\bibitem{Xing:2023wuk}
H.~Y. Xing, M.~Ding, Z.~F. Cui, A.~V. Pimikov, C.~D. Roberts, S.~M. Schmidt, {Constraining the pion distribution amplitude using Drell-Yan reactions on a proton}, Phys. Lett. B 849 (2024) 138462.
\newblock \href {http://arxiv.org/abs/2308.13695} {\path{arXiv:2308.13695}}, \href {https://doi.org/10.1016/j.physletb.2024.138462} {\path{doi:10.1016/j.physletb.2024.138462}}.

\bibitem{Zhang:2021mtn}
J.-L. Zhang, K.~Raya, L.~Chang, Z.-F. Cui, J.~M. Morgado, C.~D. Roberts, J.~Rodr\'\i{}guez-Quintero, {Measures of pion and kaon structure from generalised parton distributions}, Phys. Lett. B 815 (2021) 136158.
\newblock \href {http://arxiv.org/abs/2101.12286} {\path{arXiv:2101.12286}}, \href {https://doi.org/10.1016/j.physletb.2021.136158} {\path{doi:10.1016/j.physletb.2021.136158}}.

\bibitem{Raya:2021zrz}
K.~Raya, Z.-F. Cui, L.~Chang, J.-M. Morgado, C.~D. Roberts, J.~Rodriguez-Quintero, {Revealing pion and kaon structure via generalised parton distributions *}, Chin. Phys. C 46~(1) (2022) 013105.
\newblock \href {http://arxiv.org/abs/2109.11686} {\path{arXiv:2109.11686}}, \href {https://doi.org/10.1088/1674-1137/ac3071} {\path{doi:10.1088/1674-1137/ac3071}}.

\bibitem{Xing:2023eed}
Z.~Xing, M.~Ding, K.~Raya, L.~Chang, {Fresh look at the generalized parton distributions of light pseudoscalar mesons}, Eur. Phys. J. A 60~(2) (2024) 33.
\newblock \href {http://arxiv.org/abs/2301.02958} {\path{arXiv:2301.02958}}, \href {https://doi.org/10.1140/epja/s10050-024-01256-z} {\path{doi:10.1140/epja/s10050-024-01256-z}}.

\bibitem{Bourrely:2023yzi}
C.~Bourrely, F.~Buccella, W.-C. Chang, J.-C. Peng, {Extraction of kaon partonic distribution functions from Drell-Yan and J/\ensuremath{\psi} production data}, Phys. Lett. B 848 (2024) 138395.
\newblock \href {http://arxiv.org/abs/2305.18117} {\path{arXiv:2305.18117}}, \href {https://doi.org/10.1016/j.physletb.2023.138395} {\path{doi:10.1016/j.physletb.2023.138395}}.

\bibitem{Brommel:2005ee}
D.~Brommel, M.~Diehl, M.~Gockeler, P.~Hagler, R.~Horsley, D.~Pleiter, P.~E.~L. Rakow, A.~Schafer, G.~Schierholz, J.~M. Zanotti, {Structure of the pion from full lattice QCD}, PoS LAT2005 (2006) 360.
\newblock \href {http://arxiv.org/abs/hep-lat/0509133} {\path{arXiv:hep-lat/0509133}}, \href {https://doi.org/10.22323/1.020.0360} {\path{doi:10.22323/1.020.0360}}.

\bibitem{Brommel:2007zz}
D.~Brommel, {Pion Structure from the Lattice}, Ph.D. thesis, Regensburg U. (2007).
\newblock \href {https://doi.org/10.3204/DESY-THESIS-2007-023} {\path{doi:10.3204/DESY-THESIS-2007-023}}.

\bibitem{Hackett:2023nkr}
D.~C. Hackett, P.~R. Oare, D.~A. Pefkou, P.~E. Shanahan, {Gravitational form factors of the pion from lattice QCD}, Phys. Rev. D 108~(11) (2023) 114504.
\newblock \href {http://arxiv.org/abs/2307.11707} {\path{arXiv:2307.11707}}, \href {https://doi.org/10.1103/PhysRevD.108.114504} {\path{doi:10.1103/PhysRevD.108.114504}}.

\bibitem{Delmar:2024vxn}
J.~Delmar, C.~Alexandrou, S.~Bacchio, I.~Clo\"et, M.~Constantinou, G.~Koutsou, {Generalized form factors of the pion and kaon using twisted mass fermions}, PoS LATTICE2023 (2024) 308.
\newblock \href {http://arxiv.org/abs/2401.04080} {\path{arXiv:2401.04080}}, \href {https://doi.org/10.22323/1.453.0308} {\path{doi:10.22323/1.453.0308}}.

\bibitem{Constantinou:2020hdm}
M.~Constantinou, et~al., {Parton distributions and lattice-QCD calculations: Toward 3D structure}, Prog. Part. Nucl. Phys. 121 (2021) 103908.
\newblock \href {http://arxiv.org/abs/2006.08636} {\path{arXiv:2006.08636}}, \href {https://doi.org/10.1016/j.ppnp.2021.103908} {\path{doi:10.1016/j.ppnp.2021.103908}}.

\bibitem{Ji:1996nm}
X.-D. Ji, {Deeply virtual Compton scattering}, Phys. Rev. D 55 (1997) 7114--7125.
\newblock \href {http://arxiv.org/abs/hep-ph/9609381} {\path{arXiv:hep-ph/9609381}}, \href {https://doi.org/10.1103/PhysRevD.55.7114} {\path{doi:10.1103/PhysRevD.55.7114}}.

\bibitem{Radyushkin:1997ki}
A.~V. Radyushkin, {Nonforward parton distributions}, Phys. Rev. D 56 (1997) 5524--5557.
\newblock \href {http://arxiv.org/abs/hep-ph/9704207} {\path{arXiv:hep-ph/9704207}}, \href {https://doi.org/10.1103/PhysRevD.56.5524} {\path{doi:10.1103/PhysRevD.56.5524}}.

\bibitem{Balitsky:1997mj}
I.~I. Balitsky, A.~V. Radyushkin, {Light ray evolution equations and leading twist parton helicity dependent nonforward distributions}, Phys. Lett. B 413 (1997) 114--121.
\newblock \href {http://arxiv.org/abs/hep-ph/9706410} {\path{arXiv:hep-ph/9706410}}, \href {https://doi.org/10.1016/S0370-2693(97)01095-2} {\path{doi:10.1016/S0370-2693(97)01095-2}}.

\bibitem{Radyushkin:1998es}
A.~V. Radyushkin, {Double distributions and evolution equations}, Phys. Rev. D 59 (1999) 014030.
\newblock \href {http://arxiv.org/abs/hep-ph/9805342} {\path{arXiv:hep-ph/9805342}}, \href {https://doi.org/10.1103/PhysRevD.59.014030} {\path{doi:10.1103/PhysRevD.59.014030}}.

\bibitem{Blumlein:1997pi}
J.~Blumlein, B.~Geyer, D.~Robaschik, {On the evolution kernels of twist-2 light ray operators for unpolarized and polarized deep inelastic scattering}, Phys. Lett. B 406 (1997) 161--170.
\newblock \href {http://arxiv.org/abs/hep-ph/9705264} {\path{arXiv:hep-ph/9705264}}, \href {https://doi.org/10.1016/S0370-2693(97)00680-1} {\path{doi:10.1016/S0370-2693(97)00680-1}}.

\bibitem{Blumlein:1999sc}
J.~Blumlein, B.~Geyer, D.~Robaschik, {The Virtual Compton amplitude in the generalized Bjorken region: twist-2 contributions}, Nucl. Phys. B 560 (1999) 283--344.
\newblock \href {http://arxiv.org/abs/hep-ph/9903520} {\path{arXiv:hep-ph/9903520}}, \href {https://doi.org/10.1016/S0550-3213(99)00418-6} {\path{doi:10.1016/S0550-3213(99)00418-6}}.

\bibitem{Bertone:2022frx}
V.~Bertone, H.~Dutrieux, C.~Mezrag, J.~M. Morgado, H.~Moutarde, {Revisiting evolution equations for generalised parton distributions}, Eur. Phys. J. C 82~(10) (2022) 888.
\newblock \href {http://arxiv.org/abs/2206.01412} {\path{arXiv:2206.01412}}, \href {https://doi.org/10.1140/epjc/s10052-022-10793-0} {\path{doi:10.1140/epjc/s10052-022-10793-0}}.

\bibitem{Diakonov:1979nj}
D.~Diakonov, V.~Y. Petrov, {Quark Propagator and Chiral Condensate in an Instanton Vacuum}, Sov. Phys. JETP 62 (1985) 204--214.

\bibitem{Diakonov:1984vw}
D.~Diakonov, V.~Y. Petrov, {CHIRAL CONDENSATE IN THE INSTANTON VACUUM}, Phys. Lett. B 147 (1984) 351--356.
\newblock \href {https://doi.org/10.1016/0370-2693(84)90132-1} {\path{doi:10.1016/0370-2693(84)90132-1}}.

\bibitem{Diakonov:1985eg}
D.~Diakonov, V.~Y. Petrov, {A Theory of Light Quarks in the Instanton Vacuum}, Nucl. Phys. B 272 (1986) 457--489.
\newblock \href {https://doi.org/10.1016/0550-3213(86)90011-8} {\path{doi:10.1016/0550-3213(86)90011-8}}.

\bibitem{Diakonov:1985fjw}
D.~Diakonov, V.~Y. Petrov, {Meson Current Correlation Functions in Instanton Vacuum}, Sov. Phys. JETP 62 (1985) 431--437.

\bibitem{Diakonov:1997vc}
D.~Diakonov, V.~Y. Petrov, P.~V. Pobylitsa, M.~V. Polyakov, C.~Weiss, {Unpolarized and polarized quark distributions in the large N(c) limit}, Phys. Rev. D 56 (1997) 4069--4083.
\newblock \href {http://arxiv.org/abs/hep-ph/9703420} {\path{arXiv:hep-ph/9703420}}, \href {https://doi.org/10.1103/PhysRevD.56.4069} {\path{doi:10.1103/PhysRevD.56.4069}}.

\bibitem{Son:2014sna}
H.-D. Son, H.-C. Kim, {Stability of the pion and the pattern of chiral symmetry breaking}, Phys. Rev. D 90~(11) (2014) 111901.
\newblock \href {http://arxiv.org/abs/1410.1420} {\path{arXiv:1410.1420}}, \href {https://doi.org/10.1103/PhysRevD.90.111901} {\path{doi:10.1103/PhysRevD.90.111901}}.

\bibitem{Polyakov:1999gs}
M.~V. Polyakov, C.~Weiss, {Skewed and double distributions in pion and nucleon}, Phys. Rev. D 60 (1999) 114017.
\newblock \href {http://arxiv.org/abs/hep-ph/9902451} {\path{arXiv:hep-ph/9902451}}, \href {https://doi.org/10.1103/PhysRevD.60.114017} {\path{doi:10.1103/PhysRevD.60.114017}}.

\bibitem{Belitsky:1998gc}
A.~V. Belitsky, D.~Mueller, {Broken conformal invariance and spectrum of anomalous dimensions in QCD}, Nucl. Phys. B 537 (1999) 397--442.
\newblock \href {http://arxiv.org/abs/hep-ph/9804379} {\path{arXiv:hep-ph/9804379}}, \href {https://doi.org/10.1016/S0550-3213(98)00677-4} {\path{doi:10.1016/S0550-3213(98)00677-4}}.

\bibitem{Belitsky:1999gu}
A.~V. Belitsky, D.~Mueller, A.~Freund, {Reconstruction of nonforward evolution kernels}, Phys. Lett. B 461 (1999) 270--279.
\newblock \href {http://arxiv.org/abs/hep-ph/9904477} {\path{arXiv:hep-ph/9904477}}, \href {https://doi.org/10.1016/S0370-2693(99)00837-0} {\path{doi:10.1016/S0370-2693(99)00837-0}}.

\bibitem{Belitsky:1999fu}
A.~V. Belitsky, D.~Mueller, {Exclusive evolution kernels in two loop order: Parity even sector}, Phys. Lett. B 464 (1999) 249--256.
\newblock \href {http://arxiv.org/abs/hep-ph/9906409} {\path{arXiv:hep-ph/9906409}}, \href {https://doi.org/10.1016/S0370-2693(99)01003-5} {\path{doi:10.1016/S0370-2693(99)01003-5}}.

\bibitem{Belitsky:1999hf}
A.~V. Belitsky, A.~Freund, D.~Mueller, {Evolution kernels of skewed parton distributions: Method and two loop results}, Nucl. Phys. B 574 (2000) 347--406.
\newblock \href {http://arxiv.org/abs/hep-ph/9912379} {\path{arXiv:hep-ph/9912379}}, \href {https://doi.org/10.1016/S0550-3213(00)00012-2} {\path{doi:10.1016/S0550-3213(00)00012-2}}.

\bibitem{Braun:2019qtp}
V.~M. Braun, A.~N. Manashov, S.~Moch, M.~Strohmaier, {Two-loop evolution equations for flavor-singlet light-ray operators}, JHEP 02 (2019) 191.
\newblock \href {http://arxiv.org/abs/1901.06172} {\path{arXiv:1901.06172}}, \href {https://doi.org/10.1007/JHEP02(2019)191} {\path{doi:10.1007/JHEP02(2019)191}}.

\bibitem{Vinnikov:2006xw}
A.~V. Vinnikov, {Code for prompt numerical computation of the leading order GPD evolution} (4 2006).
\newblock \href {http://arxiv.org/abs/hep-ph/0604248} {\path{arXiv:hep-ph/0604248}}.

\bibitem{Bertone:2017gds}
V.~Bertone, {APFEL++: A new PDF evolution library in C++}, PoS DIS2017 (2018) 201.
\newblock \href {http://arxiv.org/abs/1708.00911} {\path{arXiv:1708.00911}}, \href {https://doi.org/10.22323/1.297.0201} {\path{doi:10.22323/1.297.0201}}.

\bibitem{Broniowski:1999bz}
W.~Broniowski, {Gauging nonlocal quark models}, in: {Mini-Workshop Bled 1999}: {Hadrons as Solitons}, 1999, pp. 17--26.
\newblock \href {http://arxiv.org/abs/hep-ph/9909438} {\path{arXiv:hep-ph/9909438}}.

\bibitem{Dorokhov:2000gu}
A.~E. Dorokhov, L.~Tomio, {Pion structure function within the instanton model}, Phys. Rev. D 62 (2000) 014016.
\newblock \href {https://doi.org/10.1103/PhysRevD.62.014016} {\path{doi:10.1103/PhysRevD.62.014016}}.

\bibitem{Kim:2004hd}
H.-C. Kim, M.~Musakhanov, M.~Siddikov, {Magnetic susceptibility of the QCD vacuum}, Phys. Lett. B 608 (2005) 95--106.
\newblock \href {http://arxiv.org/abs/hep-ph/0411181} {\path{arXiv:hep-ph/0411181}}, \href {https://doi.org/10.1016/j.physletb.2004.12.080} {\path{doi:10.1016/j.physletb.2004.12.080}}.

\bibitem{Saclay-CERN-CollegedeFrance-EcolePoly-Orsay:1980fhh}
J.~Badier, et~al., {Measurement of the $K^- / \pi^-$ Structure Function Ratio Using the {Drell-Yan} Process}, Phys. Lett. B 93 (1980) 354--356.
\newblock \href {https://doi.org/10.1016/0370-2693(80)90530-4} {\path{doi:10.1016/0370-2693(80)90530-4}}.

\bibitem{Polyakov:2018zvc}
M.~V. Polyakov, P.~Schweitzer, {Forces inside hadrons: pressure, surface tension, mechanical radius, and all that}, Int. J. Mod. Phys. A 33~(26) (2018) 1830025.
\newblock \href {http://arxiv.org/abs/1805.06596} {\path{arXiv:1805.06596}}, \href {https://doi.org/10.1142/S0217751X18300259} {\path{doi:10.1142/S0217751X18300259}}.

\bibitem{Polyakov:2002yz}
M.~V. Polyakov, {Generalized parton distributions and strong forces inside nucleons and nuclei}, Phys. Lett. B 555 (2003) 57--62.
\newblock \href {http://arxiv.org/abs/hep-ph/0210165} {\path{arXiv:hep-ph/0210165}}, \href {https://doi.org/10.1016/S0370-2693(03)00036-4} {\path{doi:10.1016/S0370-2693(03)00036-4}}.

\bibitem{Perevalova:2016dln}
I.~A. Perevalova, M.~V. Polyakov, P.~Schweitzer, {On LHCb pentaquarks as a baryon-$\psi$(2S) bound state: prediction of isospin-$\frac3{2}$ pentaquarks with hidden charm}, Phys. Rev. D 94~(5) (2016) 054024.
\newblock \href {http://arxiv.org/abs/1607.07008} {\path{arXiv:1607.07008}}, \href {https://doi.org/10.1103/PhysRevD.94.054024} {\path{doi:10.1103/PhysRevD.94.054024}}.

\bibitem{Ji:1997gm}
X.-D. Ji, W.~Melnitchouk, X.~Song, {A Study of off forward parton distributions}, Phys. Rev. D 56 (1997) 5511--5523.
\newblock \href {http://arxiv.org/abs/hep-ph/9702379} {\path{arXiv:hep-ph/9702379}}, \href {https://doi.org/10.1103/PhysRevD.56.5511} {\path{doi:10.1103/PhysRevD.56.5511}}.

\bibitem{Goeke:2007fp}
K.~Goeke, J.~Grabis, J.~Ossmann, M.~V. Polyakov, P.~Schweitzer, A.~Silva, D.~Urbano, {Nucleon form-factors of the energy momentum tensor in the chiral quark-soliton model}, Phys. Rev. D 75 (2007) 094021.
\newblock \href {http://arxiv.org/abs/hep-ph/0702030} {\path{arXiv:hep-ph/0702030}}, \href {https://doi.org/10.1103/PhysRevD.75.094021} {\path{doi:10.1103/PhysRevD.75.094021}}.

\bibitem{Cebulla:2007ei}
C.~Cebulla, K.~Goeke, J.~Ossmann, P.~Schweitzer, {The Nucleon form-factors of the energy momentum tensor in the Skyrme model}, Nucl. Phys. A 794 (2007) 87--114.
\newblock \href {http://arxiv.org/abs/hep-ph/0703025} {\path{arXiv:hep-ph/0703025}}, \href {https://doi.org/10.1016/j.nuclphysa.2007.08.004} {\path{doi:10.1016/j.nuclphysa.2007.08.004}}.

\bibitem{Jung:2013bya}
J.-H. Jung, U.~Yakhshiev, H.-C. Kim, {Energy\textendash{}momentum tensor form factors of the nucleon within a \ensuremath{\pi}\textendash{}\ensuremath{\rho}\textendash{}\ensuremath{\omega} soliton model}, J. Phys. G 41 (2014) 055107.
\newblock \href {http://arxiv.org/abs/1310.8064} {\path{arXiv:1310.8064}}, \href {https://doi.org/10.1088/0954-3899/41/5/055107} {\path{doi:10.1088/0954-3899/41/5/055107}}.

\bibitem{Pasquini:2014vua}
B.~Pasquini, M.~V. Polyakov, M.~Vanderhaeghen, {Dispersive evaluation of the D-term form factor in deeply virtual Compton scattering}, Phys. Lett. B 739 (2014) 133--138.
\newblock \href {http://arxiv.org/abs/1407.5960} {\path{arXiv:1407.5960}}, \href {https://doi.org/10.1016/j.physletb.2014.10.047} {\path{doi:10.1016/j.physletb.2014.10.047}}.

\bibitem{Burkert:2018bqq}
V.~D. Burkert, L.~Elouadrhiri, F.~X. Girod, {The pressure distribution inside the proton}, Nature 557~(7705) (2018) 396--399.
\newblock \href {https://doi.org/10.1038/s41586-018-0060-z} {\path{doi:10.1038/s41586-018-0060-z}}.

\bibitem{LHPC:2007blg}
P.~Hagler, et~al., {Nucleon Generalized Parton Distributions from Full Lattice QCD}, Phys. Rev. D 77 (2008) 094502.
\newblock \href {http://arxiv.org/abs/0705.4295} {\path{arXiv:0705.4295}}, \href {https://doi.org/10.1103/PhysRevD.77.094502} {\path{doi:10.1103/PhysRevD.77.094502}}.

\bibitem{Hackett:2023rif}
D.~C. Hackett, D.~A. Pefkou, P.~E. Shanahan, {Gravitational Form Factors of the Proton from Lattice QCD}, Phys. Rev. Lett. 132~(25) (2024) 251904.
\newblock \href {http://arxiv.org/abs/2310.08484} {\path{arXiv:2310.08484}}, \href {https://doi.org/10.1103/PhysRevLett.132.251904} {\path{doi:10.1103/PhysRevLett.132.251904}}.

\bibitem{Voloshin:1982eb}
M.~B. Voloshin, A.~D. Dolgov, {ON GRAVITATIONAL INTERACTION OF THE GOLDSTONE BOSONS}, Sov. J. Nucl. Phys. 35 (1982) 120--121.

\bibitem{Leutwyler:1989tn}
H.~Leutwyler, M.~A. Shifman, {GOLDSTONE BOSONS GENERATE PECULIAR CONFORMAL ANOMALIES}, Phys. Lett. B 221 (1989) 384--388.
\newblock \href {https://doi.org/10.1016/0370-2693(89)91730-9} {\path{doi:10.1016/0370-2693(89)91730-9}}.

\bibitem{Donoghue:1991qv}
J.~F. Donoghue, H.~Leutwyler, {Energy and momentum in chiral theories}, Z. Phys. C 52 (1991) 343--351.
\newblock \href {https://doi.org/10.1007/BF01560453} {\path{doi:10.1007/BF01560453}}.

\bibitem{Hudson:2017xug}
J.~Hudson, P.~Schweitzer, {D term and the structure of pointlike and composed spin-0 particles}, Phys. Rev. D 96~(11) (2017) 114013.
\newblock \href {http://arxiv.org/abs/1712.05316} {\path{arXiv:1712.05316}}, \href {https://doi.org/10.1103/PhysRevD.96.114013} {\path{doi:10.1103/PhysRevD.96.114013}}.

\bibitem{Won:2023ial}
H.-Y. Won, H.-C. Kim, J.-Y. Kim, {Role of strange quarks in the D-term and cosmological constant term of the proton}, Phys. Rev. D 108~(9) (2023) 094018.
\newblock \href {http://arxiv.org/abs/2307.00740} {\path{arXiv:2307.00740}}, \href {https://doi.org/10.1103/PhysRevD.108.094018} {\path{doi:10.1103/PhysRevD.108.094018}}.

\bibitem{ParticleDataGroup:2024cfk}
S.~Navas, et~al., {Review of particle physics}, Phys. Rev. D 110~(3) (2024) 030001.
\newblock \href {https://doi.org/10.1103/PhysRevD.110.030001} {\path{doi:10.1103/PhysRevD.110.030001}}.

\end{thebibliography}

\end{document}